\documentclass[aps,twocolumn,superscriptaddress,showpacs,floatfix]{revtex4}
\usepackage[latin9]{inputenc}
\usepackage{amsmath}
\usepackage{amssymb}
\usepackage{graphicx}
\usepackage{esint}
\usepackage{epstopdf}
\usepackage{braket}
\usepackage{hyperref}
\usepackage{color}

\makeatletter

\begin{document}
\unitlength = 1mm

\title{Time-dependent study of a black-hole laser in a flowing atomic condensate}

\author{J. R. M. de Nova}

\affiliation{Departamento de Física de Materiales, Universidad Complutense de
Madrid, E-28040 Madrid, Spain}
\affiliation{Department of Physics, Technion-Israel Institute of Technology, Technion City, Haifa 32000, Israel}
\author{S. Finazzi}
\affiliation{INO-CNR BEC Center and Dipartimento di Fisica, Universit\`{a} di Trento, I-38123 Povo, Italy}
\affiliation{Laboratoire Mat\'eriaux et Ph\'enom\`enes Quantiques, Universit\'e Paris Diderot-Paris 7 and CNRS,
B\^atiment Condorcet, 10 rue Alice Domon et L\'eonie Duquet, 75205 Paris Cedex 13, France
}
\affiliation{GiViTI Coordinating Center, Laboratory of Clinical Epidemiology, IRCCS - Istituto di Ricerche Farmacologiche Mario Negri, Villa Camozzi, Ranica (BG), Italy}

\author{I. Carusotto}
\affiliation{INO-CNR BEC Center and Dipartimento di Fisica, Universit\`{a} di Trento, I-38123 Povo, Italy}

\begin{abstract}

We numerically study the temporal evolution of a black-hole laser configuration
displaying a pair of black and white hole horizons in a flowing atomic
condensate. This configuration is initially prepared starting from a
homogeneous flow via a suitable space-dependent change of the
interaction constant and the evolution is then followed up to long
times. Depending on the values of the system parameters, the system typically either converges to
the lowest energy solution by evaporating away the horizons or displays a continuous and periodic coherent emission of
solitons. By making a physical comparison with optical laser
devices, we identify the latter regime of continuous emission of solitons as the proper black-hole laser effect. We include some movies of the temporal evolution of the spatial density and velocity profiles in the most significant cases.
\end{abstract}


\date{\today}

\maketitle

\section{Introduction}

Since the pioneering work by Corley and Jacobson~\cite{Corley1999}, gravitational configurations showing a pair of neighboring horizons have attracted a great deal of attention: for a quantum Bose field with a superluminal dispersion, the negative energy partner of the Hawking emission by the outer horizon can bounce on the inner horizon and travel back to the outer one, so to stimulate further Hawking emission. In suitable conditions, this process may give rise to a dynamical instability, the so-called {\em black-hole laser} instability, and then to the exponential growth of a self-amplifying coherent Hawking emission. Most remarkably, this coherent emission is expected to have very different statistical properties from the spontaneous Hawking radiation from a single horizon, which originates from the parametric amplification of zero-point quantum fluctuations and whose two-mode squeezed state nature reduces to a simple thermal state upon tracing out the Hawking partner emission~\cite{QuantumNoise,QuantumOptics,Petruccione,MandelWolf}.

While this intriguing emission seems to have little chances to be observed in astrophysics, it is central to present-day studies of optical and condensed matter analogues of gravitational systems. Such analog models, first proposed by Unruh~\cite{Unruh1981} as table-top systems where to study the analog Hawking emission process, are presently attracting a great interest from both the theoretical and experimental points of view. Among the specific systems under most active study, we can mention nonlinear optical systems~\cite{Belgiorno2010}, surface waves in water tanks~\cite{WeinfurtnerPRL2011,euve2015wave}, quantum fluids of light~\cite{RMPCarusotto2013,nguyenPRL2015} and Bose-Einstein condensates of ultracold atoms~\cite{garay2000sonic,Carusotto2008}. A clear evidence of the presence of a black-hole horizon has been obtained in both fluids of light~\cite{nguyenPRL2015} and atomic gases~\cite{Lahav2010} and strong experimental efforts are presently being devoted to the detection of the spontaneous Hawking emission. A remarkable first experimental evidence of a black-hole laser instability has been recently reported in~\cite{Steinhauer2014} by looking at the fast growth in time of a complex density modulation pattern in between the black- and white-hole horizons.

%Among the different optical and condensed-matter analog models presently under study, a special attention has been devoted to black-hole laser configurations in flowing atomic condensates for which theoretically accessible multi-horizon configurations have been identified along the lines of~\cite{Balbinot2008,Carusotto2008}.
Whereas the first works on black-hole laser configurations in astrophysics~\cite{Corley1999} and in analog models~\cite{Barcelo2006,leonhardt2007a,Coutant2010,Finazzi2010} have focussed on the linear dynamics at early times after the onset of the instability, the most interesting physics occurs at late times when the exponentially growing black-hole laser emission has become strong enough to exert a measurable back-reaction effect onto the horizons. Focusing on flowing atomic condensates for which theoretically accessible multi-horizon configurations were proposed along the lines of~\cite{Balbinot2008,Carusotto2008}, pioneering recent works~\cite{Michel2013,Michel2015} have identified some most remarkable behaviors that the system can take in response to the back-reaction effect. The black-hole laser configuration can quickly relax towards a horizonless sub-sonic flow by ``evaporating'' away the horizons, and, in strongly unstable configurations, spontaneous oscillations can appear and eventually result in the periodic emission of solitons.

In this Article we extend those pioneering works with an extensive campaign of numerical simulations of the long-time dynamics of atomic condensates starting from a black-hole laser configuration with a pair of neighboring black- and white-hole horizons. Our simulations are based on the numerical integration of the one-dimensional time-dependent Gross-Pitaevskii equation describing the condensate dynamics at mean-field level. Besides confirming the qualitative behaviors anticipated in Refs.~\cite{Michel2015}, a thorough classification of the different regimes that the system can display depending on the system parameters is presented and some physical explanation is given for the more complex and intriguing ones.

A special attention is devoted to the spontaneously oscillating regimes where the classical dynamics of the system quickly forgets its initial condition and approaches a limit cycle attractor at late times; as a consequence of these oscillations, solitonic sound waves keep being emitted into the surrounding condensate for indefinite times in a continuous and periodic way. Even though the ``black-hole laser'' expression has been used in the past to refer to different aspects of this physics, a detailed comparison to continuous-wave optical laser devices suggests that this self-oscillating regime is the closest counterpart of laser operation in quantum optics, where the periodic oscillation of the electromagnetic field in a laser cavity leads to the emission of monochromatic radiation~\cite{QuantumNoise,QuantumOptics,Petruccione,MandelWolf}. The clarification of the relation between the laser operation in quantum optics and the black-hole laser phenomenon in gravitation and analog models is another main contribution of this work.

The scheme of the paper is the following. In Sec. \ref{sec:BHLaserintro} we revisit the basic theory of black-hole lasers in Bose-Einstein condensates. In Sec. \ref{sec:numresults} we present our numerical results and we classify the different regimes according to the number of linearly unstable modes in the initial condition (Secs. \ref{subsec:n=0} and Sec. \ref{subsec:n=1}). The continuous soliton emission is then discussed in Sec. \ref{subsec:CES}. Comparison between this mechanism of emission of solitons with the operation of optical laser devices is given in Sec. \ref{sec:LaserComparison}. Conclusions and outlook are finally given in Sec.\ref{sec:conclu}. Technical details about the method used for the numerical calculations are presented in Appendix \ref{app:NumericalScheme}. A summary of the movies and of the parameters used in each of them is given in Appendix \ref{app:Movies}.

%[[Garay citation of solitons. No hair theorem????]] [[Plots??]] [[Atomtronics]]
\section{Black-hole laser configurations}\label{sec:BHLaserintro}

In this section, we briefly revise the basic concepts of the theory of black-hole lasers in flowing Bose-Einstein condensates showing two neighboring horizons of opposite black- and white-hole nature. These concepts are instrumental to physically understand the results of the numerical simulations that will be discussed in the following sections. For further details, we refer the reader to the recent literature on the topic, in particular Refs. \cite{leonhardt2007a,Coutant2010,Finazzi2010,Michel2013,Michel2015}.

\subsection{The system and the theoretical model}
\label{subsec:setup}

We consider a flowing atomic one-dimensional Bose-Einstein condensate~\cite{Pitaevskii2003} near $T=0$ in the so-called 1D mean-field regime \cite{Menotti2002,Leboeuf2001} where the transverse confinement length is much larger than the $\it s$-wave atom-atom scattering length and the transverse wavefunction is not distorted from its Gaussian shape in the absence of interactions. Under these assumptions, the condensate dynamics is accurately described in terms of a macroscopic wave function which evolves according to a time-dependent one-dimensional Gross-Pitaevskii (GP) equation:
\begin{multline}
i\hbar\frac{\partial\Psi(x,t)}{\partial t} = \left[-\frac{\hbar^{2}}{2m}\frac{\partial^2}{\partial x^2} +V(x)+\right. \\ +\left. g(x)|\Psi(x,t)|^{2}\right]\Psi(x,t)\, ,\label{eq:TDGP}
\end{multline}
where we have allowed for a spatial dependence of the external potential $V(x)$ and of the effective one-dimensional atom-atom interaction constant $g(x)$. In experiments, both quantities can be controlled and manipulated using standard tools of atomic physics~\cite{Pitaevskii2003}.

\begin{figure}[!htb]\includegraphics[width=\columnwidth]{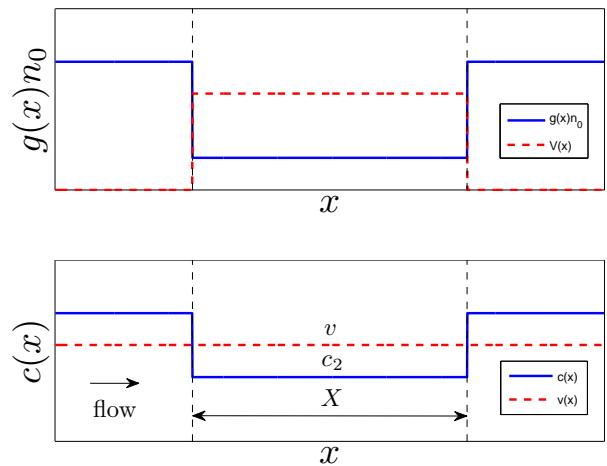}
\caption{Scheme of the theoretical model here considered for studying analog black-hole laser configurations, where the stationary GP wave function is a plane wave of the form $\Psi(x)=\sqrt{n_0}e^{iqx}$. Upper panel: Scheme of the spatial profile of the constant coupling $g(x)$ (solid blue) and the external potential $V(x)$ (dashed red), which are piecewise functions satisfying Eq. (\ref{eq:matchingcondition}). Lower panel: Scheme of the velocity profile. The local speed of sound, $c(x)=\sqrt{g(x)n_0/m}$, (solid blue line) crosses twice the uniform flow velocity, $v(x)=v=\hbar q/m>0$, directed in the rightwards direction (red dashed line). The size of the internal supersonic region between the two horizons is $X$ and the supersonic sound speed is $c_2<v$.}
\label{fig:BHScheme}
\end{figure}

Among the different configurations proposed and/or experimentally realized to implement analog models of gravity in flowing atomic condensates, a theoretically very convenient one (although very idealized from the experimental point of view) was introduced in~\cite{Balbinot2008,Carusotto2008} and then widely used for analytical and numerical studies of black holes and black-hole lasers~\cite{Recati2009,Coutant2010,Finazzi2010,Michel2013,Michel2015}. The idea is to have piecewise constant $V$ and $g$ satisfying at all points the condition
\begin{equation}\label{eq:matchingcondition}
g(x)n_0+V(x)=E_b~,
\end{equation}
$E_b$ being a constant. Under this condition, a solution of the GP equation exists with the condensate keeping at all times a plane wave shape
\begin{equation}
\Psi(x,t)=\sqrt{n_0}e^{iqx}e^{-i\mu t/\hbar}
\label{eq:planewave}
\end{equation}
with a homogeneous density $n_0$, a speed $v=\hbar q/m$, and a chemical potential $\mu=E_b+\hbar^2 q^2/2m$.

As illustrated in Fig. \ref{fig:BHScheme}, we adopt a piecewise homogeneous configuration with $g(x)=g_1$ and $V(x)=V_1$ for $|x|>X/2$ and $g(x)=g_2$ and $V(x)=V_2$ for $|x|<X/2$. The spatial variation of $g(x)$ results in a corresponding variation of the sound speed,
\begin{equation}\label{eq:piecewise}
c(x)=c_{1,2}=\sqrt{\frac{g_{1,2}n_0}{m}}~,
\end{equation}
for respectively $|x|\gtrless X/2$. A black-hole laser configuration appears when the external regions are subsonic and the internal region of size $X$ is supersonic, i.e., $c_2<v<c_1$. In this case, two sonic horizons are in fact formed, the upstream one being located at $x=-X/2$ with a black-hole nature, the downstream one being located at $x=X/2$ with a white-hole nature. The matching condition (\ref{eq:matchingcondition}) imposes that $V_1+g_1n_0=V_2+g_2n_0$.

All our numerical simulations of the GP equation start from the spatially homogeneous plane wave state (\ref{eq:planewave}): in an actual experiment, such an initial condition could be implemented by applying a spatially-selective sudden jump of the interaction constant and of the external potential in the inner region $|x|<X/2$ of an initially homogeneously flowing condensate. A similar configuration has been adopted in recent numerical studies of Hawking radiation from black hole configurations~\cite{Carusotto2008}. Even though this configuration might be extremely challenging to realize in the lab, it allows a simpler theoretical understanding of the underlying physics easier since it eliminates unwanted emission channels as condition (\ref{eq:matchingcondition}) guarantees that the horizons do not generate any deterministic excitations of the condensate, e.g. Bogoliubov-Cherenkov waves~\cite{Carusotto:PRL2006,Tettamanti2016}. For the same reason, we will also assume that the condensate is infinitely extended on either side of the supersonic region.

For computational convenience, hereafter we rescale the wave-function by the homogeneous density $n_0$,  $\Psi(x,t)\rightarrow\sqrt{n_0}\Psi(x,t)$, and adopt units such that $\hbar=m=c_1$. Within this convention, the initial homogeneous condensate density $|\Psi(x,t=0)|^2$ and the sound speed $c_1$ are $1$, the flow velocity is $v=q$ and hence the system is completely characterized by the three values of $(c_2,v,X)$.

\subsection{Unstable modes}
\label{subsec:unstablemodes}

To introduce the most relevant features of the system investigated in this Article and to facilitate the interpretation of the outcome of our numerical simulations, we will give a quick summary of the results of Refs.~\cite{Michel2013,Michel2015}. In this subsection, we briefly describe the structure of the (linear) dynamical instabilities governing the transient dynamics during the initial stages of the black-hole lasing. In the following subsection we shall present the structure of the different nonlinear stationary states of the GP equation for the given spatial profiles of $g(x)$ and $V(x)$. These states play a central role in determining the long time behavior of our system.

To study the dynamical stability of the initial plane wave stationary configuration, one can consider linear perturbations of the wave-function around the stationary plane wave solution,
\begin{equation}
\Psi(x,t)=\left[1+\delta\Psi(x,t)\right]e^{ivx}e^{-i\mu t},
\end{equation}
whose dynamics is described by the Bogoliubov-de Gennes (BdG) equations~\cite{Pitaevskii2003}. In our piecewise homogeneous geometry, the solutions of the BdG equations for the perturbation $\delta \Psi(x,t)$ can be written within each subsonic or supersonic region as linear combinations of plane waves satisfying in the laboratory frame the Bogoliubov dispersion relation
\begin{equation}\label{eq:dispersionrelation}
(\omega-vk)^2=c^2_ik^2+\frac{k^4}{4},~i=1,2
\end{equation}
For a given frequency $\omega$, the global solutions to the BdG equations are obtained by matching the plane-wave solutions at the boundaries between the external and internal regions.

While single-horizon configurations of both black- and white-hole nature reduce to a scattering problem on a stable horizon~\cite{Barcelo2006,Recati2009,Mayoral2010}, the physics of two-horizon configurations is much more intriguing \cite{Barcelo2006,Coutant2010,Finazzi2010,Michel2013,Michel2015}: if the size $X$ of the supersonic region is sufficiently large, dynamical instabilities occur, signalled by the appearance of solutions with a complex frequency $\gamma_n=\omega_n+i\Gamma_n$ with a growth rate $\Gamma_n>0$. The appearance of such dynamically unstable modes provides the black-hole laser effect.

As $X$ increases further, more and more unstable modes appear, labeled by the quantum number $n=0,1,2...$. In particular, the critical length at which a new dynamical instability appears for given $(v,c_2)$ is \cite{Michel2013}:
\begin{equation}\label{Eq:UnstableLength}
X_n=X_0+n\lambda_0
\end{equation}
where
\begin{equation}
X_0=\frac{\arctan\sqrt{\frac{1-v^2}{v^2-c^2_2}}}{\sqrt{{v^2-c^2_2}}} \textrm{\;\;and\;\;} \lambda_0=\frac{\pi}{\sqrt{{v^2-c^2_2}}}.
\end{equation}
The intermediate values $X=X_{n+1/2}$ separate qualitatively different unstable behaviors: while for $X_n<X<X_{n+1/2}$ the complex frequency $\gamma_n$ is purely imaginary, for $X_{n+1/2}<X<X_{n+1}$ it acquires a finite oscillation frequency $\omega_n\neq 0$. Remarkably, the instability rate drops to $\Gamma_n=0$ right at the transition point $X=X_{n+1/2}$~\cite{Michel2013,Michel2015}.

\subsection{Non-linear stationary solutions}\label{subsec:nonlinearsolutions}

While the short-time dynamics of the system is well captured by the BdG equations, the linear approximation underlying them eventually breaks down at some point after the onset of the instability and one needs to go back to the full time-dependent GP equation (\ref{eq:TDGP}). In particular, a step of crucial importance to understand the long-time behavior of the system is the identification of the different stationary states of the system.

We review now the inhomogeneous non-linear GP stationary solutions that exist for the same piecewise profile of $g(x),V(x)$ considered above. Following Ref. \cite{Michel2013}, we specifically focus on those solutions that asymptotically match the homogeneous plane wave solution, $\Psi(x,t)=e^{ivx}e^{-i\mu t}$. The quest of such particular set of solutions is motivated by the fact that the total energy and number of particles are conserved during the growth of the instabilities and, thus, only these solutions present a finite energy and particle number difference with respect to the initial plane wave. Hence, these solutions are good candidates as ending points of the time evolution of the black-hole laser.

Writing the stationary GP solution in the form
\begin{equation}\label{eq:nonlinearsolutions}
\Psi(x,t)=\Psi(x)e^{-i\mu t}\;\; \textrm{with}\:\: \Psi(x)=A(x)e^{i\phi(x)}
\end{equation}
gives us two equations for respectively the phase and the amplitude:
\begin{eqnarray}\label{eq:EquationAmplitude}
J & =  &A^2(x)\phi'(x)\\
\nonumber \mu A(x)&=&-\frac{A''(x)}{2}+\frac{J^2}{2A^3(x)}+ \\ &+&V(x)A(x)+g(x)A^3(x)\, ,\label{eq:EquationPhase}
\end{eqnarray}
where the conserved current $J$ is constant and the prime symbol $'$ denotes spatial derivative. For inhomogeneous solutions, we can define local flow and sound velocities, given by $v(x)\equiv\phi'(x)$ and $c(x)\equiv\sqrt{g(x)\rho(x)}$, with $\rho(x)=A^2(x)$ the density. As the solution is asymptotically a plane wave, one has $J=v$.

Within each (subsonic or supersonic) region, the coupling constant and the potential are homogeneous and we can integrate (\ref{eq:EquationAmplitude}-\ref{eq:EquationPhase}) to obtain:
\begin{equation}\label{eq:MechanicalEnergyConservation}
\frac{A'^2}{2}+\frac{v^2}{2A^2}-g_i\frac{A^4}{2}+\mu_i A^2=\frac{C_i}{2}
\end{equation}
where $C_i$ is a conserved quantity for each $i=1,2$ region and $\mu_i\equiv\mu-V_i$, with $V_i$ the constant value of the external potential in the region $i$. In our units, $g_i=c_i^2$ and $\mu_i=g_i+v^2/2$, see Eq. (\ref{eq:matchingcondition}). In particular, we have $g_1=1$, $\mu_1=1+v^2/2$ and $C_1=1+2v^2$.

\begin{figure}[!htb]\includegraphics[width=\columnwidth]{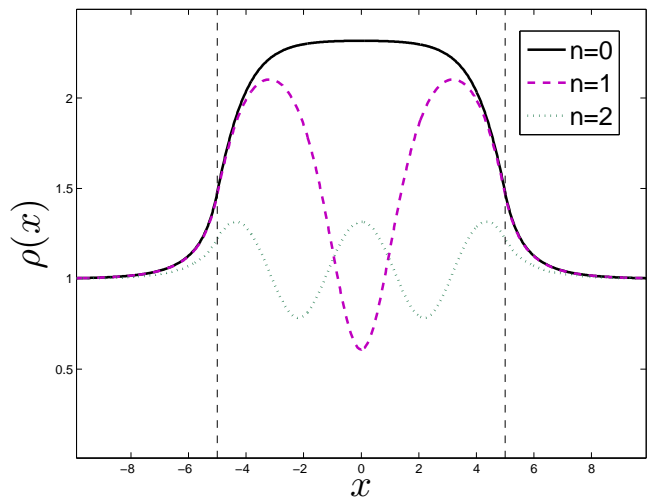}
\caption{Density profile of the $n=0,1,2$ nonlinear stationary solutions of the GP equation, Eq. (\ref{eq:groundnonlinearstates}), for parameters $v=0.9$, $c_2=0.5$ and $X=10$. Dashed vertical black lines represent the limits of the internal supersonic region.}
\label{fig:NonLinearSols}
\end{figure}

\begin{figure*}[htb]
\begin{tabular}{@{}ccc@{}}
    \includegraphics[width=0.6\columnwidth]{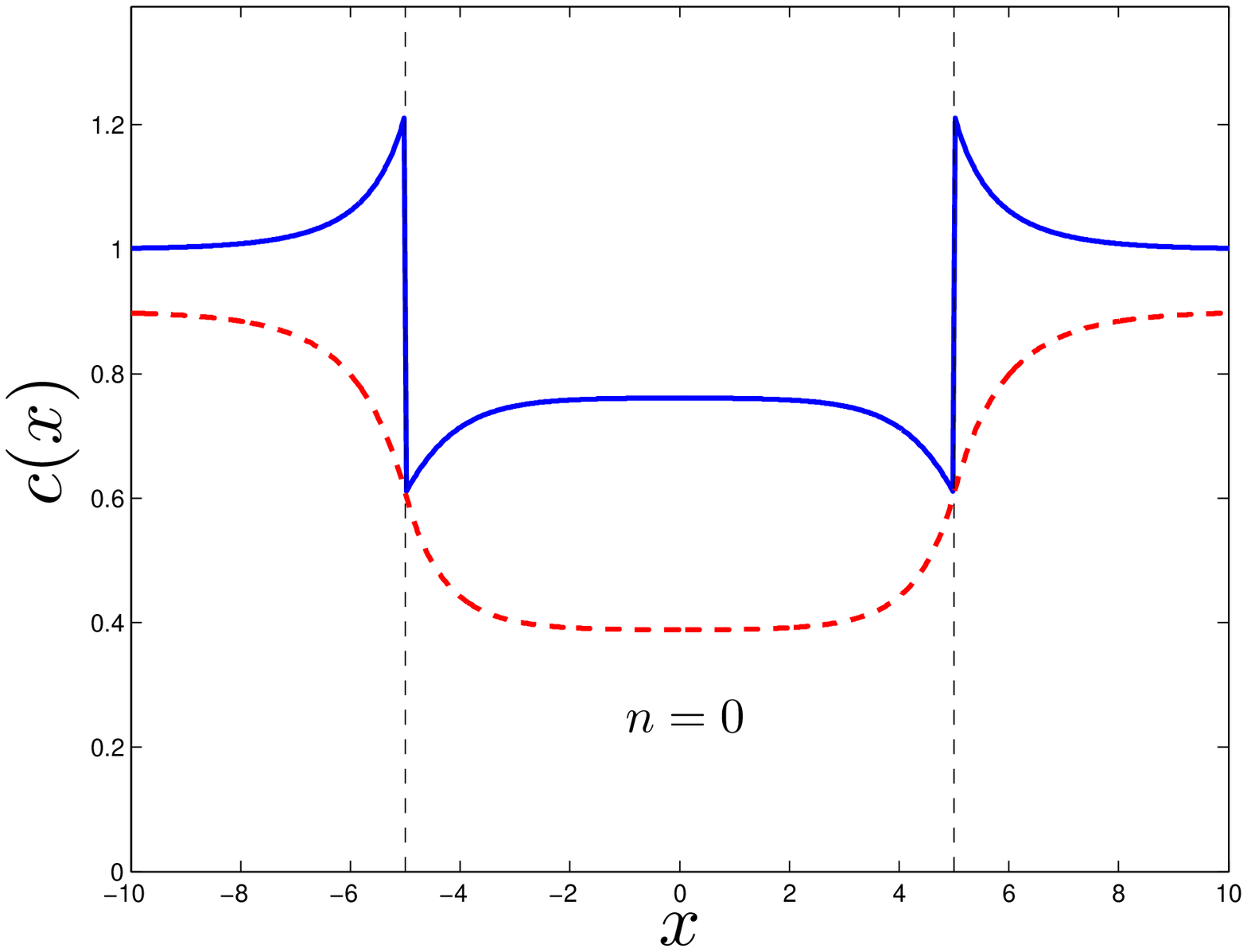} &
    \includegraphics[width=0.6\columnwidth]{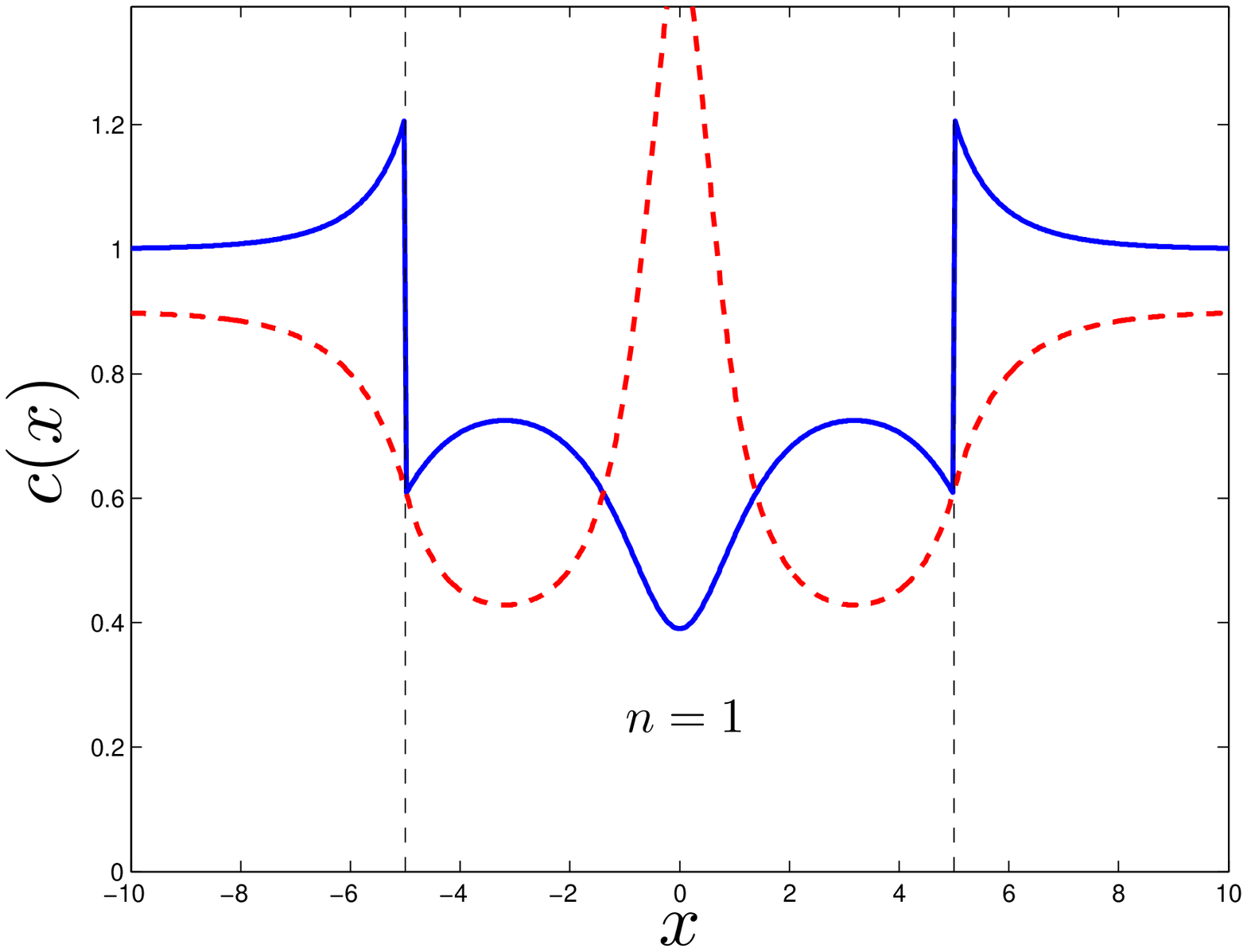} &
    \includegraphics[width=0.6\columnwidth]{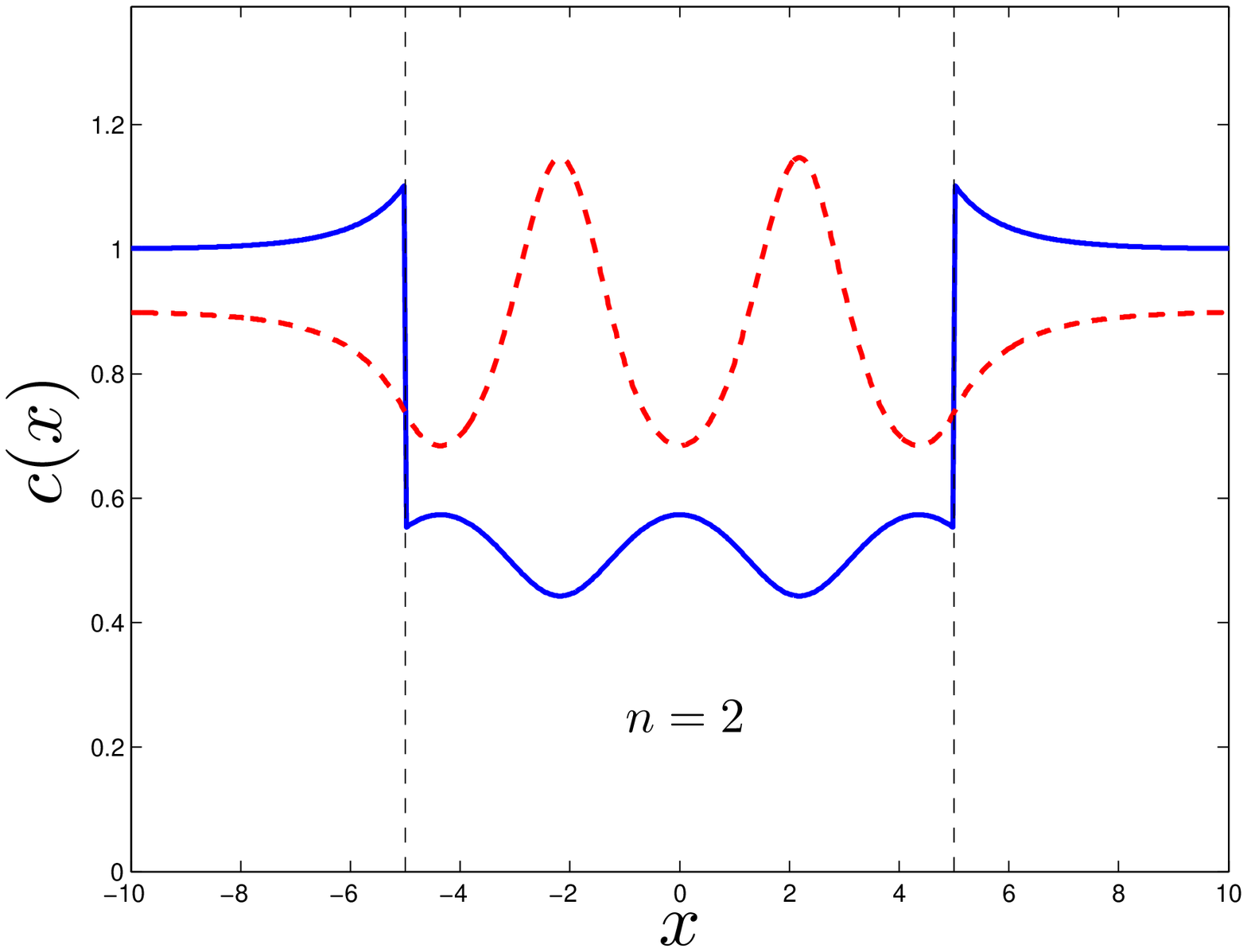}
\end{tabular}
\caption{Sound (solid blue) and flow velocity (dashed red) profiles corresponding to the $n=0,1,2$ stationary GP solutions of Fig. \ref{fig:NonLinearSols}.}
\label{fig:NonLinearCVs}
\end{figure*}

As a criterion to choose among the many possible solutions of these equations, it is reasonable to expect that the initially dynamically unstable system will evolve towards a new stationary solution which minimizes its grand-canonical energy defined as $\Omega=H-\mu N$ in terms of the the Hamiltonian  $H$ and the total number of particles $N$. The expression for the functional $\Omega[\Psi]$ for a given wave function $\Psi$ is:
\begin{equation}\label{eq:grandcanonicalenergy}
\Omega[\Psi]=n_0\int\mathrm{d}x~\frac{1}{2}\partial_x\Psi^*\partial_x\Psi+(V(x)-\mu)|\Psi|^2+\frac{g(x)}{2}|\Psi|^4
\end{equation}
When $\Psi$ is a stationary solution of the form (\ref{eq:nonlinearsolutions}), it is an extreme of the previous functional. In particular, the energy difference of a stationary solution with respect to the initial homogeneous plane wave solution (with energy $\Omega_h$) is
\begin{equation}\label{eq:grandcanonicaldiff}
\Delta \Omega=\Omega[\Psi]-\Omega_h=-n_0\int\mathrm{d}x~\frac{g(x)}{2}\left[A^4(x)-1\right]
\end{equation}
As the density $\rho(x)$ is asymptotically fixed in the external subsonic regions, an increase of the density within the internal $|x|<X/2$ region leads to an overall reduction of the grand-canonical energy. This increase of the density carries a finite increment of the number of particles:
\begin{equation}\label{eq:deltaN}
\Delta N=\int\mathrm{d}x~\left[\rho(x)-1\right]
\end{equation}
In particular, the family of stationary solutions with lowest grand-canonical energy $\Delta\Omega<0$ corresponds to the so-called sh-sh solutions \cite{Michel2013,Michel2015}. They consist of a shadow-soliton solution in the external regions and of elliptic functions in the internal region. Specifically, the stationary GP wave function $\Psi_n(x)$ of these solutions is given by:
\begin{widetext}
\begin{eqnarray}\label{eq:groundnonlinearstates}
\nonumber \Psi_n(x)&=&e^{ivx}e^{-i\Delta\phi}\left(\sqrt{1-v^2}\coth\left[\sqrt{1-v^2}(\Delta x-\frac{X}{2}-x)\right]+iv\right),~ x<-\frac{X}{2}  \\
\Psi_n(x)&=&\Psi^{\text{int}}_n(x),~|x|<X/2\\
\nonumber \Psi_n(x)&=&e^{ivx}e^{i\Delta\phi}\left(\sqrt{1-v^2}\coth\left[\sqrt{1-v^2}(\Delta x-\frac{X}{2}+x)\right]-iv\right),~ x>\frac{X}{2}  \\
\nonumber \Delta x&=&\frac{\coth^{-1}\left(\sqrt{\frac{\rho_m-v^2}{1-v^2}}\right)}{\sqrt{1-v^2}},~\rho_m=\rho(-X/2)=\rho(X/2)
\end{eqnarray}
with the phase $\Delta\phi$ chosen such the wave function is continuous at $x=\pm X/2$. The wave function in the internal region, $\Psi^{\text{int}}_n(x)$, reads
\begin{eqnarray}\label{eq:internalwavefunction}
\Psi^{\text{int}}_n(x)&=&\sqrt{\rho_n^{\text{int}}(x)}e^{i\phi_n^{\text{int}}(x)}\\
\nonumber \rho_n^{\text{int}}(x)&=&r_1+(r_2-r_1)\text{sn}^2(c_2\sqrt{r_3-r_1}x+(n+1)K(\nu),\nu),~\nu=\frac{r_2-r_1}{r_3-r_1},\\
\nonumber \phi_n^{\text{int}}(x)&=&\frac{v}{r_1c_2\sqrt{r_3-r_1}}\left[\Pi\left(\text{am}(c_2\sqrt{r_3-r_1}x+(n+1)K(\nu),\nu),1-\frac{r_2}{r_1},\nu\right)-\Pi\left(\text{am}\left((n+1)K(\nu),\nu\right),1-\frac{r_2}{r_1},\nu\right)\right]
\end{eqnarray}
\end{widetext}
In the above equations, $K(\nu)$ is the complete elliptic integral of the first kind, $\Pi(\phi,\mu,\nu)$ is the incomplete elliptic integral of the third kind and $\text{sn}$ is the corresponding Jacobi elliptic function \cite{Byrd1971,Abramowitz1988}. The quantum number $n=0,1,2\ldots$ characterizes the solution and gives the number of density minima inside the supersonic region. The parameters $0<r_1<r_2<r_3$ are the roots of the following polynomial equation for the density:
\begin{equation}\label{eq:nonlinearroots}
g_2\rho^3-2\mu_2\rho^2+C_2\rho-v^2=0
\end{equation}
The specific value of $C_2$ is obtained by imposing the continuity of the amplitude of wave function and of its derivative at $x=\pm X/2$. On one hand, after using Eq. (\ref{eq:MechanicalEnergyConservation}), we get:
\begin{equation}\label{eq:nonlinearmatching}
\rho_m=1+\sqrt{1-\frac{C_1-C_2}{1-c^2_2}}
\end{equation}
On the other hand, using Eq. (\ref{eq:groundnonlinearstates}), one finds
\begin{equation}\label{eq:constantequation}
X=\frac{2}{c_2\sqrt{r_3-r_1}}\left[(n+1)K(\nu)-\text{sn}^{-1}\left(\sqrt{\frac{\rho_m-r_1}{r_2-r_1}},\nu \right)\right]
\end{equation}
This equation gives an implicit equation for $C_2$ as $\rho_m$ and the values of the three roots $r_{1,2,3}$ depend on $C_2$. The minimum length at which a new solution appears is found in the limit $\nu\rightarrow0$, where the Jacobi elliptic functions reduce to the usual trigonometric functions, $\text{sn}(u,\nu)\backsimeq\sin(u)$. Using this relation, it can be easily shown that the minimum length at which the $n$ non-linear solution appears coincides with the onset of the dynamical instability at $X=X_n$, as given by Eq. (\ref{Eq:UnstableLength}): there is therefore a remarkable correspondence between dynamically unstable modes and non-linear solutions of the GP equation, as shown in Ref.~\cite{Michel2013}.

The density profile $\rho(x)=A^2(x)$ of the first few non-linear solutions of Eq. (\ref{eq:groundnonlinearstates}) for a given configuration is plotted in Fig. \ref{fig:NonLinearSols}. The grand-canonical energies of these solutions satisfy $\Delta \Omega_0<\Delta \Omega_1<\Delta \Omega_2\ldots<0$. As first anticipated in Refs. \cite{Michel2013,Michel2015} by studying linearized fluctuations around the stationary state, only the $n=0$ solution is dynamically stable while all other solutions are dynamically unstable: a time-dependent integration of the GP equation fully confirms this fact, see the next Section.

A qualitative argument justifying the fact that the $n=0$ solution is the most stable one can be put forward as follows: the density profile of this solution (displayed in solid black line in the left panel of Fig. \ref{fig:NonLinearSols}) shows an accumulation of atoms in the central region at $|x|<X/2$, so that the local sound velocity increases and, according to mass conservation at a fixed current, the flow velocity is reduced. As a result, the supersonic unstable character of the central region is reduced and, for sufficiently large values of $X$, is suppressed so the flow becomes everywhere subsonic, restoring in this way the full dynamical stability of the configuration. We can observe this behavior in the left plot of Fig. \ref{fig:NonLinearCVs}, where the sound and flow velocity profiles for the $n=0$ solution of the left plot are depicted.

The situation is of course completely different for the higher $n>0$ nonlinear solutions shown in the central and right panels of the same figure: the density minima that are present in these solutions in the central ($|x|<X/2$) region always correspond to a locally supersonic flow; in particular, for this specific configuration, the $n=2$ solution represented in the rightmost panel is completely supersonic in the whole central region. As usual in hydrodynamics~\cite{Frisch:PRL1992,Hakim1997,Pavloff:PRA2002}, the presence of such localized super-sonic flow regions in combination with spatial inhomogeneities is typically responsible for dynamical instabilities of the flow~\cite{Frisch:PRL1992,Hakim1997,Pavloff:PRA2002,Abad2015}.

\section{Numerical results} \label{sec:numresults}

After the brief review of the general concepts of black-hole laser physics, we are now able to start presenting our numerical results for the time evolution of the black-hole laser. As mentioned in the previous section, we consider a system with a homogeneous potential and coupling constant and then, at $t=0$, we introduce a sudden spatially-dependent quench of the coupling constant $g(x)$ and the potential $V(x)$ in such a way that they still satisfy Eq. (\ref{eq:matchingcondition}). The corresponding time evolution of the wave function for $t>0$ is governed by the following time-dependent GP equation:
\begin{equation}\label{eq:GPNumerical}
i\frac{\partial \Psi(x,t)}{\partial t}=-\frac{1}{2}\frac{\partial^2 \Psi(x,t)}{\partial x^2}+g(x)(|\Psi(x,t)|^2-1)\Psi(x,t)
\end{equation}
where the constant term $E_{b}$ of Eq. (\ref{eq:matchingcondition}) has been subtracted from the usual time-dependent GP equation (\ref{eq:TDGP}) for numerical convenience. We recall that, in our rescaled units, $E_b=g(x)+V(x)$. In particular, we will concentrate on the configuration sketched in Fig. \ref{fig:BHScheme}, where $g(x)$ and $V(x)$ are piecewise constant functions as described around Eq. (\ref{eq:piecewise}) and the initial condition is taken in the form of the stationary plane wave (\ref{eq:planewave}), $e^{ivx}$. On one hand, these choices allow us to make direct use of the general results reviewed in Secs. \ref{subsec:unstablemodes} and \ref{subsec:nonlinearsolutions}. On the other hand, we also expect that most of the main conclusions of our work should be transferrable to the experimental configuration of~\cite{Steinhauer2014}, at least for times shorter than the actual transit time of the incident condensate on the potential barrier.

To ensure that all the dynamical instabilities are duly triggered in the simulations, a weak random noise is added on top of the plane wave initial wavefunction. The size of the numerical grid is taken sufficiently large to avoid finite size effects, $L_g\sim10^3\gg X$. Undesired reflections from the boundaries of the integration box are further suppressed by implementing a diffusive term at the edges of the grid. This allows us to extend the numerical simulations up to long times $t\sim 10^4$ without suffering from numerical artifacts. In the next subsections, the main features of the time evolution of the system will be characterized as a function of the relevant parameters $(c_2,v,X)$. A more detailed description of the numerical integration scheme and of the implementation of the diffusive term is given in Appendix \ref{app:NumericalScheme}.

\subsection{Long-time stationary state}
\label{subsec:longtimestat}

As a first step, three main regimes can be identified depending on whether $X<X_0(v,c_2)$, $X_0(v,c_2)<X<X_1(v,c_2)$ or $X_1(v,c_2)<X$ with $X_{0,1}$ defined in (\ref{Eq:UnstableLength}).

The first case $X<X_0(v,c_2)$ is trivial as there are no instabilities and the time evolution reduces to the dynamics of the (very weak) noise on top of the stationary plane wave solution. In the following we therefore focus our attention on the other two cases that show the most interesting physics.

Before entering into the details, some preliminary remarks are however in order. First, we note that the limit $c_2\rightarrow0$ is ill-defined, as in this limit the $n=0$ non-linear solution presents an infinite accumulation of particles between the two horizons, which cannot be achieved starting from a finite condensate as that considered in our numerical simulations. We then restrict our simulations to finite values of $c_2$ for which the particle accumulation $\Delta N$ is much smaller than the total number of particles in the system, $\Delta N\ll N$.

As a further tool to facilitate the reader in following our discussion, we supplement the figures presented in the main text of the Article with a few Movies that highlight the main qualitative features of the black-hole laser dynamics. Links to the Movies are included in the main text of the Article, while the system parameters chosen for each of them are summarized in Appendix \ref{app:Movies}.

\subsubsection{Single unstable mode $X_0<X<X_1$}\label{subsec:n=0}

In the window $X_0(v,c_2)<X<X_1(v,c_2)$, only one unstable mode is present, associated with the appearance of the lowest energy $n=0$ nonlinear solution described by (\ref{eq:groundnonlinearstates}), whose energy lies below that of the homogeneous solution. After some transient governed by the unstable mode, we can therefore expect that the system will expel the extra energy in the form of waves propagating to $x\to \pm\infty$ and eventually relax towards this non-linear stationary solution.

\begin{figure*}[!tb]
\begin{tabular}{@{}cccc@{}}
    \includegraphics[width=0.5\columnwidth]{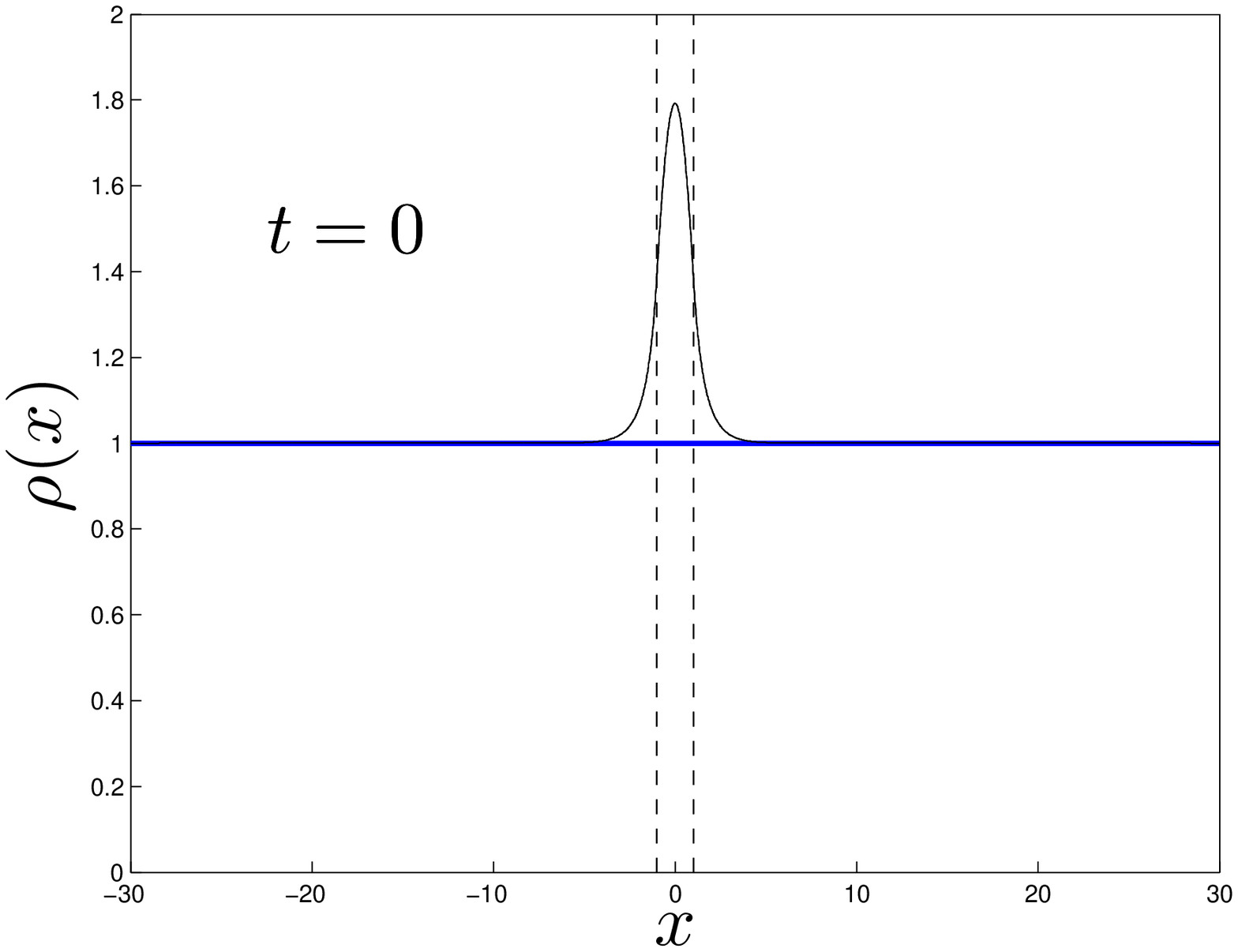} &
    \includegraphics[width=0.5\columnwidth]{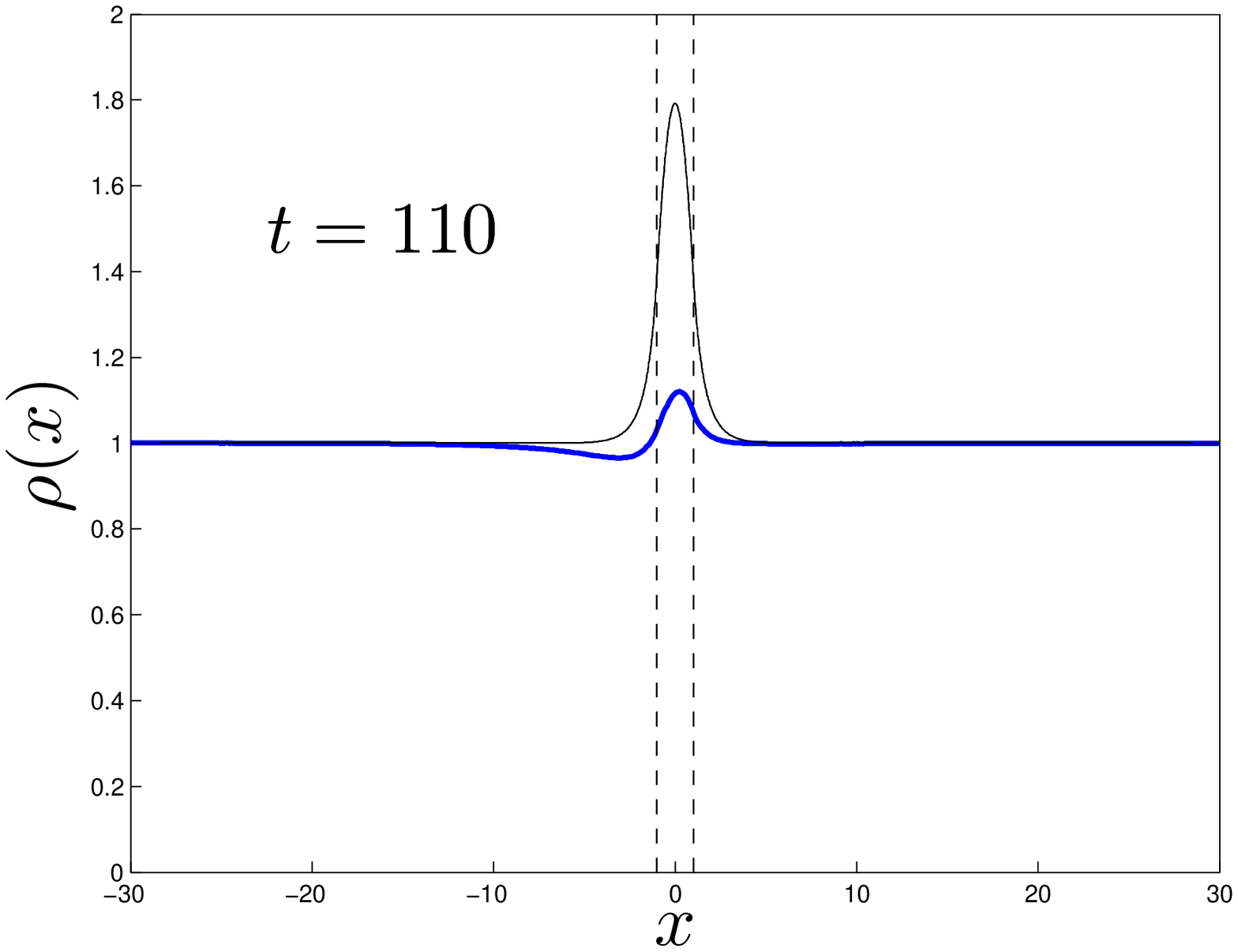} &
    \includegraphics[width=0.5\columnwidth]{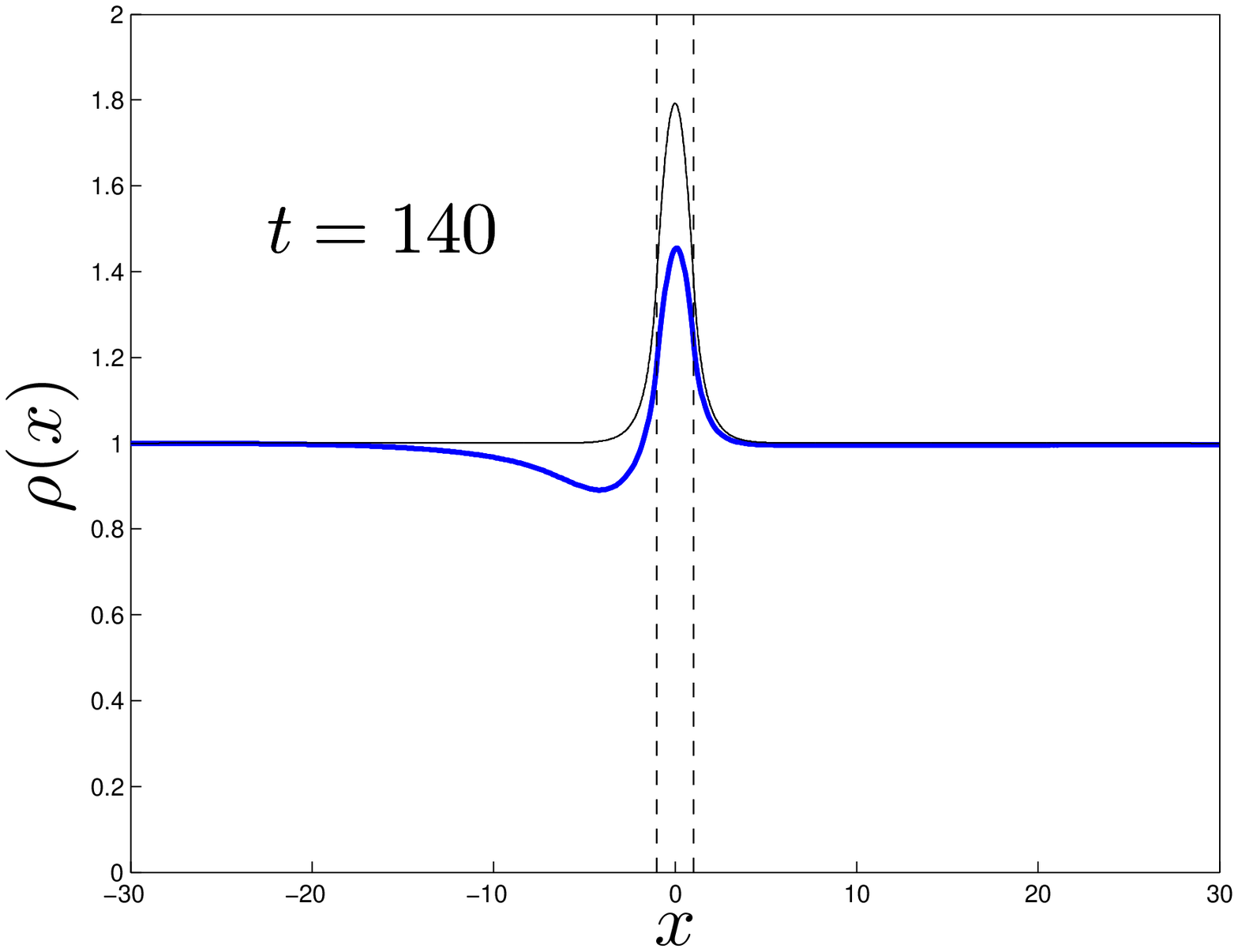} &
    \includegraphics[width=0.5\columnwidth]{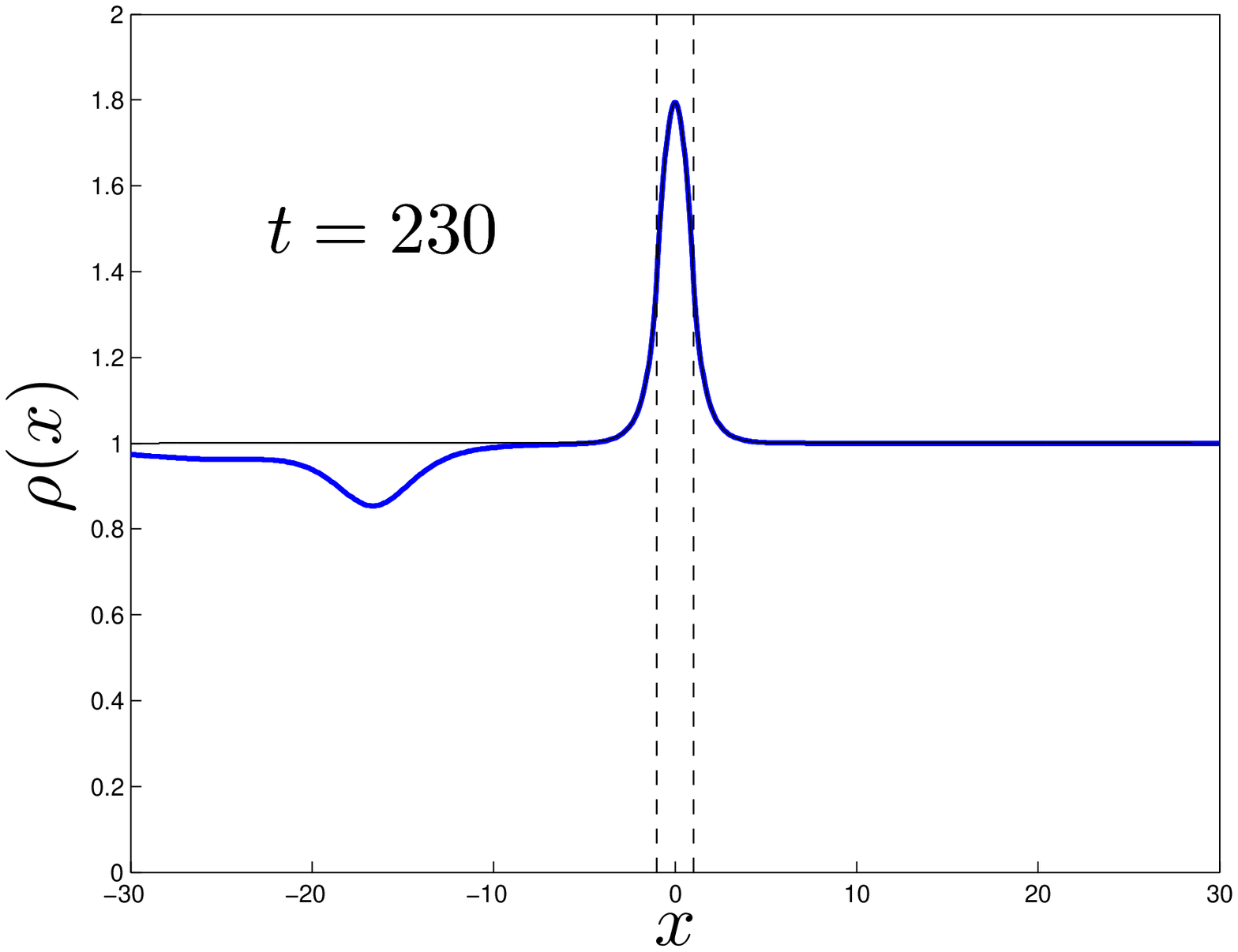} \\[2em]
    \includegraphics[width=0.5\columnwidth]{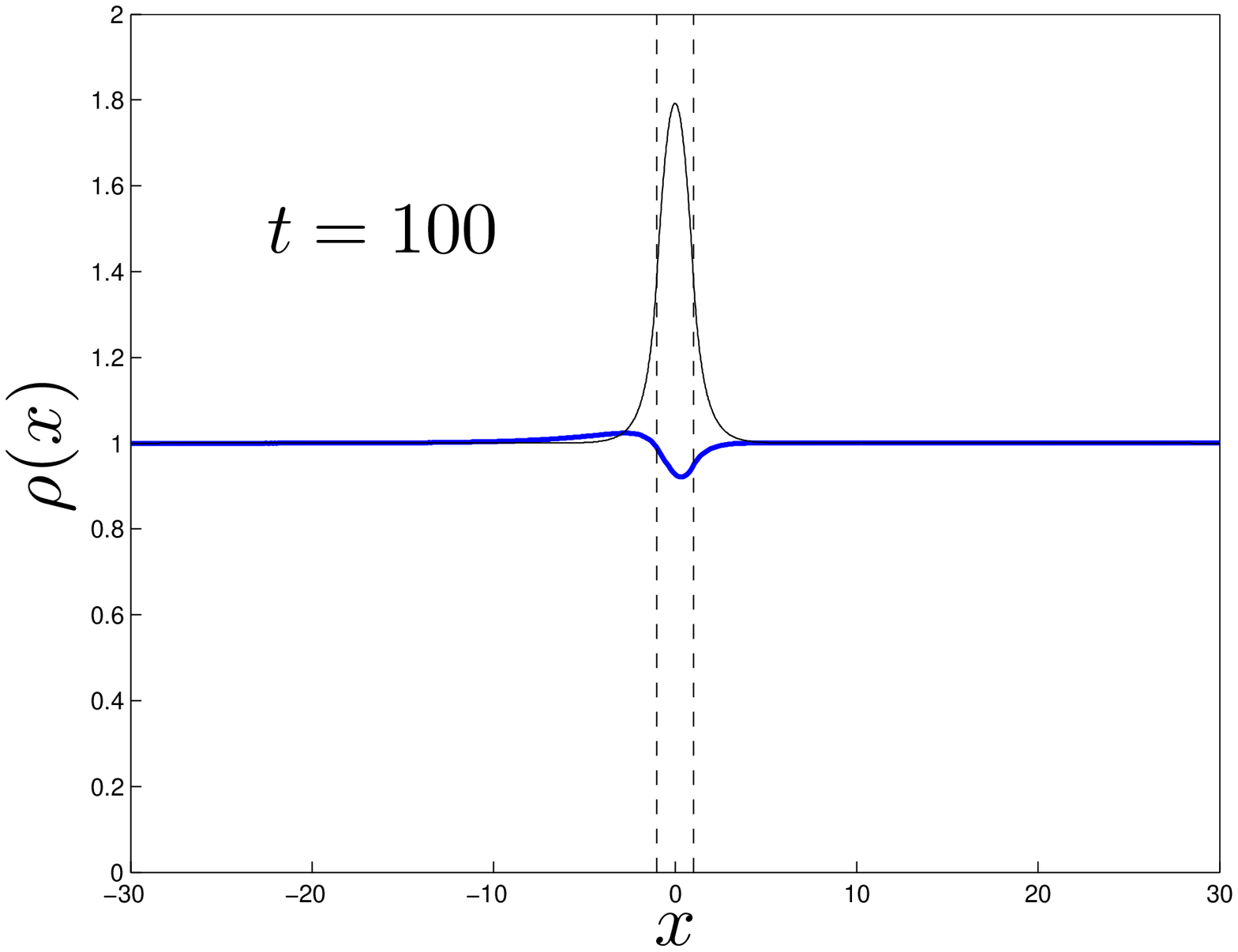} &
    \includegraphics[width=0.5\columnwidth]{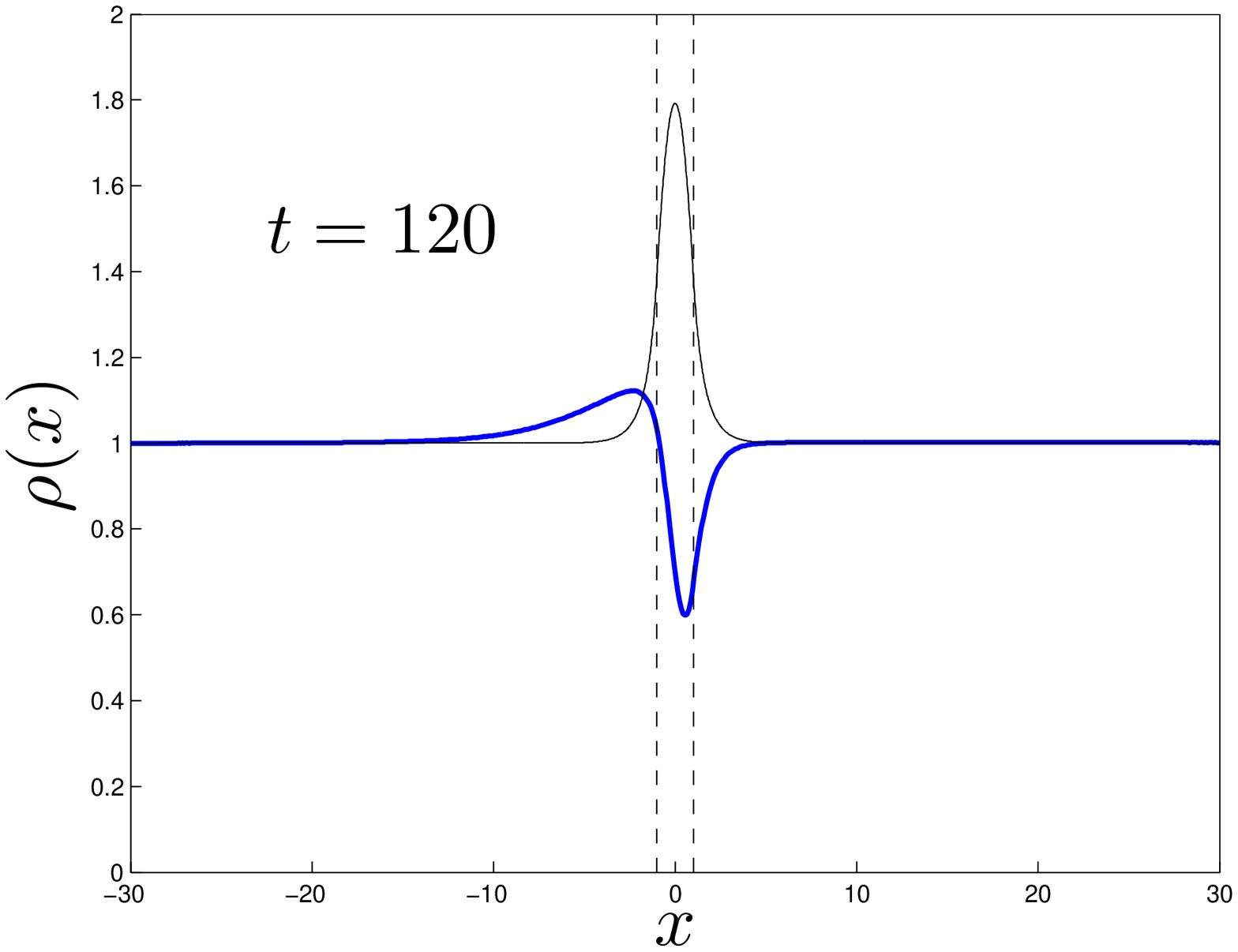} &
    \includegraphics[width=0.5\columnwidth]{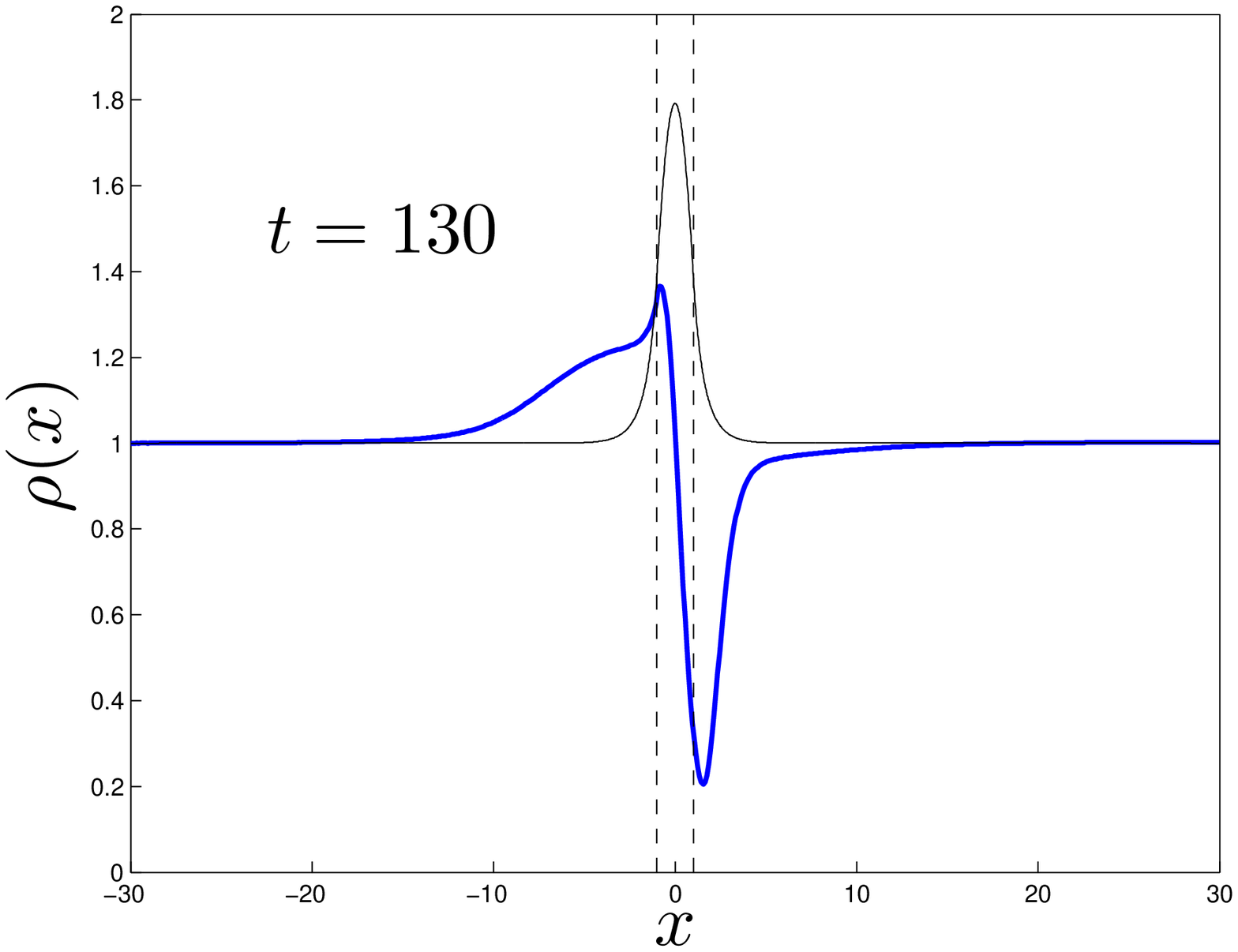} &
    \includegraphics[width=0.5\columnwidth]{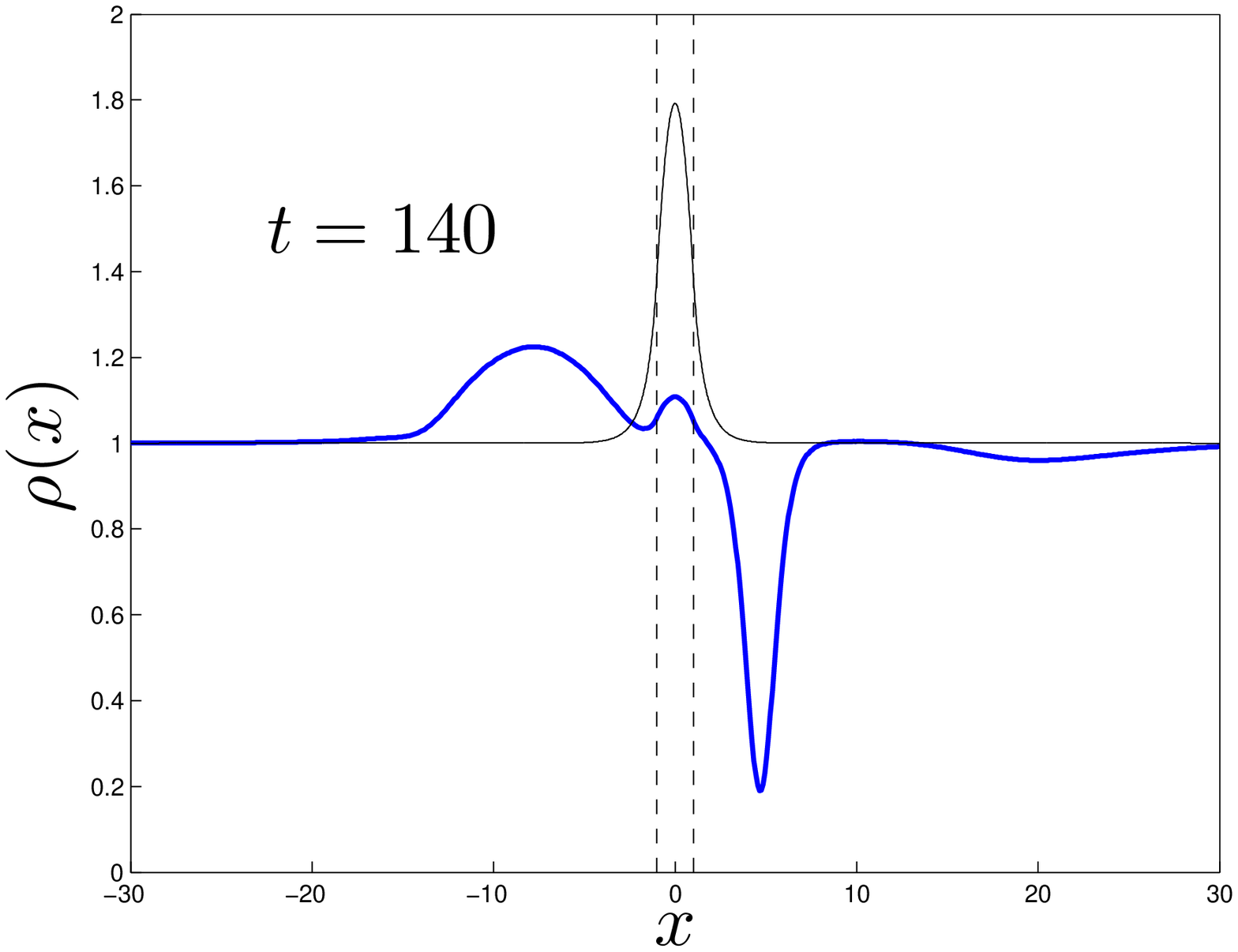} \\
    \includegraphics[width=0.5\columnwidth]{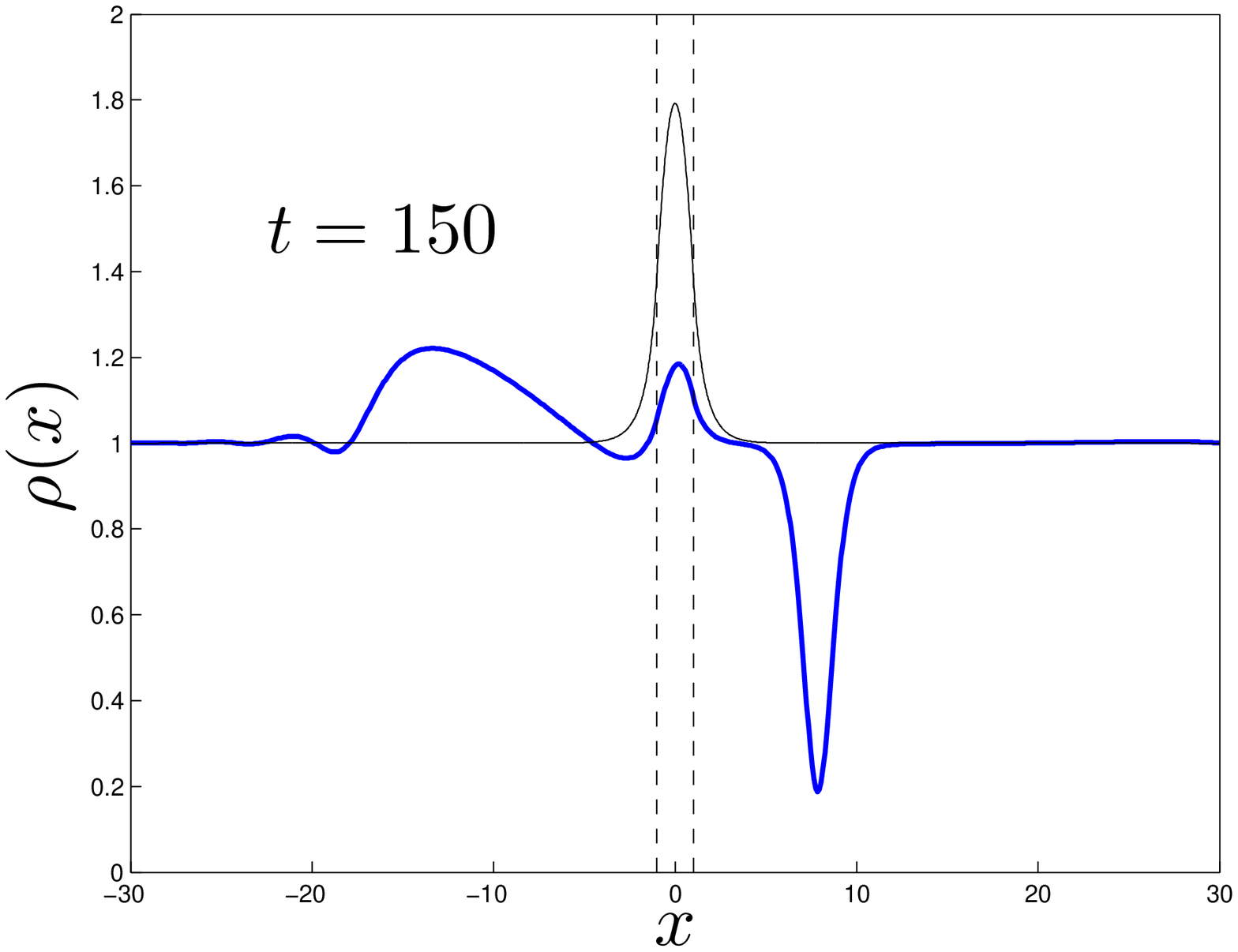} &
    \includegraphics[width=0.5\columnwidth]{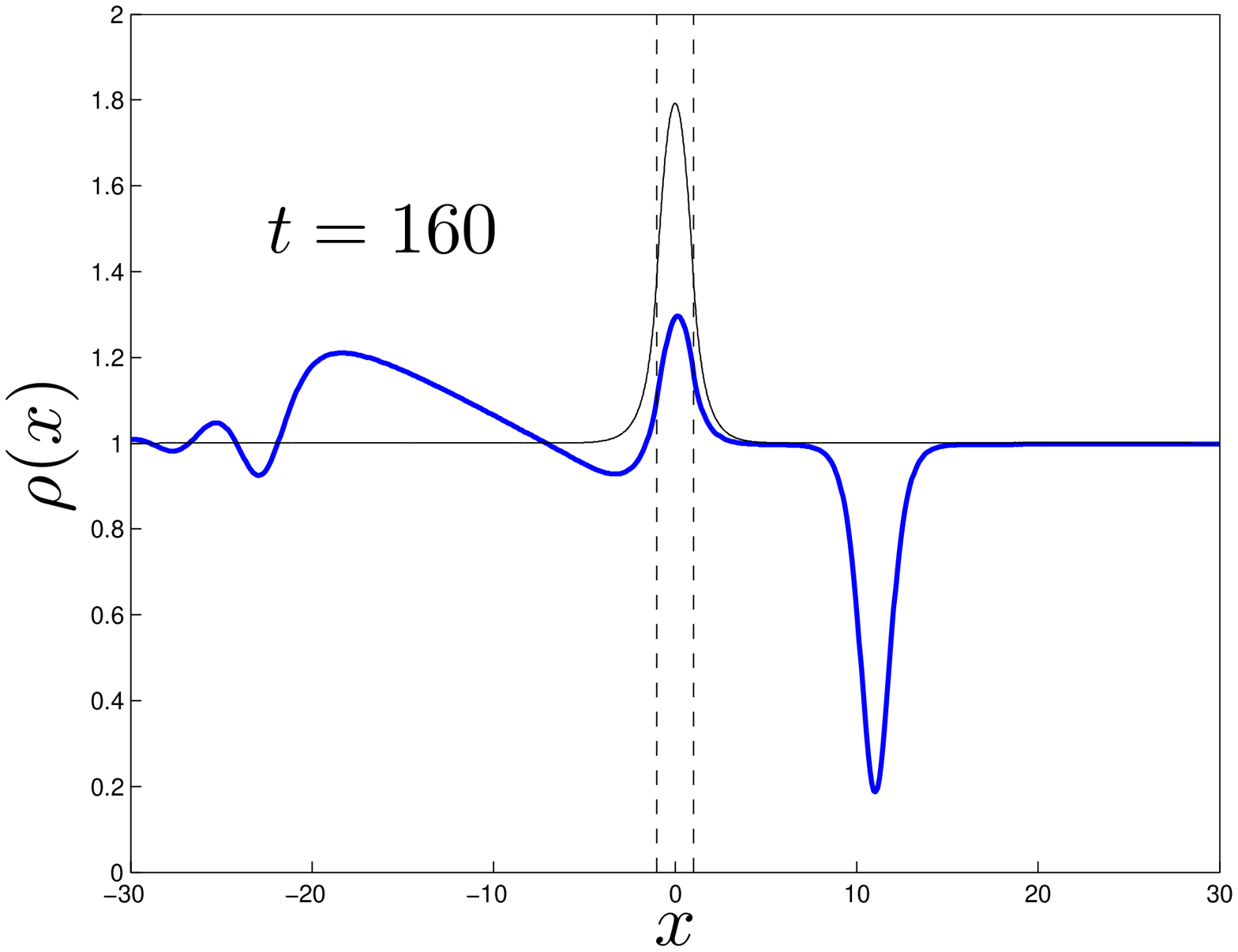} &
    \includegraphics[width=0.5\columnwidth]{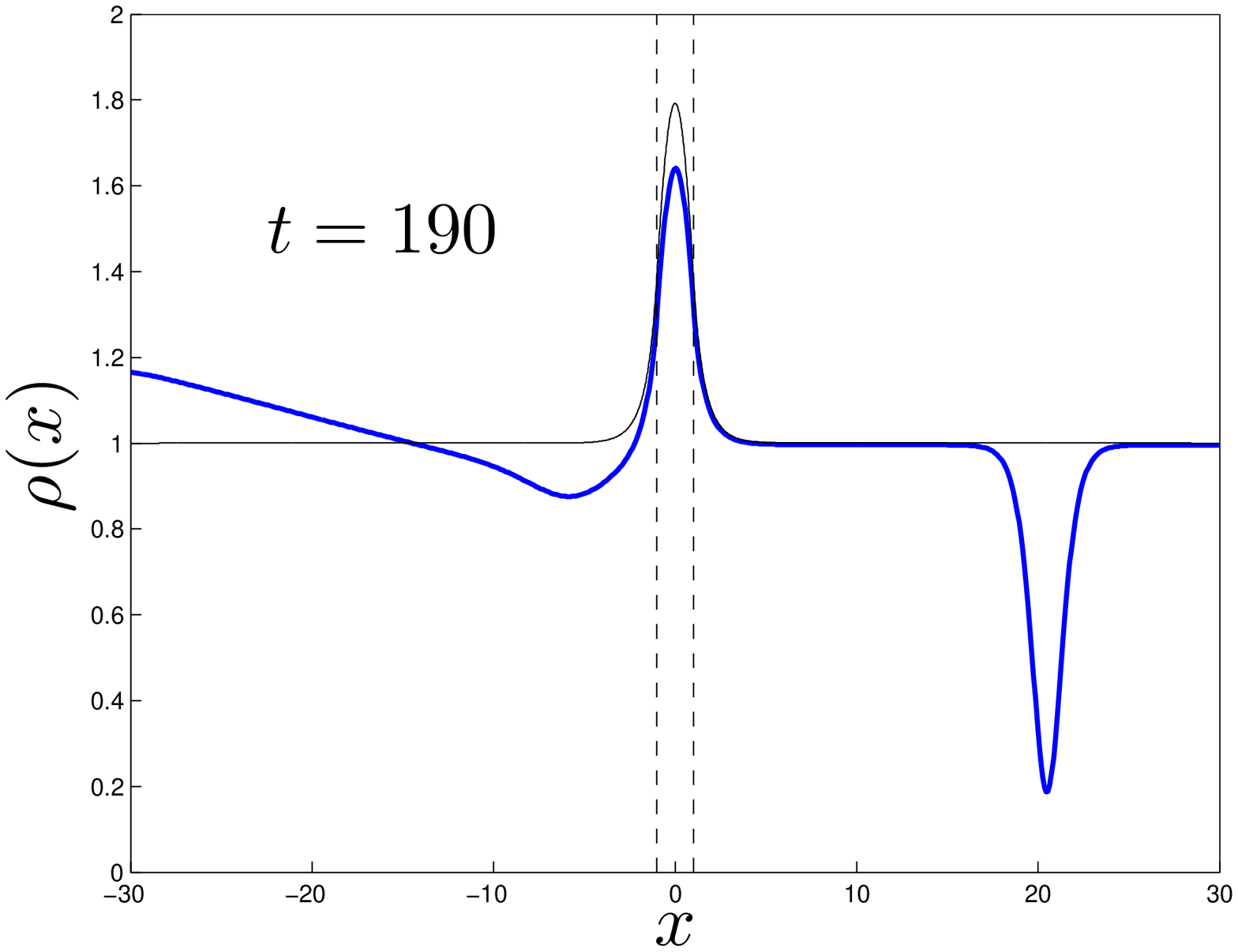} &
    \includegraphics[width=0.5\columnwidth]{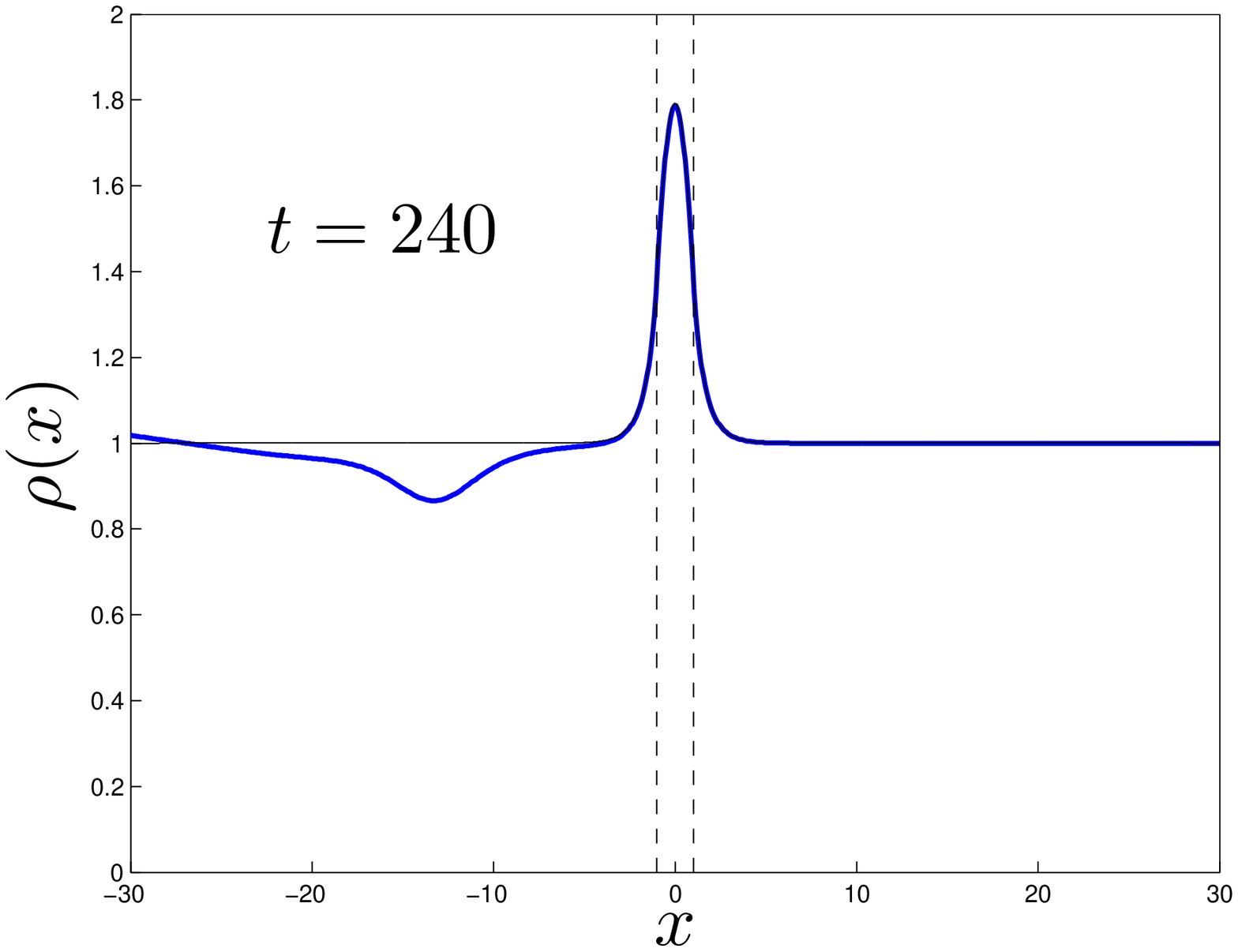} \\
\end{tabular}
\caption{Snapshots of the condensate density (thick solid blue line) at different evolution times (indicated in the panels) for a configuration with $v=0.75$, $c_2=0.3$ and $X=2$ that eventually reaches the $n=0$ stationary solution (thin solid black line). The vertical dashed lines mark the boundaries of the supersonic region. The complete time evolution is shown in \href{https://www.youtube.com/watch?v=gbt3tL9haD4}{Movie 1}. Upper row: evolution of the system when the initial linear instability grows in the direction of the $n=0$. Central and lower row: evolution of the system when the initial linear instability grows in the opposite direction. The complete time evolution is shown in \href{https://www.youtube.com/watch?v=2wPY_D7Aou8}{Movie 2}.}
\label{fig:UpDownMechanism}
\end{figure*}

Whilst this behavior was indeed observed in Ref. \cite{Michel2015}, we have noticed that it is restricted to a certain region in parameter space, corresponding to sufficiently low values of $v$ and high values of $c_2$. Outside this region, the instability of the supersonic region becomes too severe and the convergence towards the $n=0$ stationary solution is replaced by a time-dependent regime of continuous emission of solitons (CES). Importantly, we have numerically observed that the choice between the two behaviors only depends on the system parameters $(c_2,v,X)$ and not on the specific initial configuration of the noise used to trigger the instability. In this subsection, we focus our attention on the first behavior while the detailed characterization of the CES regime and its phase diagram is postponed to Sec. \ref{subsec:CES}.

For given values of $(v,c_2)$ outside the CES window, two cases $X_0(v,c_2)<X<X_{1/2}(v,c_2)$ and $X_{1/2}(v,c_2)<X<X_{1}(v,c_2)$ can be further distinguished as briefly mentioned at the end of Sec.\ref{subsec:unstablemodes}. In the first case, $X_0(v,c_2)<X<X_{1/2}(v,c_2)$, the frequency of the unstable mode is purely imaginary and the mode does not oscillate. As a result, the density shows for times $t\gtrsim1/\Gamma_0$ a monotonic exponential evolution of the form:
\begin{equation}\label{eq:densityzero}
\rho(x,t)\approx1+\alpha_0\delta\rho_0(x)e^{\Gamma_0t}
\end{equation}
where $\delta\rho_0(x)$ is the linear density perturbation corresponding to the unstable mode and $\alpha_0$ its initial amplitude. Depending on the sign of the real number $\alpha_0$, the density will either grow or decrease: as the exponential evolution is triggered by a random noise on top of the stationary solution, the probabilities of the two instances are both equal to $1/2$. This fact has been numerically checked.

When the density initially increases, the system smoothly reaches the $n=0$ ground state solution. The increase of the density in the central, initially supersonic region is associated with a small emission of waves and a small soliton to the upstream region ($x\rightarrow-\infty$) in order to conserve the total number of particles $N$. We can observe this behavior in the upper row of Fig. \ref{fig:UpDownMechanism}, which shows the result of a simulation for a choice of parameters $v=0.75$, $c_2=0.3$ and $X=2$ that falls in the $X_0(v,c_2)<X<X_{1/2}(v,c_2)$ window. The complete time evolution can be observed in \href{https://www.youtube.com/watch?v=gbt3tL9haD4}{Movie 1}.

On the other hand, when the density initially decreases, the system has to emit a larger soliton to the downstream region ($x\rightarrow\infty$) in order to compensate the initial decrease in the particle density. Once the soliton has been emitted, the system is again free to evolve to the $n=0$ solution by locally increasing the density in the central region. This scenario is shown in the central and lower row of Fig. \ref{fig:UpDownMechanism}, which corresponds to a simulation generated with the same parameters $(c_2,v,X)$ but a different configuration of initial noise; the complete evolution can be observed in \href{https://www.youtube.com/watch?v=2wPY_D7Aou8}{Movie 2}.

\begin{figure*}[!tb]
\begin{tabular}{@{}cccc@{}}
    \includegraphics[width=0.5\columnwidth]{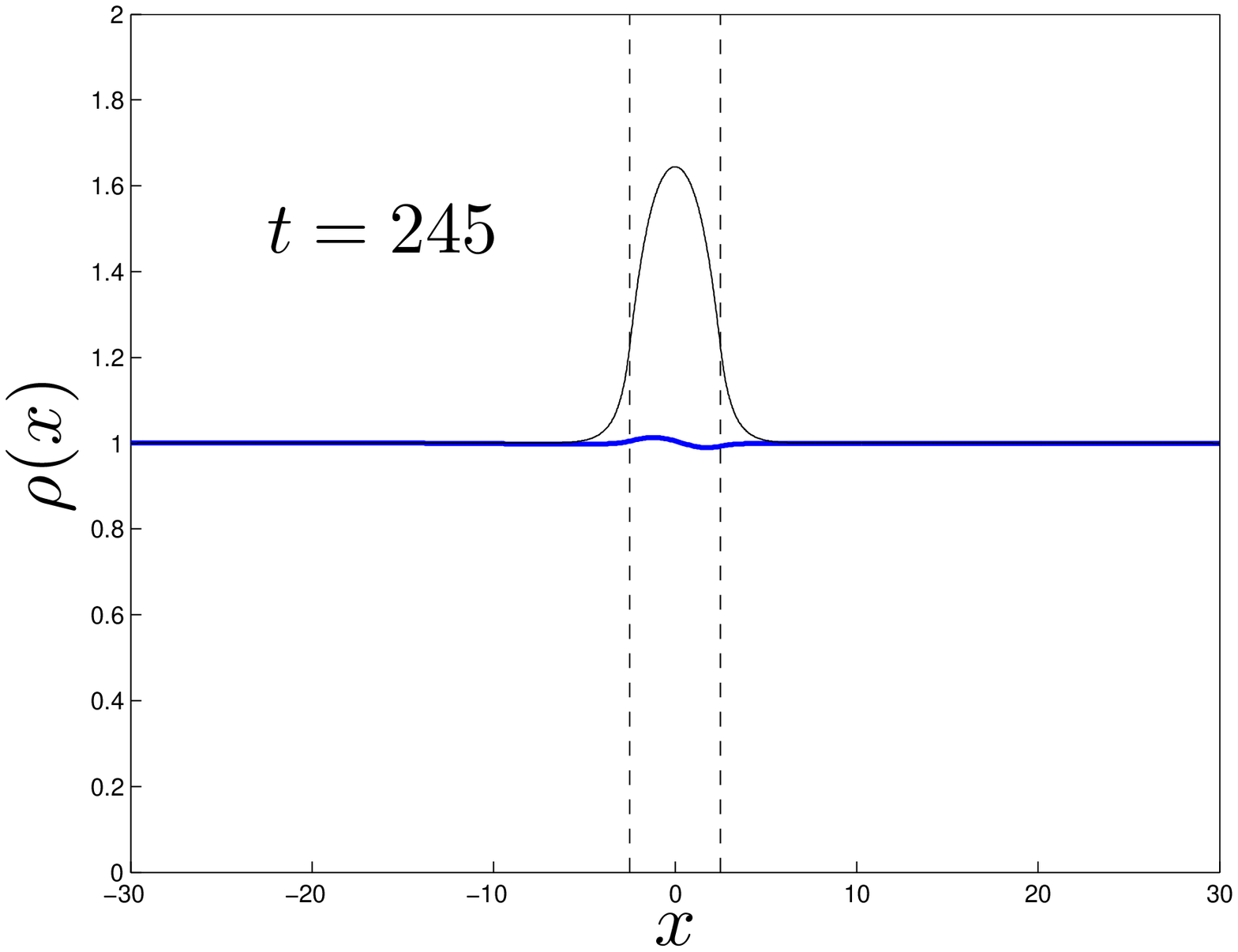} &
    \includegraphics[width=0.5\columnwidth]{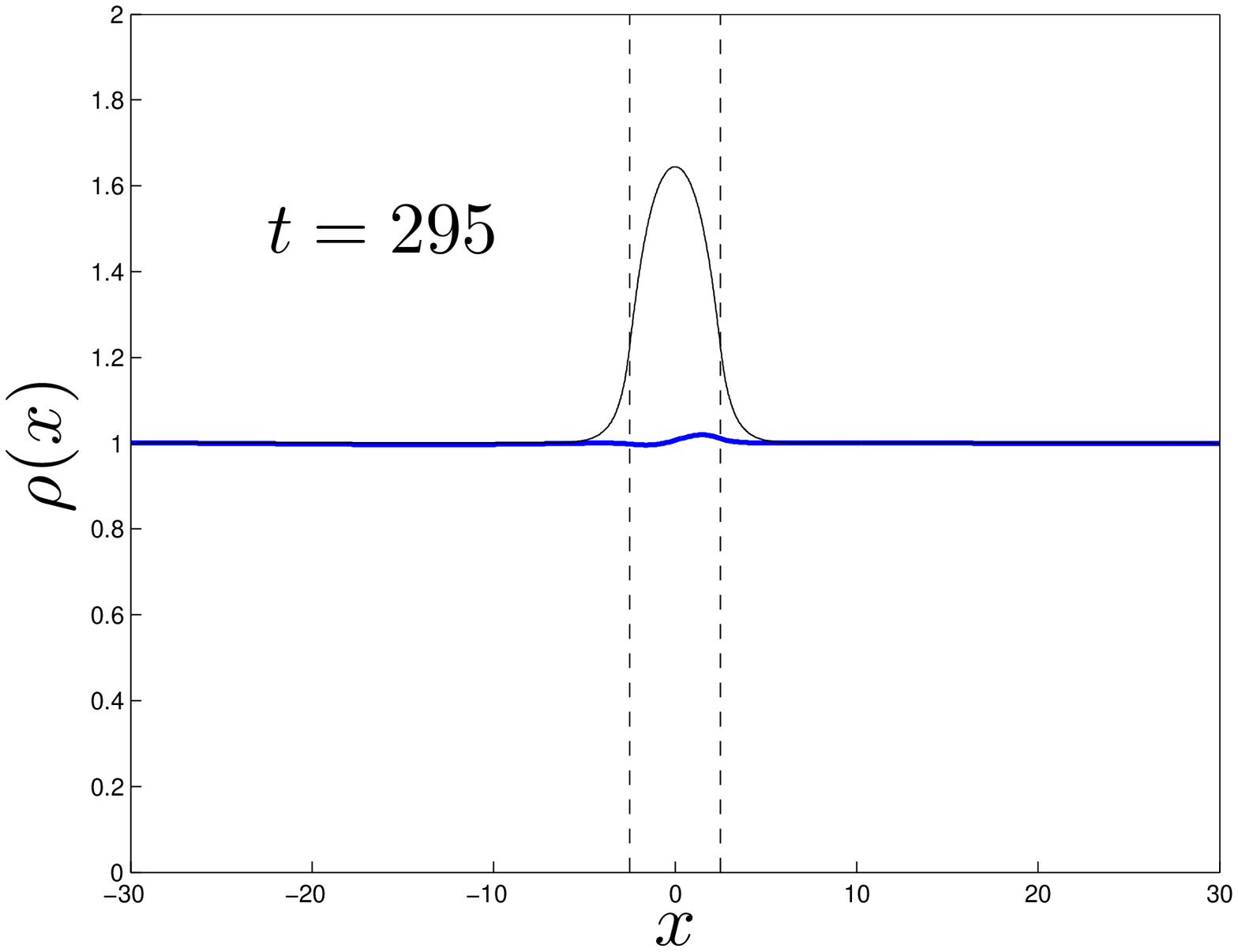} &
    \includegraphics[width=0.5\columnwidth]{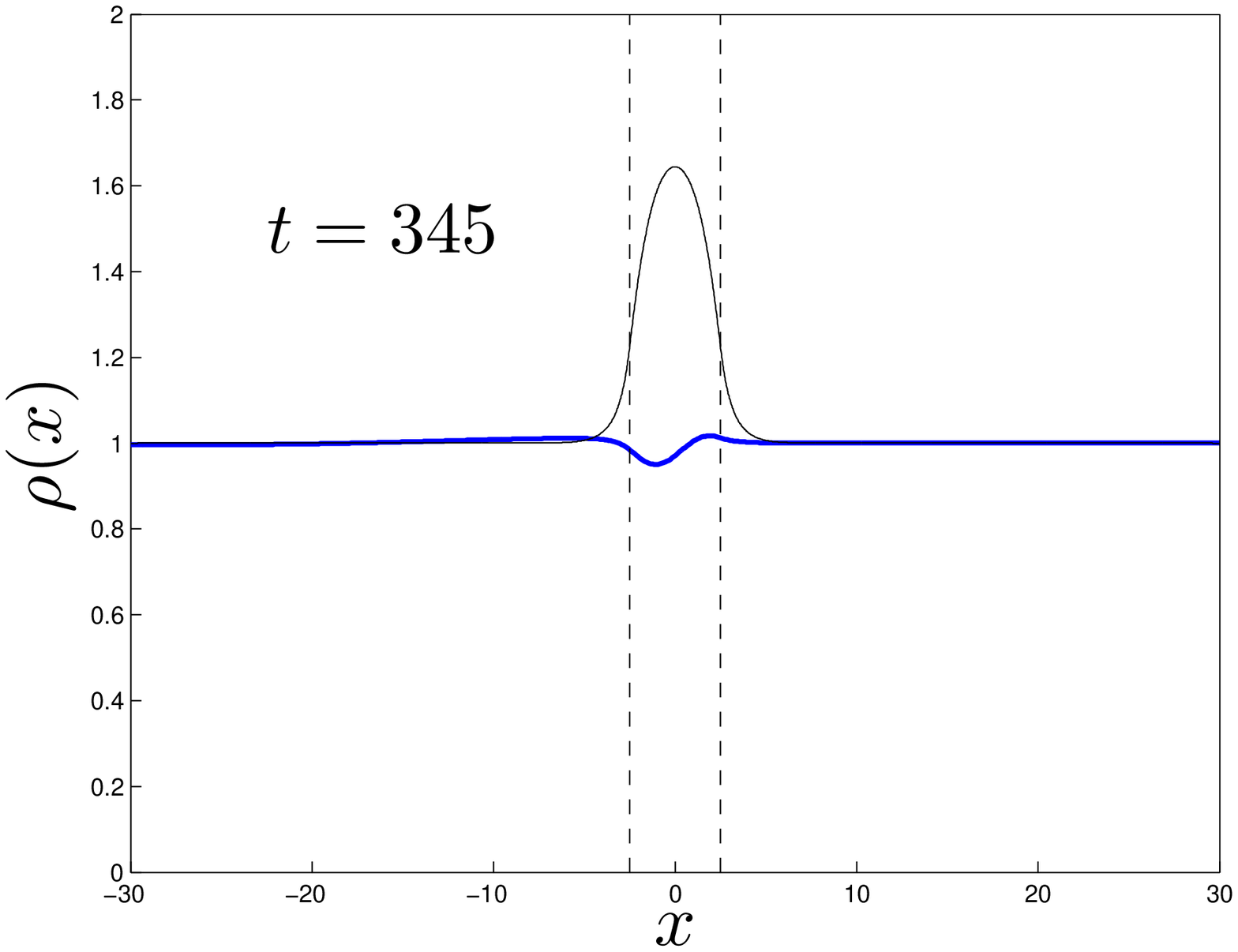} &
    \includegraphics[width=0.5\columnwidth]{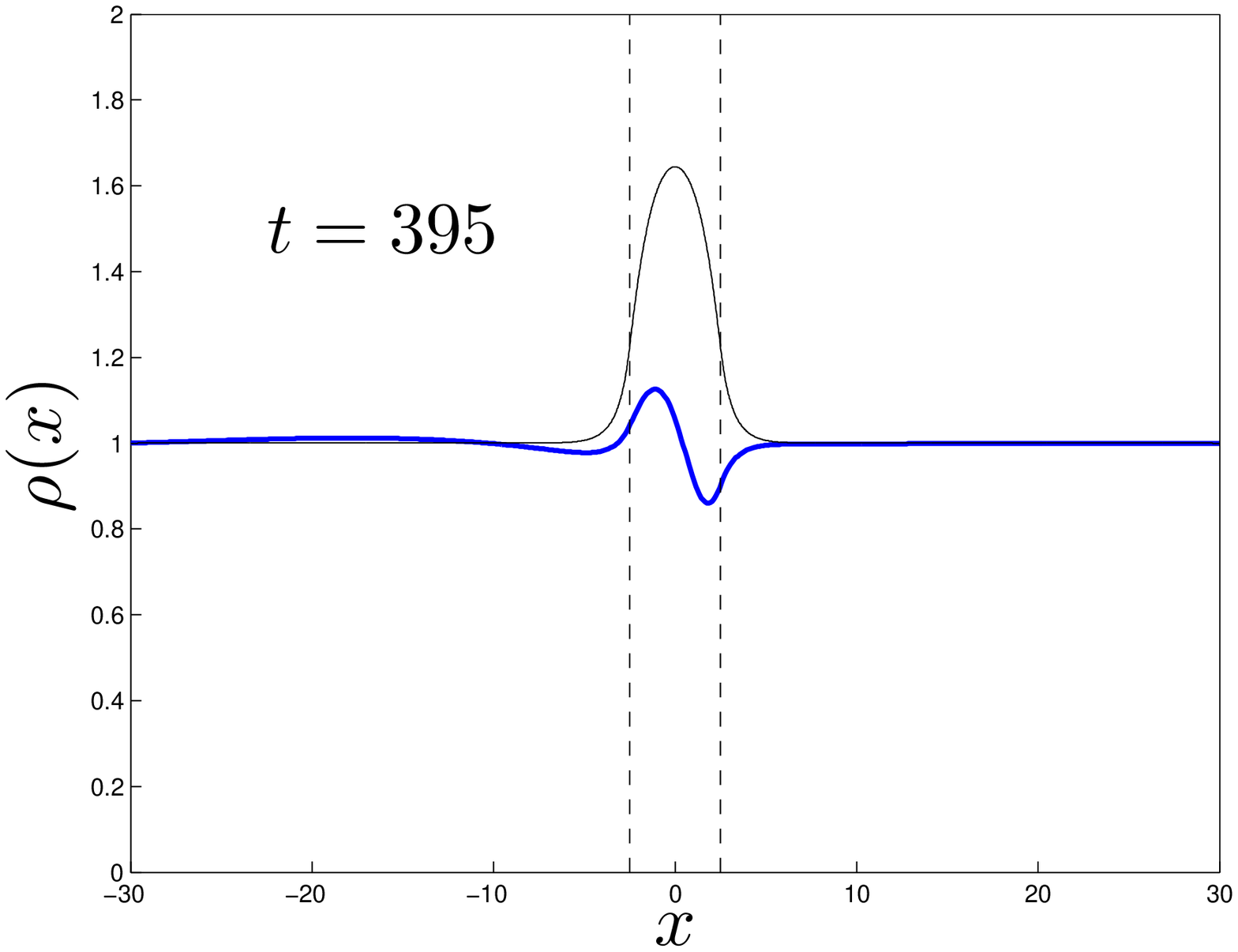} \\
    \includegraphics[width=0.5\columnwidth]{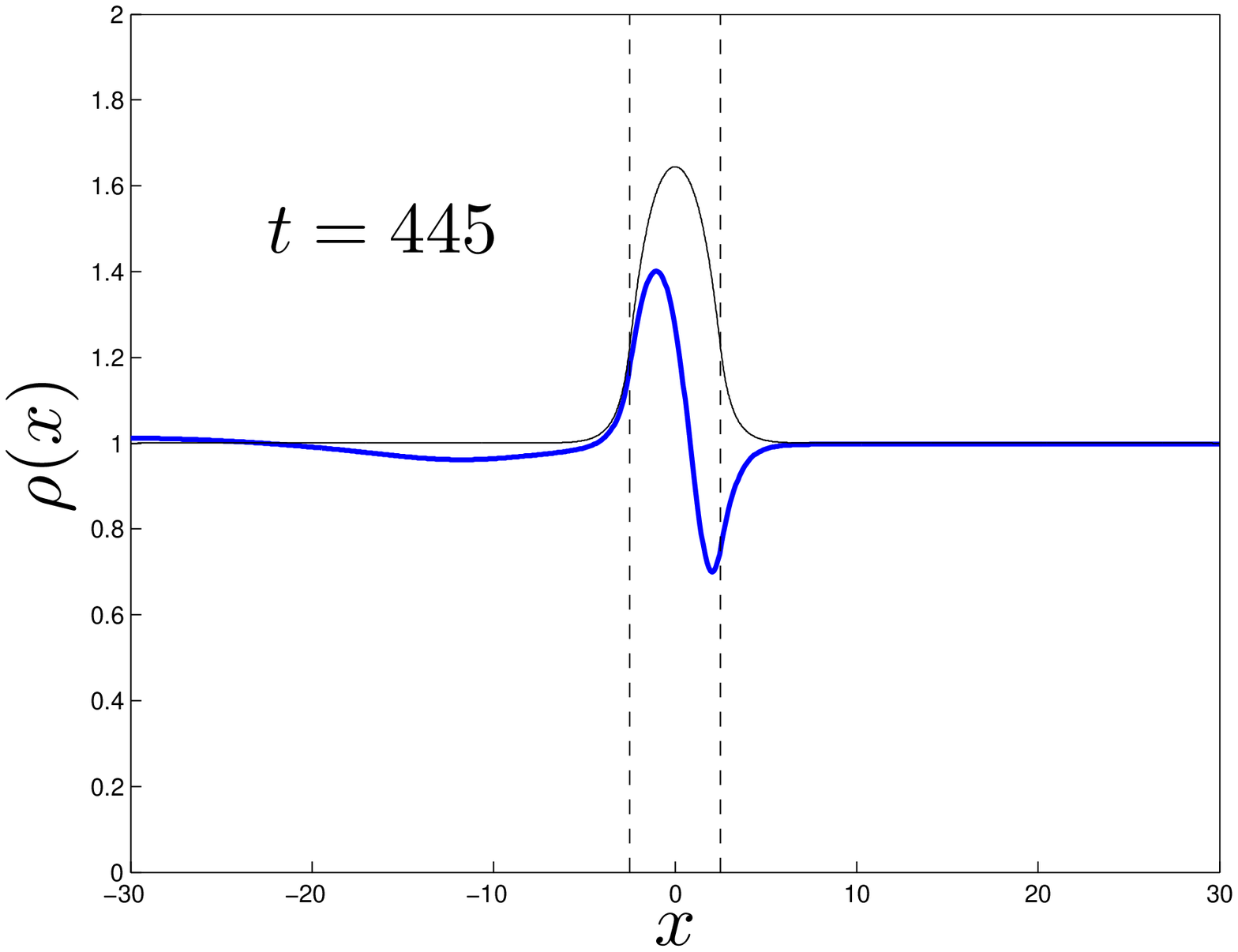} &
    \includegraphics[width=0.5\columnwidth]{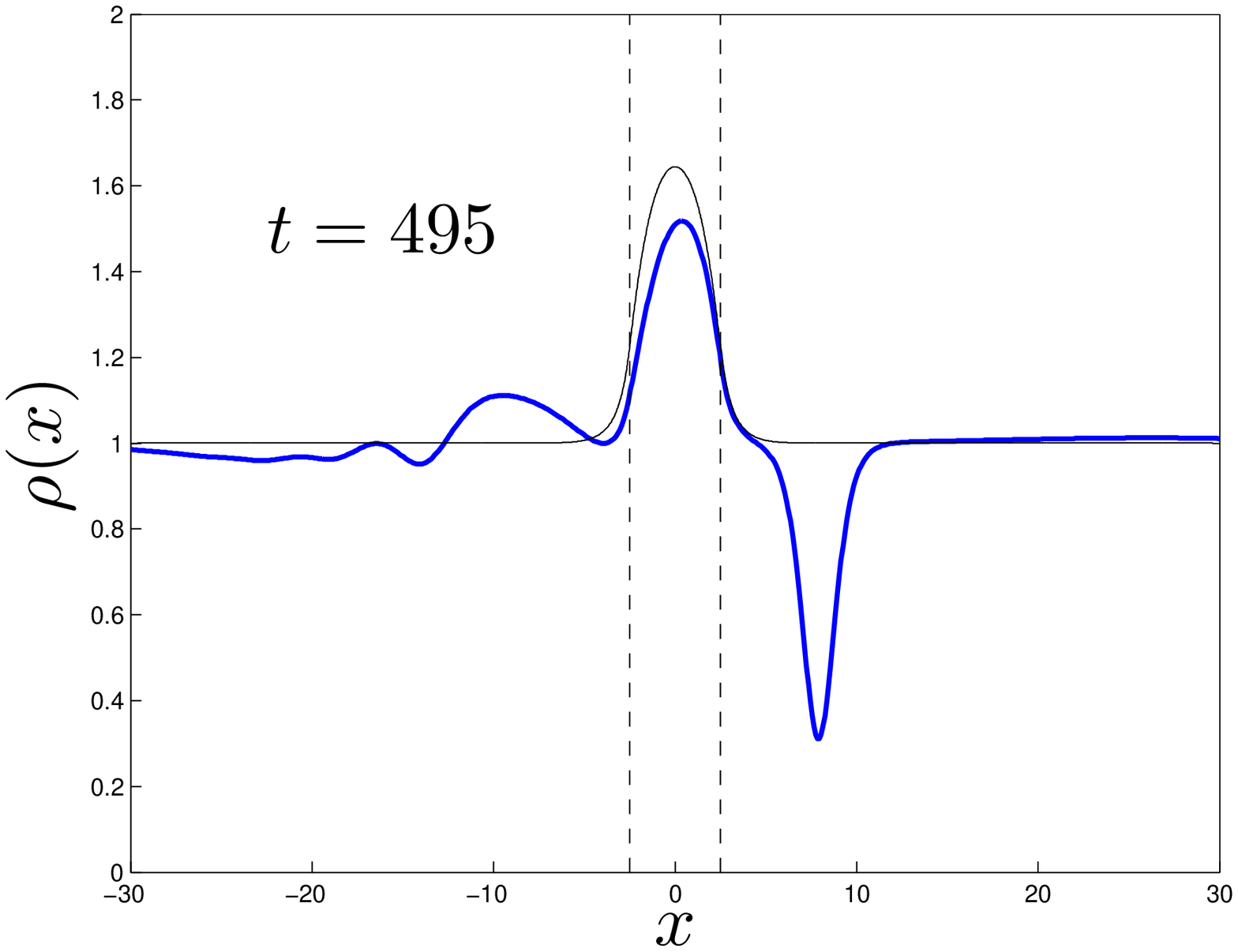} &
    \includegraphics[width=0.5\columnwidth]{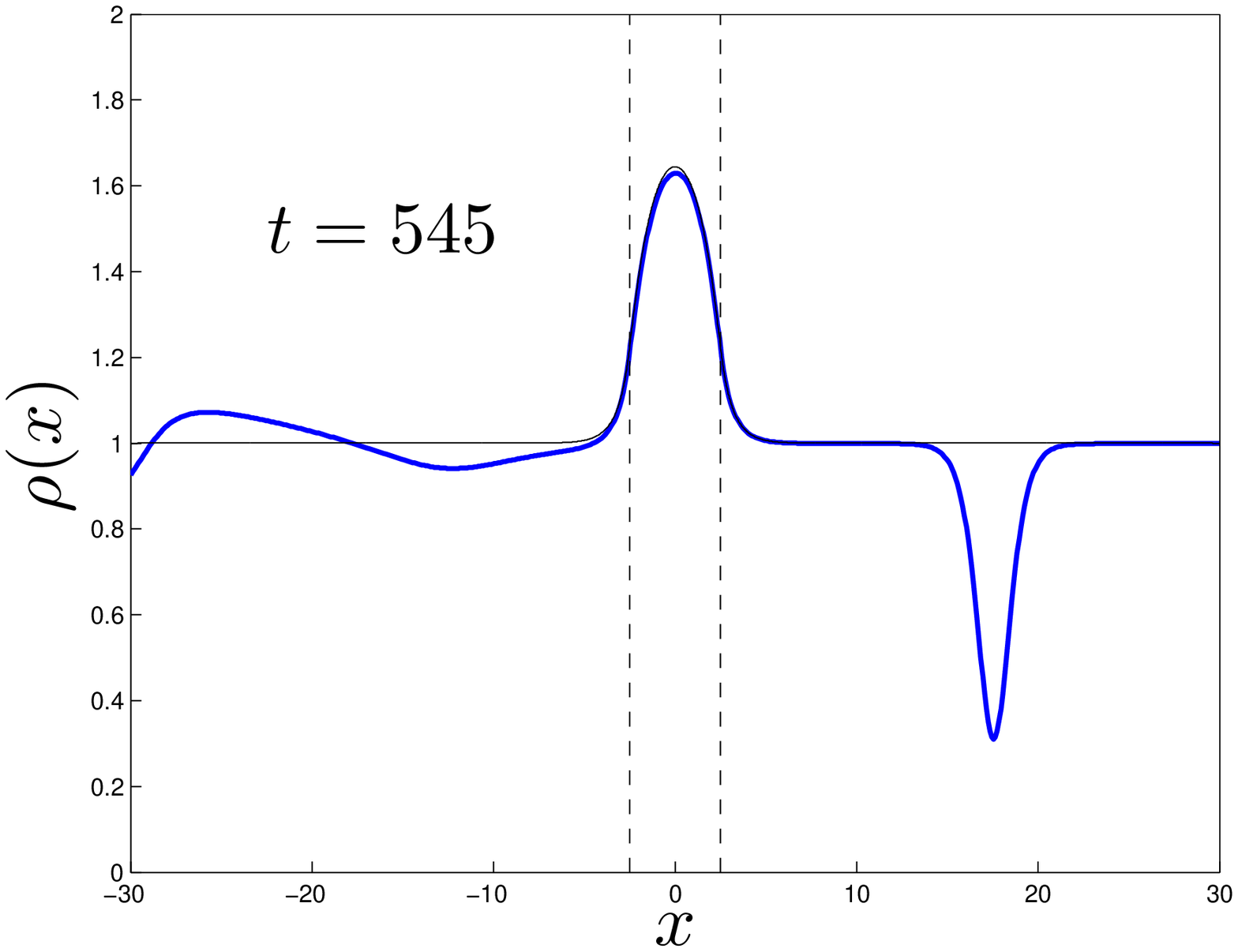} &
    \includegraphics[width=0.5\columnwidth]{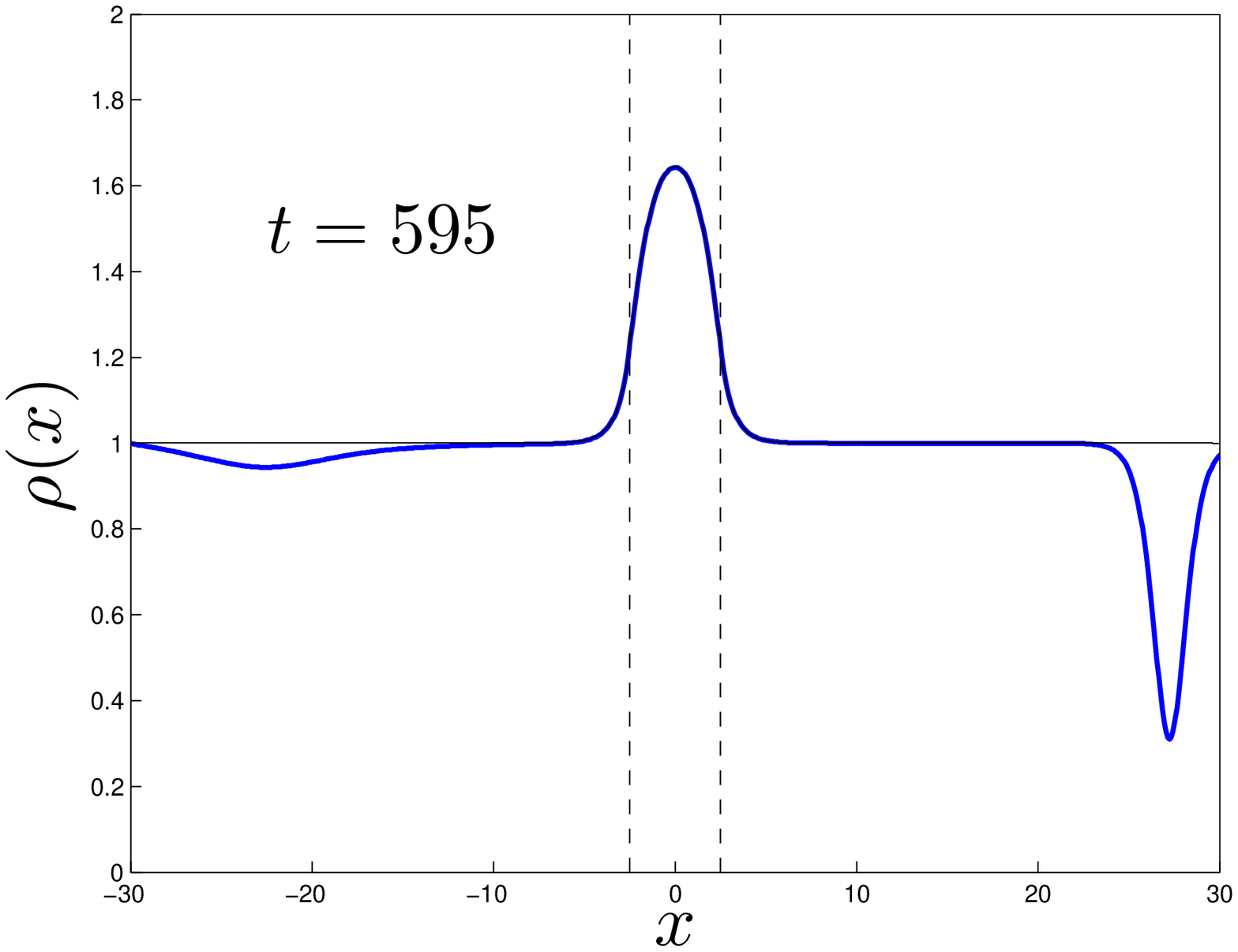} \\
\end{tabular}
\caption{Similar to Fig. \ref{fig:UpDownMechanism} but now for a system with $v=0.75$, $c_2=0.5$ and $X=5$, where the unstable mode presents a non-zero real part of the frequency so it oscillates while growing. The complete evolution is shown in \href{https://www.youtube.com/watch?v=xI6LE5FXLnk}{Movie 3}.}
\label{fig:HalfMechanism}
\end{figure*}

For $X_{1/2}(v,c_2)<X<X_{1}(v,c_2)$, the system also evolves towards the $n=0$ stationary solution. The only difference is the transient: as the instability keeps oscillating while growing, the dependence on the initial noise condition is no longer relevant and no distinction between initially increasing and decreasing cases can be made any longer. This behavior is shown in Fig. \ref{fig:HalfMechanism}; see also \href{https://www.youtube.com/watch?v=xI6LE5FXLnk}{Movie 3} for the complete picture.

Our numerical results then confirm that in this stable regime the (very different) transient dynamics does not play any significant role on the long time behavior of the system, which is only determined by the values of $(c_2,v,X)$ and in all cases eventually tends to the ground-state $n=0$ solution.

\subsubsection{Several unstable modes $X_1<X$}\label{subsec:n=1}

For larger sizes $X$ of the central region, several unstable modes are present and, because of that, the dynamics is potentially much more complicated. From Eq. (\ref{Eq:UnstableLength}), we see that the minimum length at which a new unstable mode appears is given by:
\begin{equation}\label{eq:minimum10length}
X_n^{min}=X_n(v=1,c_2=0)=n\pi
\end{equation}
Also in this case, two different behaviors are observed depending on the degree of instability: for sufficiently slow speeds $v$ and high values of $c_2$, the long-time evolution after a (sometimes complex) transient tends to a stationary state. On the other hand, for sufficiently high speeds $v$ and low values of $c_2$, a CES regime appears. In this subsection we restrict ourselves to the former case, leaving the discussion of the CES regime for the next subsection.

It was shown in Ref. \cite{Michel2013} that the unstable mode with largest value of $n$ is typically the dominant one in the early evolution given its fastest growth rate $\Gamma_n$. This feature has been qualitatively confirmed in our simulations by looking at the spatial profile of the density modulation at intermediate times, at its oscillating/non-oscillating temporal dependence, and at the instability growth rate.

\begin{figure*}[!tb]
\begin{tabular}{@{}cccc@{}}
    \includegraphics[width=0.5\columnwidth]{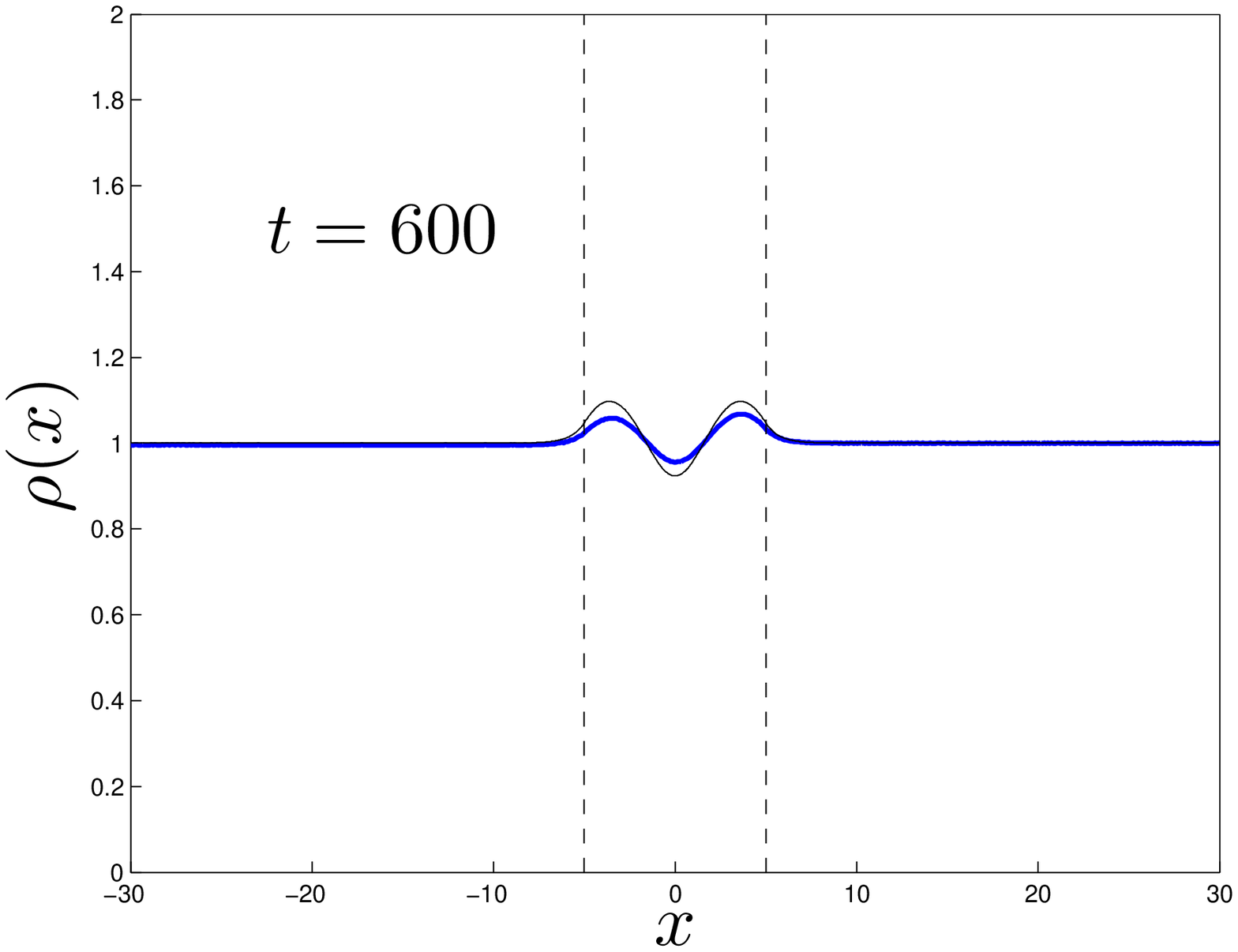} &
    \includegraphics[width=0.5\columnwidth]{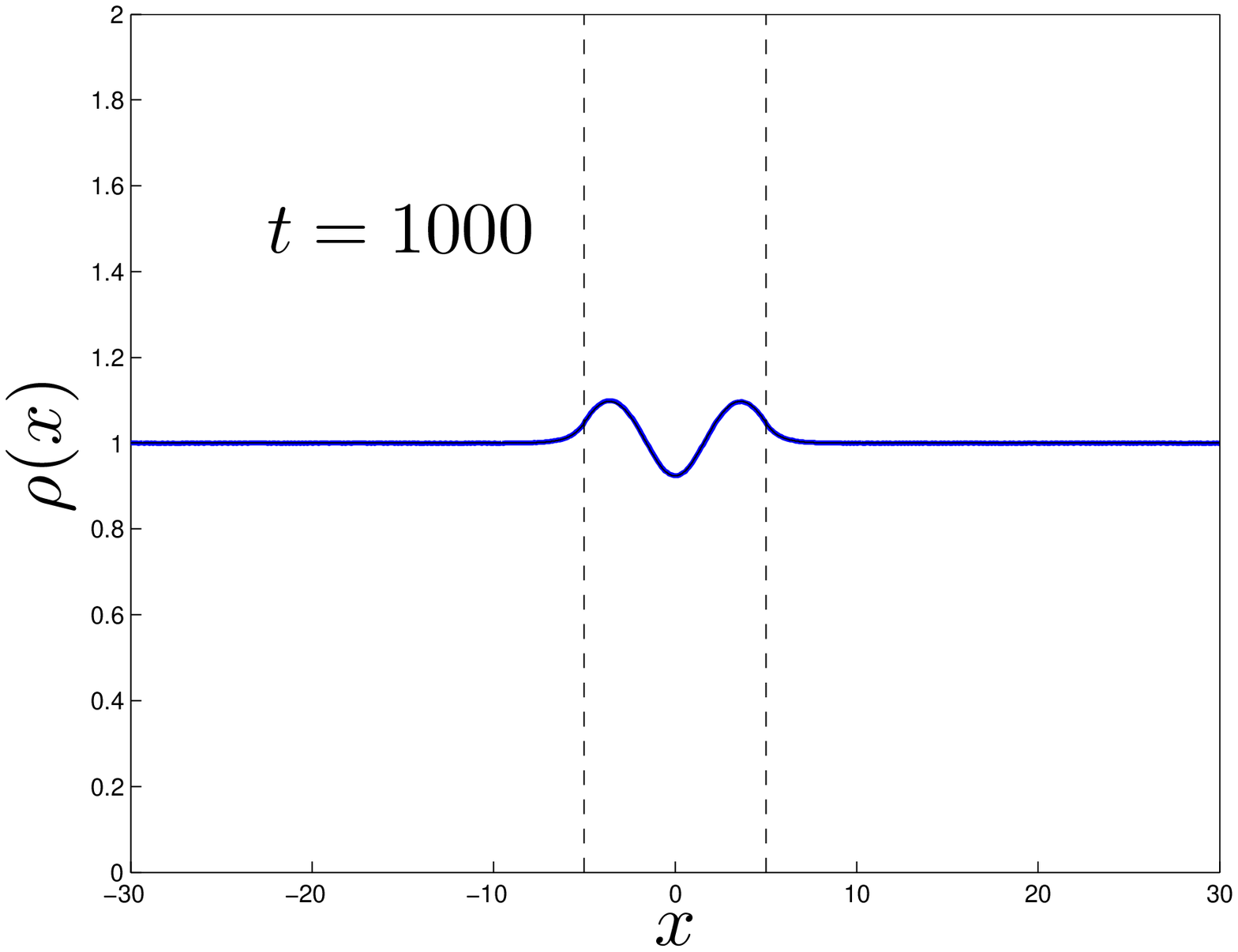} &
    \includegraphics[width=0.5\columnwidth]{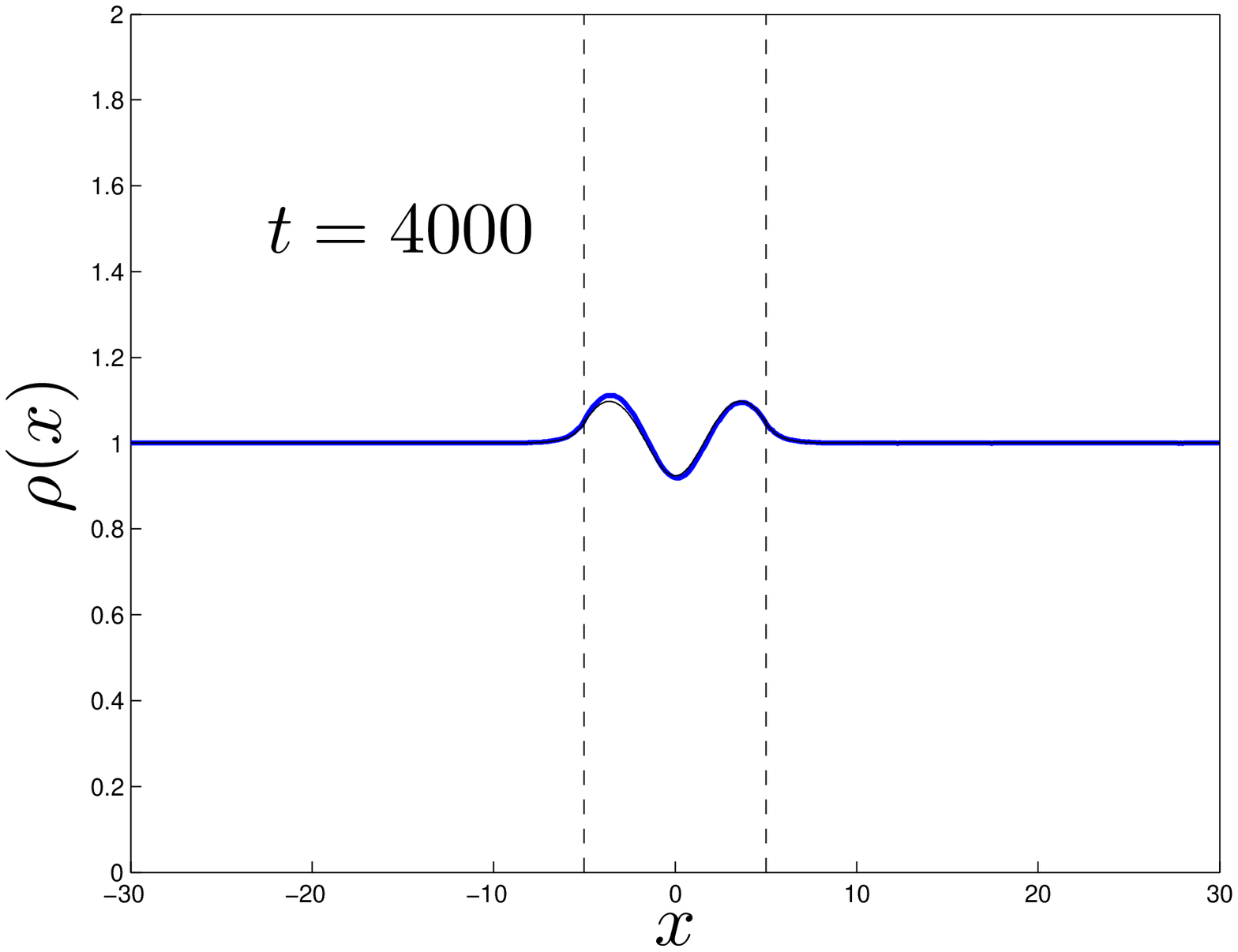} &
    \includegraphics[width=0.5\columnwidth]{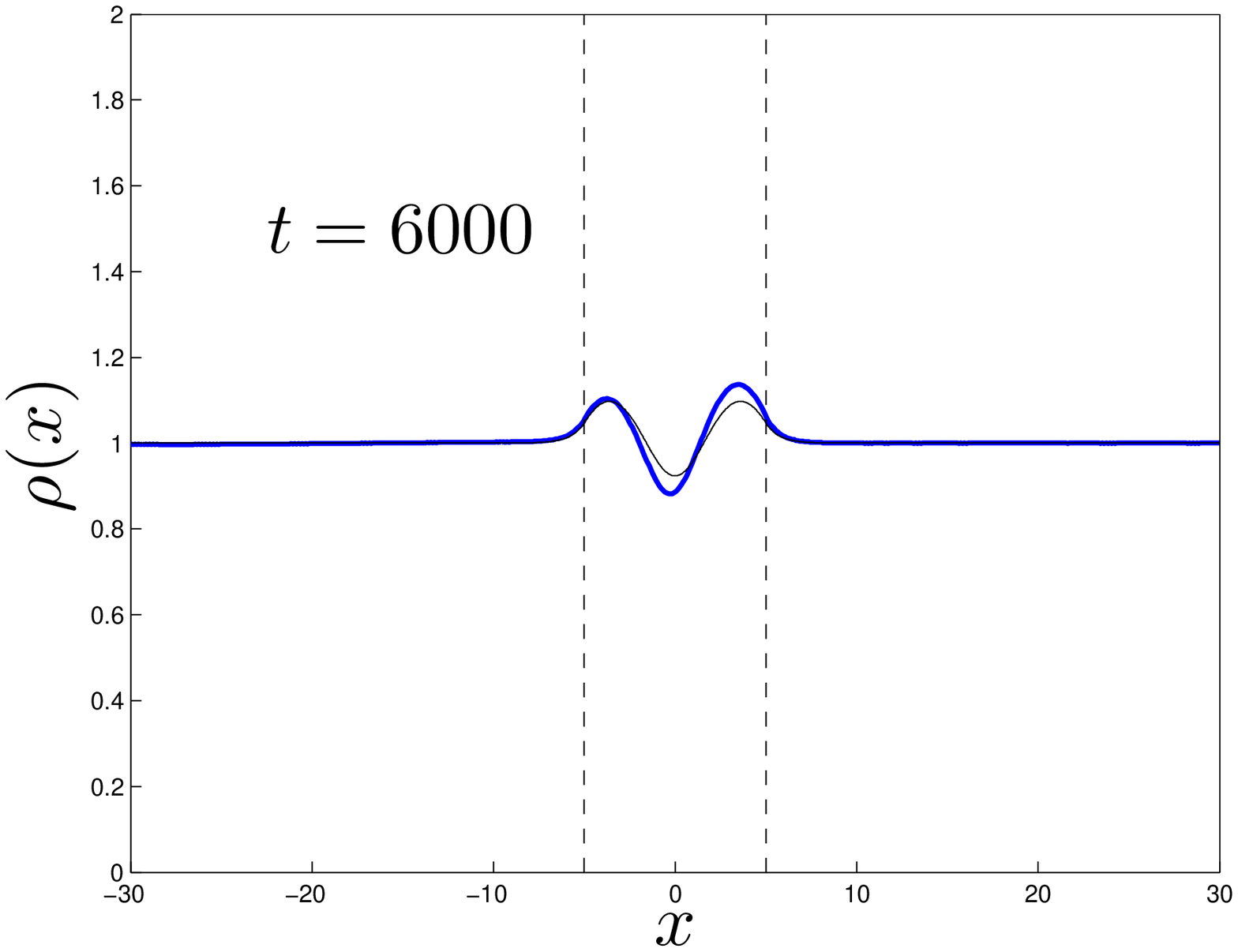} \\
    \includegraphics[width=0.5\columnwidth]{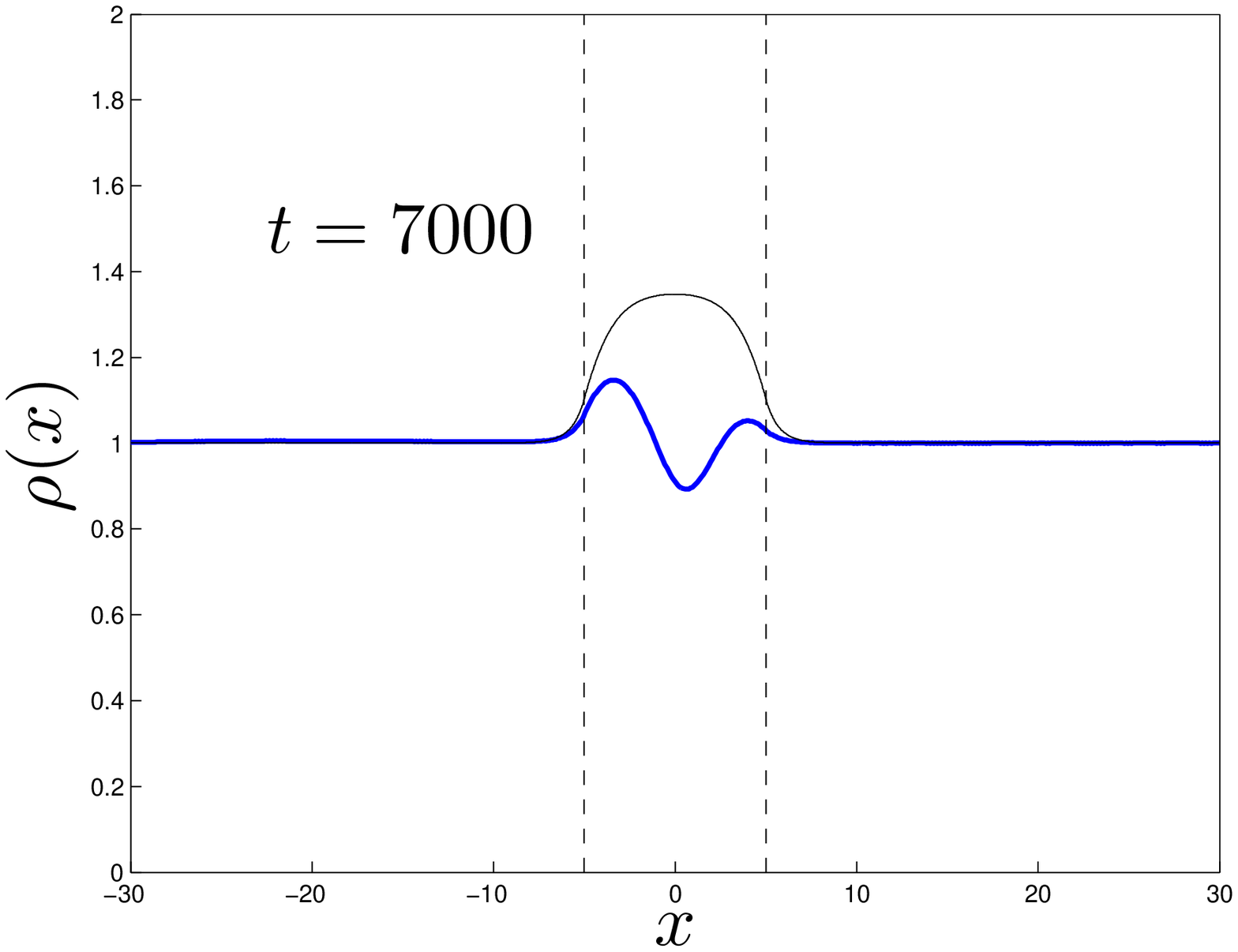} &
    \includegraphics[width=0.5\columnwidth]{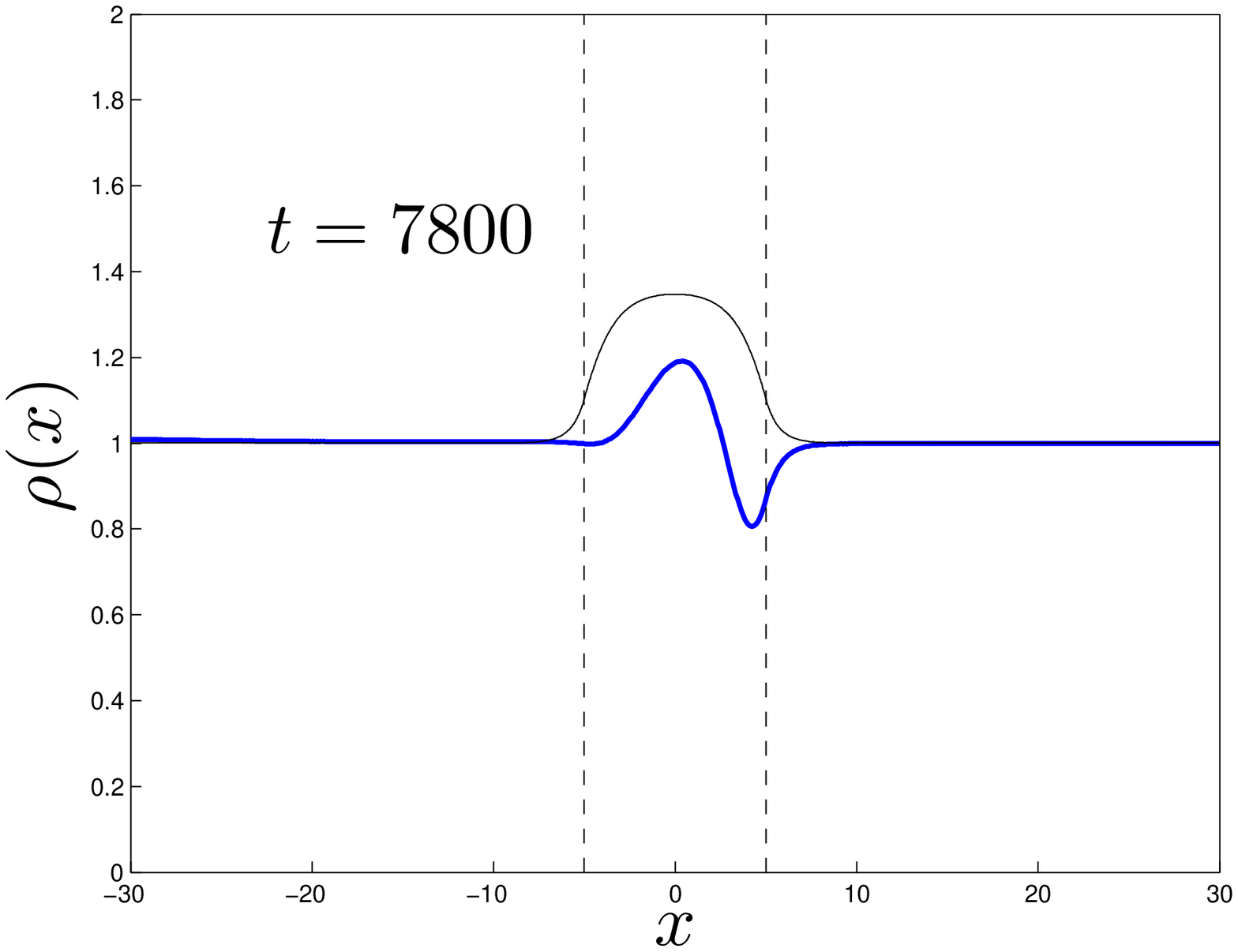} &
    \includegraphics[width=0.5\columnwidth]{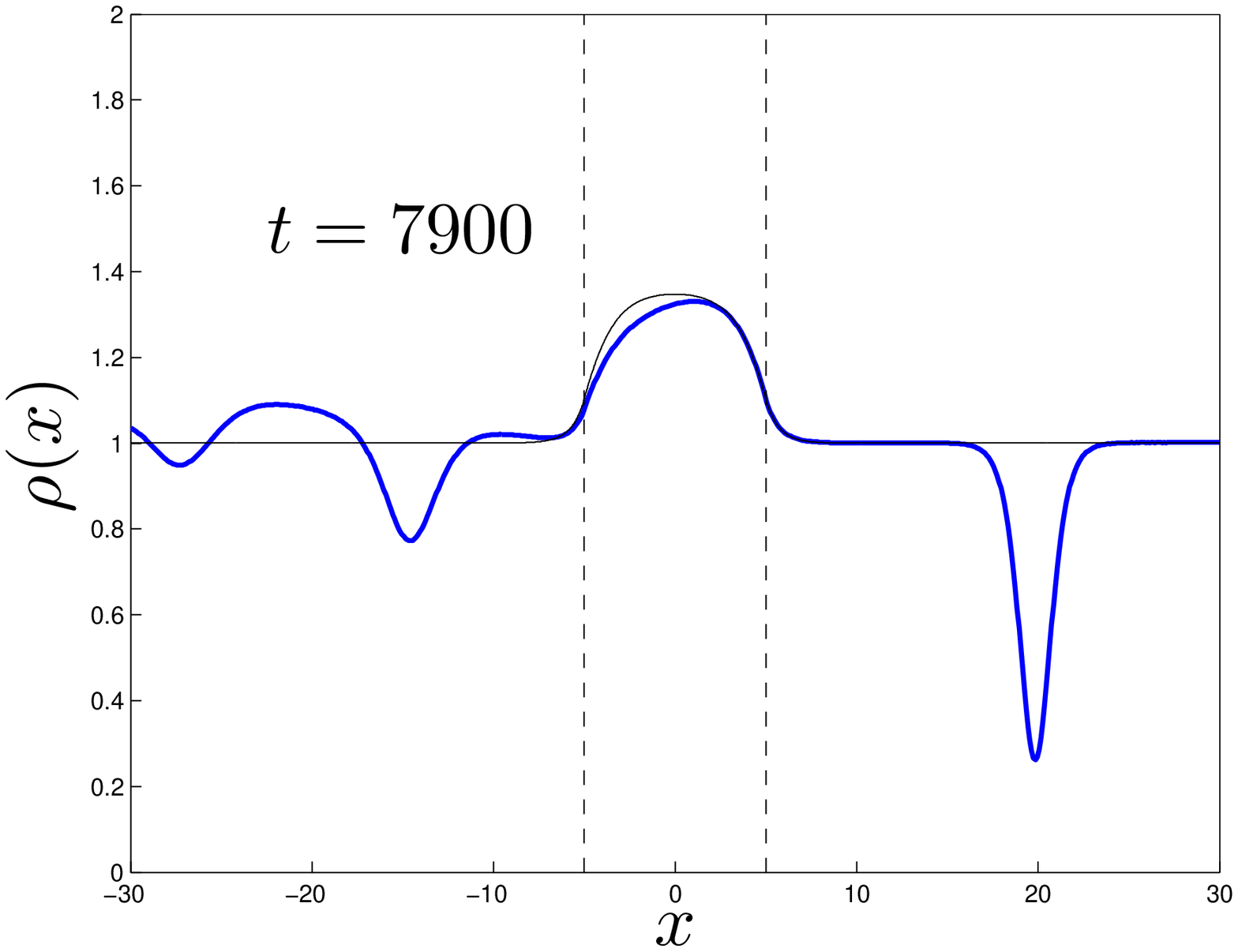} &
    \includegraphics[width=0.5\columnwidth]{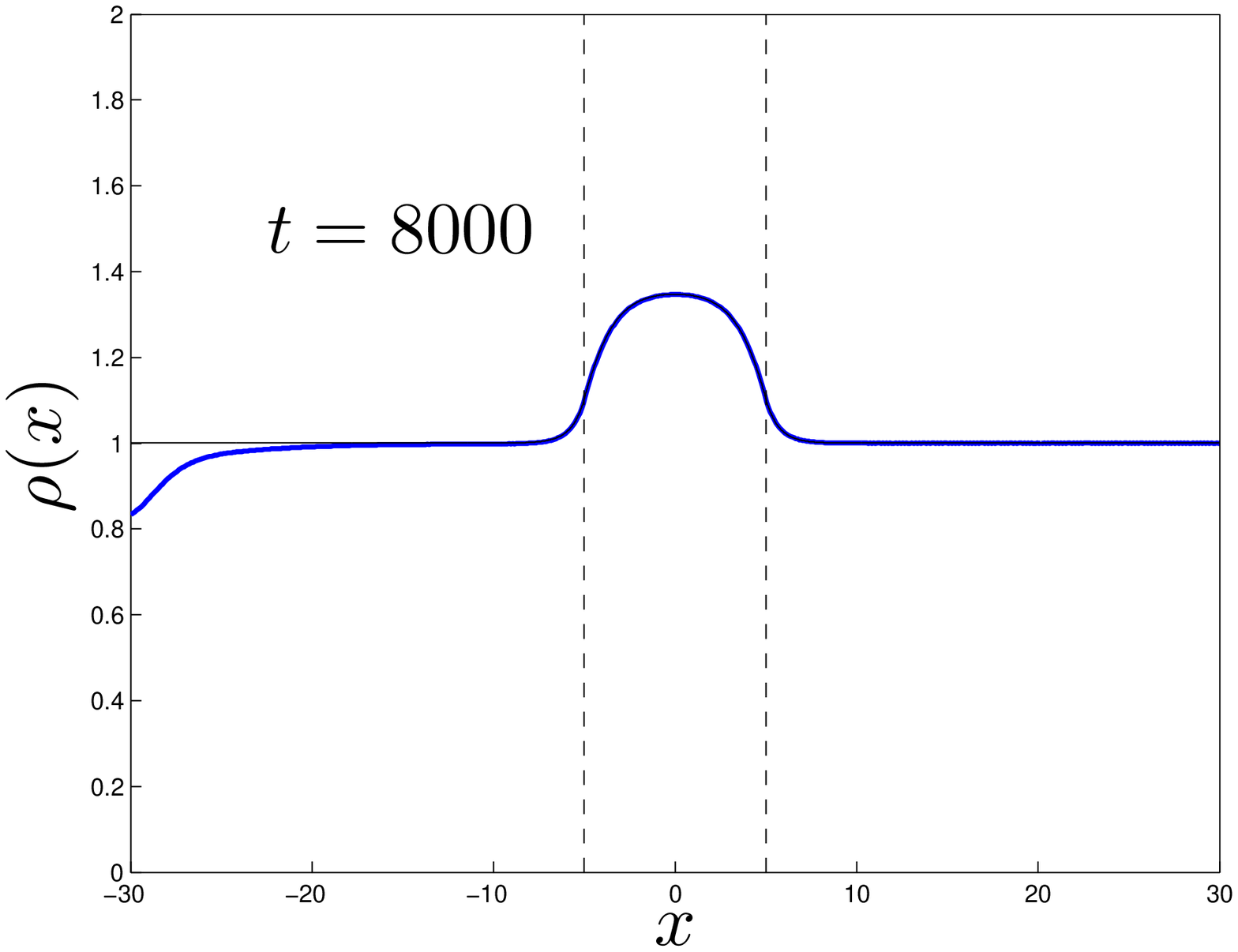} \\[2em]
    \includegraphics[width=0.5\columnwidth]{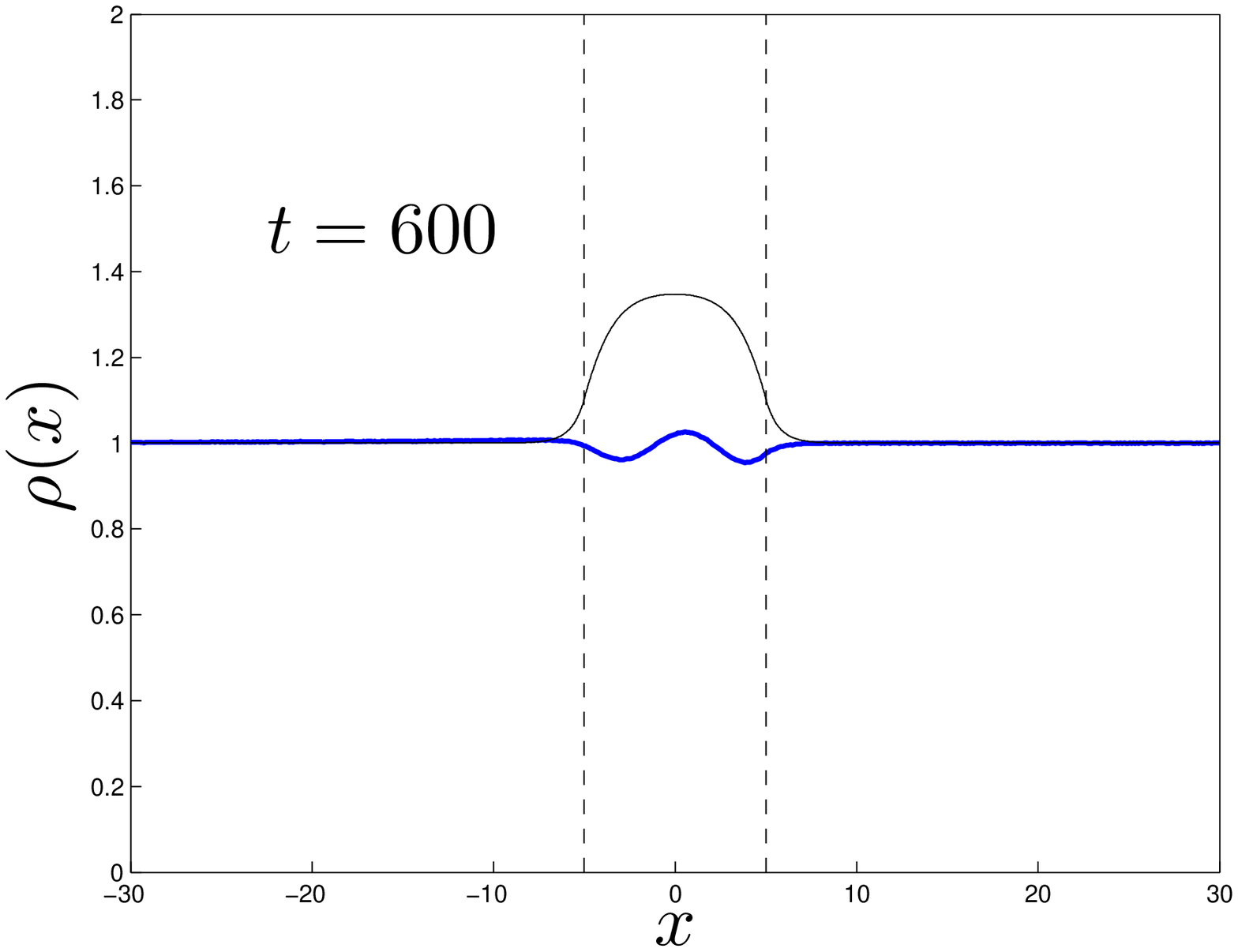} &
    \includegraphics[width=0.5\columnwidth]{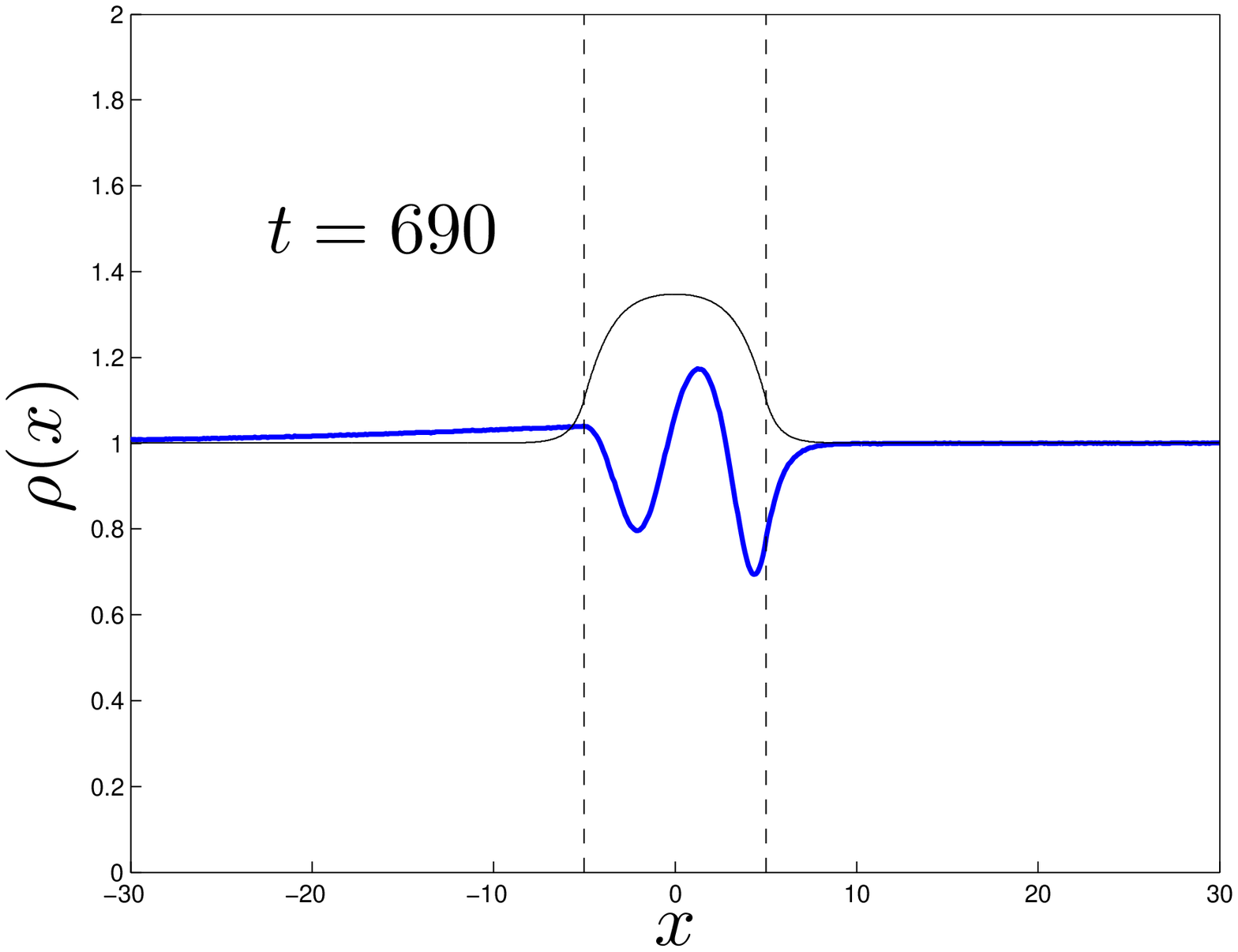} &
    \includegraphics[width=0.5\columnwidth]{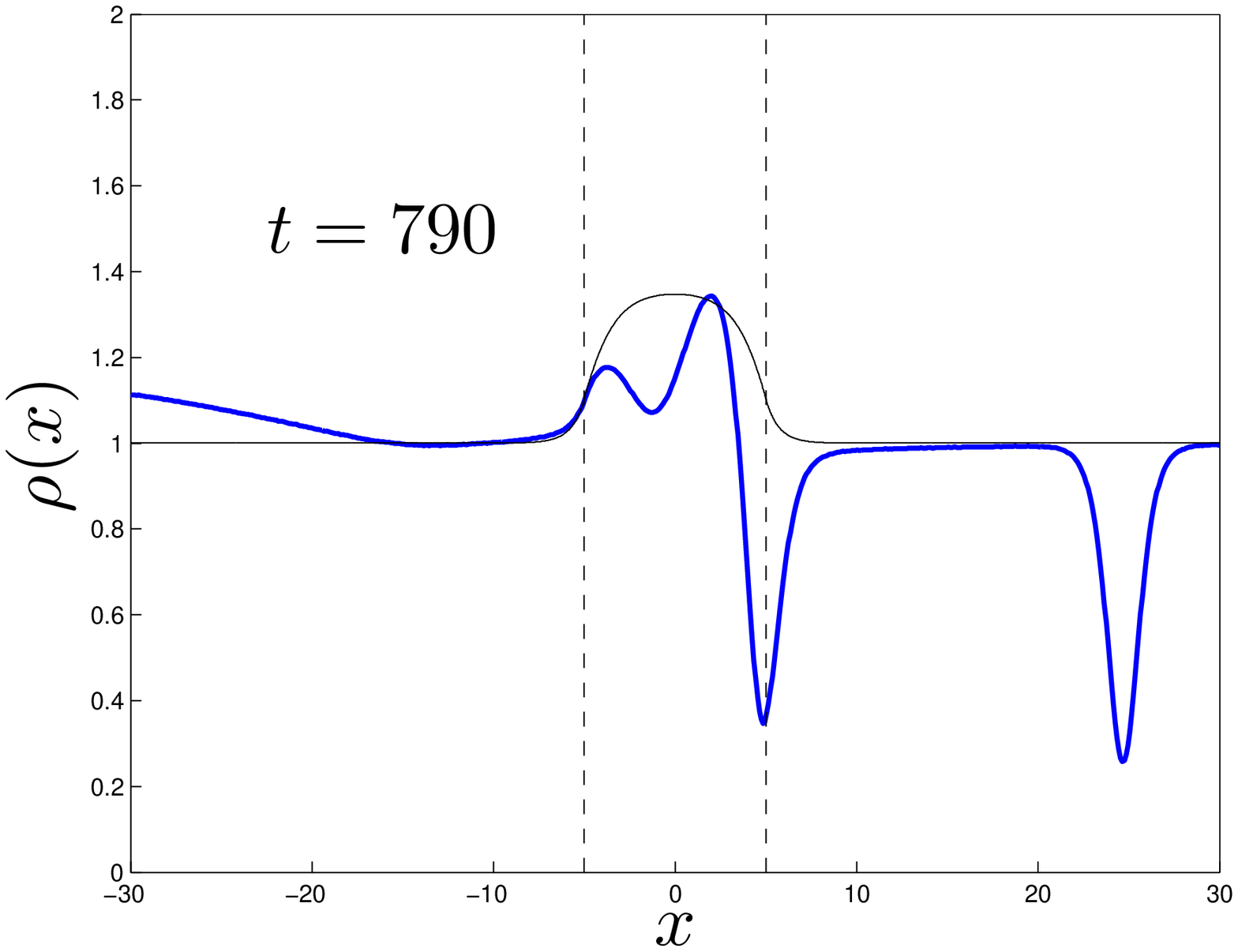} &
    \includegraphics[width=0.5\columnwidth]{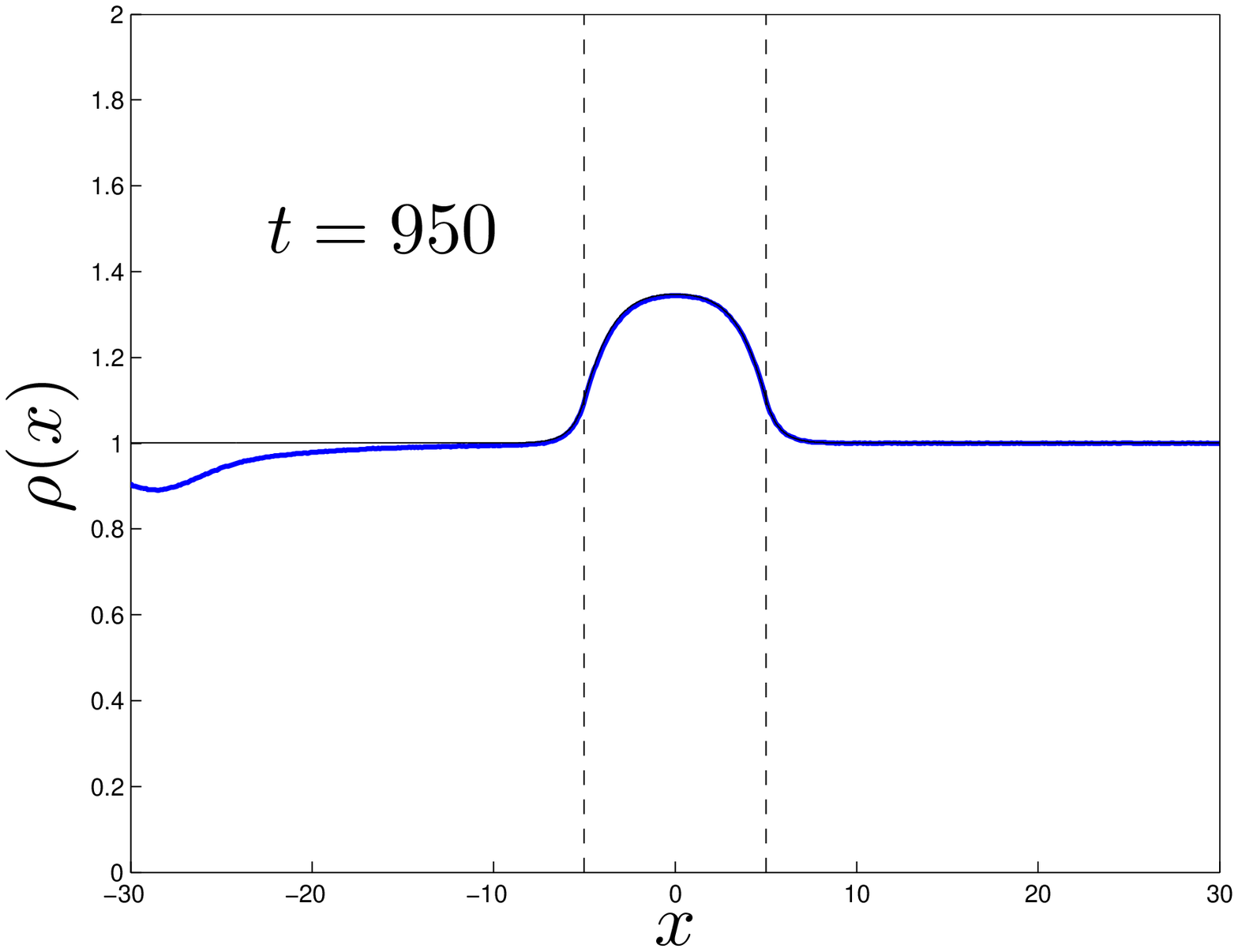}
\end{tabular}
\caption{Analog of Fig. \ref{fig:UpDownMechanism} but for a system with $v=0.75$, $c_2=0.6$ and $X=10$ so there are two unstable modes. Upper and central rows: evolution of the system when the linear unstable mode $n=1$ grows towards the $n=1$ solution (depicted in a thin black line in the upper row) and remains there for some time until it collapses and evolves towards the $n=0$ ground state (thin black line in the central row). The complete evolution is displayed in \href{https://www.youtube.com/watch?v=kX0z8AZCzNk}{Movie 4}. Lower row: evolution of the system when the linear unstable mode $n=1$ grows in the opposite direction, emitting some waves and solitons before reaching the $n=0$ solution. The complete evolution is shown in \href{https://www.youtube.com/watch?v=hqcxajEjDoE}{Movie 5}.}
\label{fig:N1UpMechanism}
\end{figure*}

For what concerns the later dynamics, the situation is quite similar to the one discussed in the previous section, the main difference being that several stationary solutions with $\Delta\Omega<0$ are now available. While on the long run the system typically tries to relax to the lowest energy $n=0$ solution (which is the only dynamically stable one) by emitting the extra energy and particles in the form of solitons and small waves, at intermediate times it may or may not approach a higher $n>0$ stationary solution and remain for a quite macroscopic time in its vicinity. As such solutions are dynamically unstable, the system must eventually depart from it and finally converge to the stable $n=0$ solution after another stage of soliton and wave emission.

When $X_n(v,c_2)<X<X_{n+1/2}(v,c_2)$, some useful information about whether the system does or does not spend time in the vicinity of a higher $n>0$ stationary solution can be obtained along the lines of the discussion around Eq. \ref{eq:densityzero} of the previous subsection: depending on the sign of the projection of the initial noise configuration onto the $n$ dominant unstable mode, the system can smoothly evolve towards the $n$ non-linear solution or it can start evolving in the opposite direction. In this latter case, more solitons and waves need emitting to conserve the total number of particles so the evolution is more chaotic. Both trends are presented in Fig. \ref{fig:N1UpMechanism}, where we consider a simple configuration with only two unstable ($n=0,1$) modes. In the upper and central rows, we show the case where the system smoothly evolves to the $n=1$ solution, remaining there for a long time until it collapses and then reaches the ground state $n=0$. In the lower row, we show the evolution when the state grows in the opposite direction, where after emitting some waves and solitons, the system reaches the ground state in a much shorter time scale. In both cases, after a (possibly very long) transient, the simulation ends up in the lowest energy $n=0$ stationary solution. The complete evolution of both situations is displayed in \href{https://www.youtube.com/watch?v=kX0z8AZCzNk}{Movie 4} and in \href{https://www.youtube.com/watch?v=hqcxajEjDoE}{Movie 5}, respectively.

This dramatic dependence on the initial conditions is a simplest example of the potentially chaotic dynamics of the system for $X_1<X$ when many unstable modes are present. As another remarkable example of non-trivial behaviour, \href{https://www.youtube.com/watch?v=OfUsjMDDMJU}{Movie 6} shows a case when the system happens to intercept some other nonlinear stationary solution, so that it remains trapped in a sort of metastable state in its vicinity for a macroscopically long time and the $n=0$ solution is not yet reached for very long times on the order of $t=10^4$. Even though all $n>0$ stationary solutions are dynamically unstable and the system must eventually depart from them, we cannot exclude that there exist regions of the $(X,c_2,v)$ parameter space for which the time scale of the decay to the $n=0$ stationary state might take arbitrarily long.

\subsection{Continuous emission of solitons}\label{subsec:CES}

This subsection expands the discussion of the CES mechanism briefly mentioned in Secs. \ref{subsec:n=0} and \ref{subsec:n=1} and reports some among the main results of this work, in the direction of understanding such an intriguing regime. In contrast to the Bogoliubov-Cherenkov emission of sound waves that takes place in the upstream direction~\cite{Carusotto2008}, the continuous emission of solitons occurs in the downstream direction towards $x=+\infty$~\cite{Leboeuf2001,Pavloff:PRA2002}. Similar scenarios of soliton train emission have been predicted in Refs. \cite{Frisch:PRL1992,Hakim1997,Pavloff:PRA2002,Wuester2007} and experimentally observed in~\cite{engels2007}. In all these works the soliton emission process takes place in the vicinity of a localized defect potential where the condensate density is locally depleted and a localized supersonic flow appears. An interpretation of this physics in terms of a black-hole laser instability has been proposed in~\cite{Abad2015}. Related, but somehow different soliton emission mechanisms were discussed in~\cite{Kamchatnov2002,Kamchatnov2012}.

\begin{figure*}[!htb]
\begin{tabular}{@{}cc@{}}
    \includegraphics[width=\columnwidth]{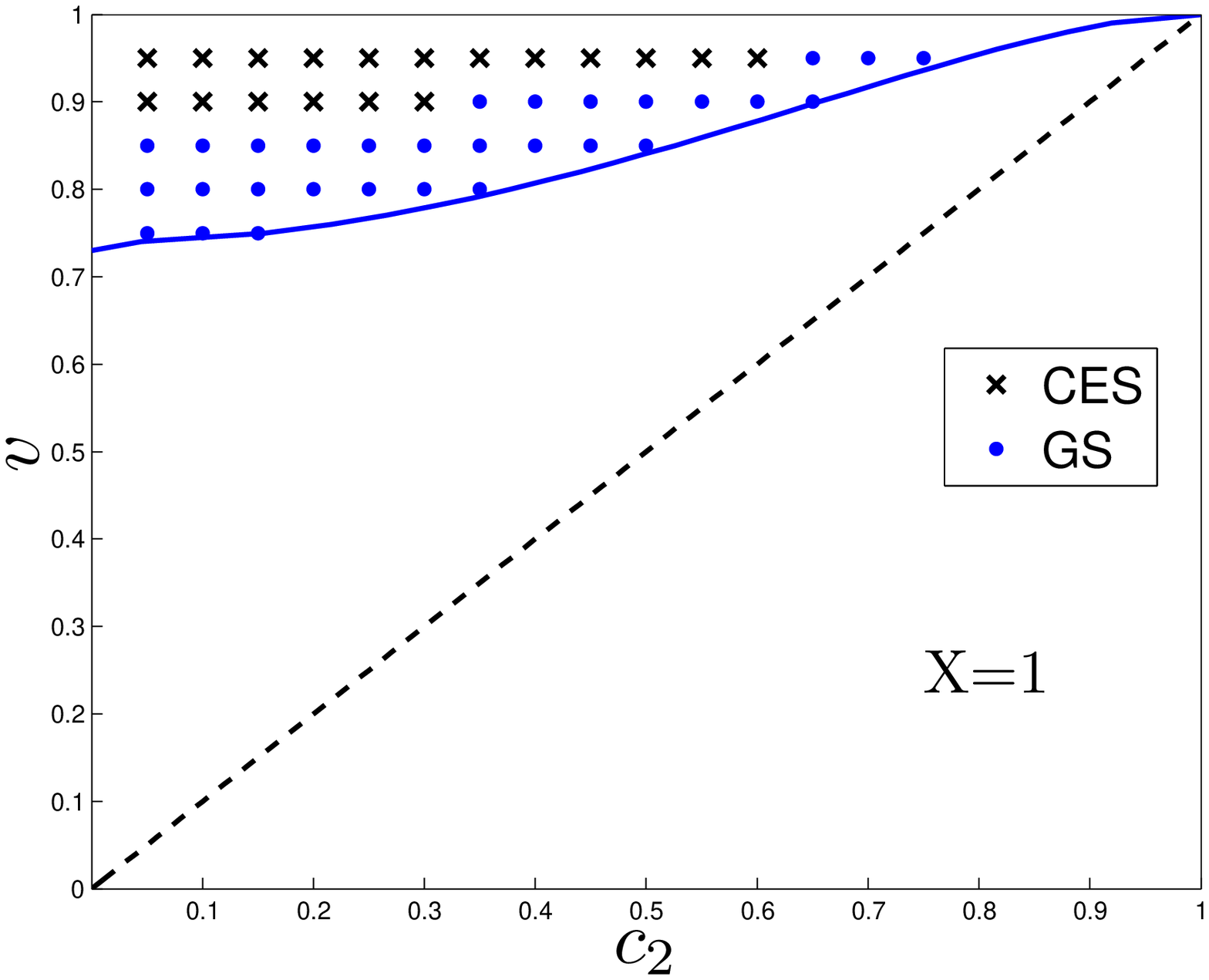} &
    \includegraphics[width=\columnwidth]{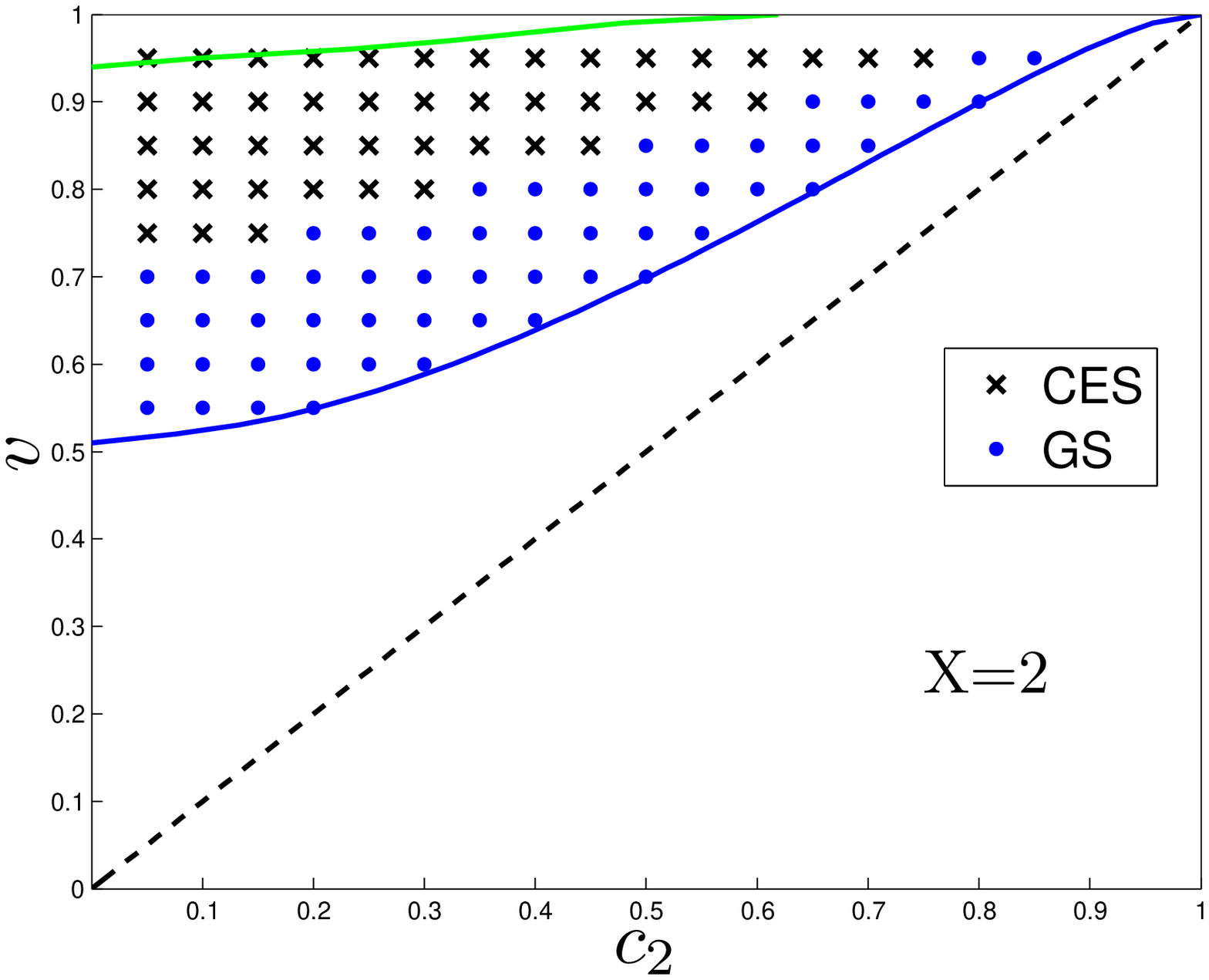} \\
    \includegraphics[width=\columnwidth]{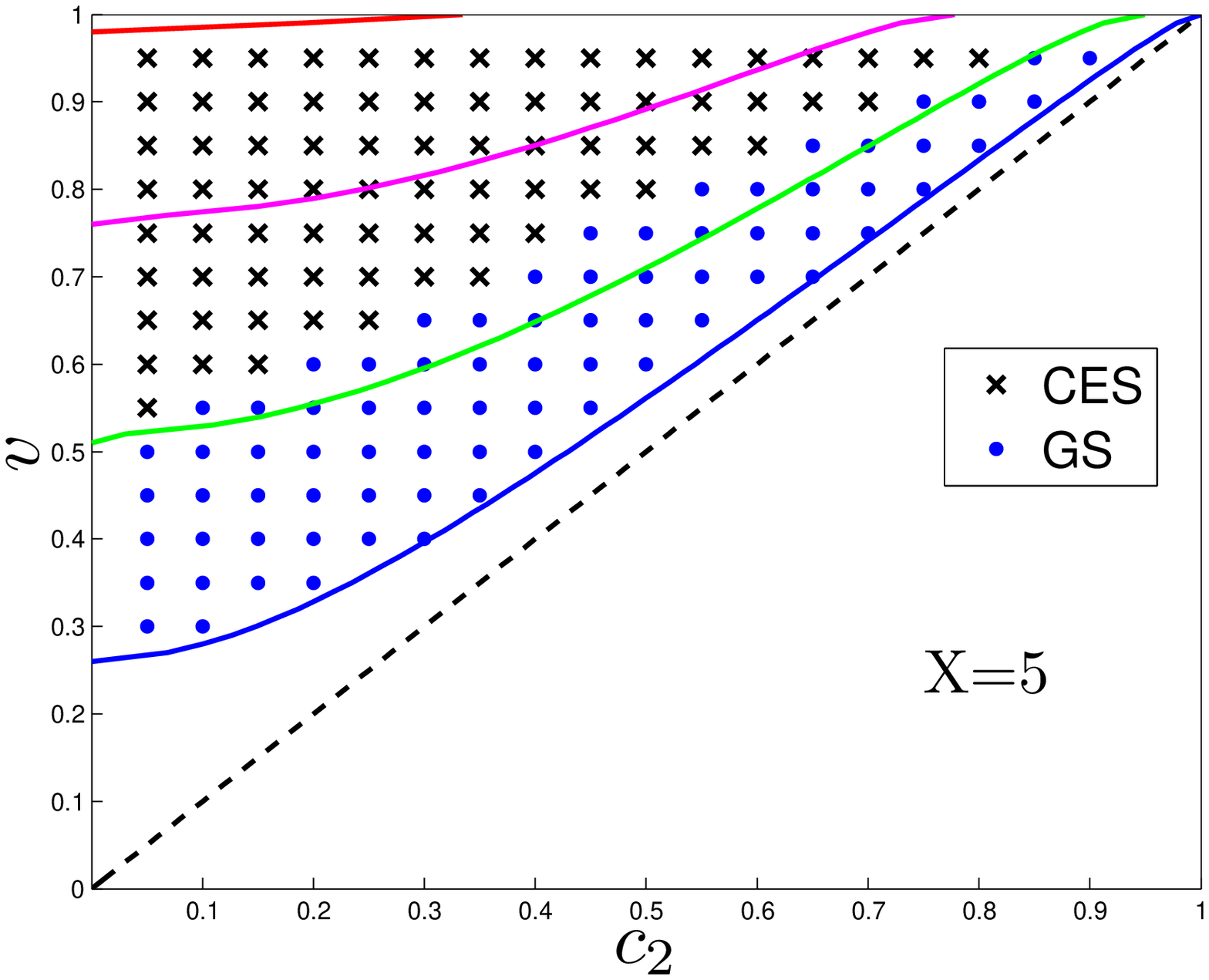} &
    \includegraphics[width=\columnwidth]{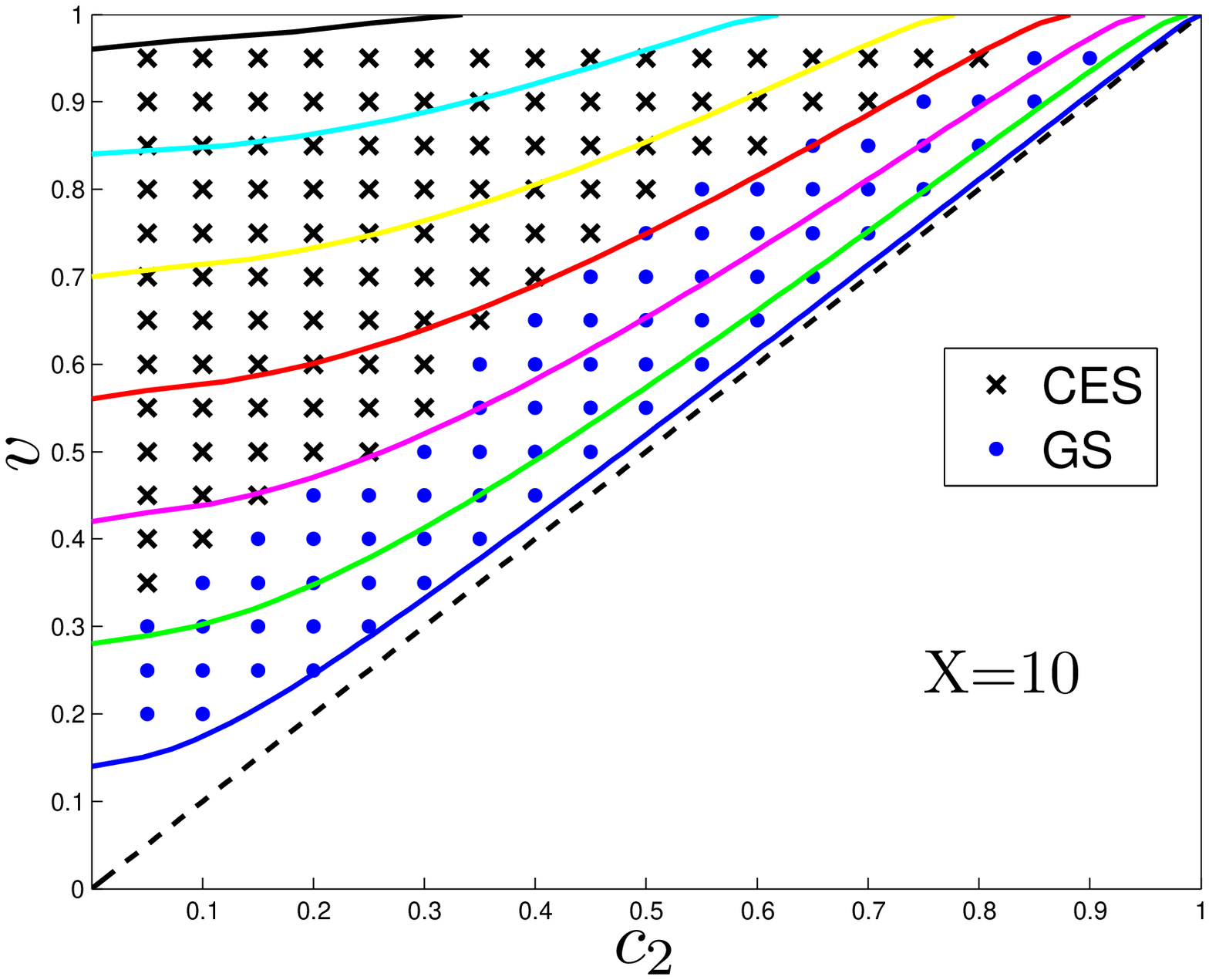} \\
\end{tabular}
\caption{Phase diagram in the $(c_2,v)$ plane for different values of $X$. The blue dots represents numerical simulations that end in the $n=0$ ground state (labeled as GS in the legend). The black crosses represents simulations in which CES has been observed. The solid curves from bottom to top represent the lower boundary of the linear instability regions defined in Eq. \ref{Eq:UnstableLength} with $n=0$ (blue), $1/2$ (green), $1$ (red), $3/2$ (pink), etc. The dashed oblique straight line is the upper boundary of the region where the flow is everywhere subsonic.}
\label{fig:CESregion}
\end{figure*}

The first step in the direction of characterizing the CES regime is to identify in the $(X,c_2,v)$ parameter space the regions of CES: cuts of such phase diagram along the $(c_2,v)$ plane are shown in the different panels of Fig. \ref{fig:CESregion} for growing values of $X$. The region where CES happens is indicated by black crosses and extends for growing $X$ starting from the upper-left corner (i.e. the region with high $v$ and low $c_2$). The CES regime is indeed qualitatively related to the degree of instability of the central supersonic region, which is favored by a large flow speed $v$ and a low $c_2$. Even though a chaotic CES can be observed in very strongly unstable systems with many unstable modes, we will not focus our attention on it and we will restrict the use of the CES expression to periodic soliton emission processes.

\begin{figure*}[!htb]
\begin{tabular}{@{}cccc@{}}
    \includegraphics[width=0.5\columnwidth]{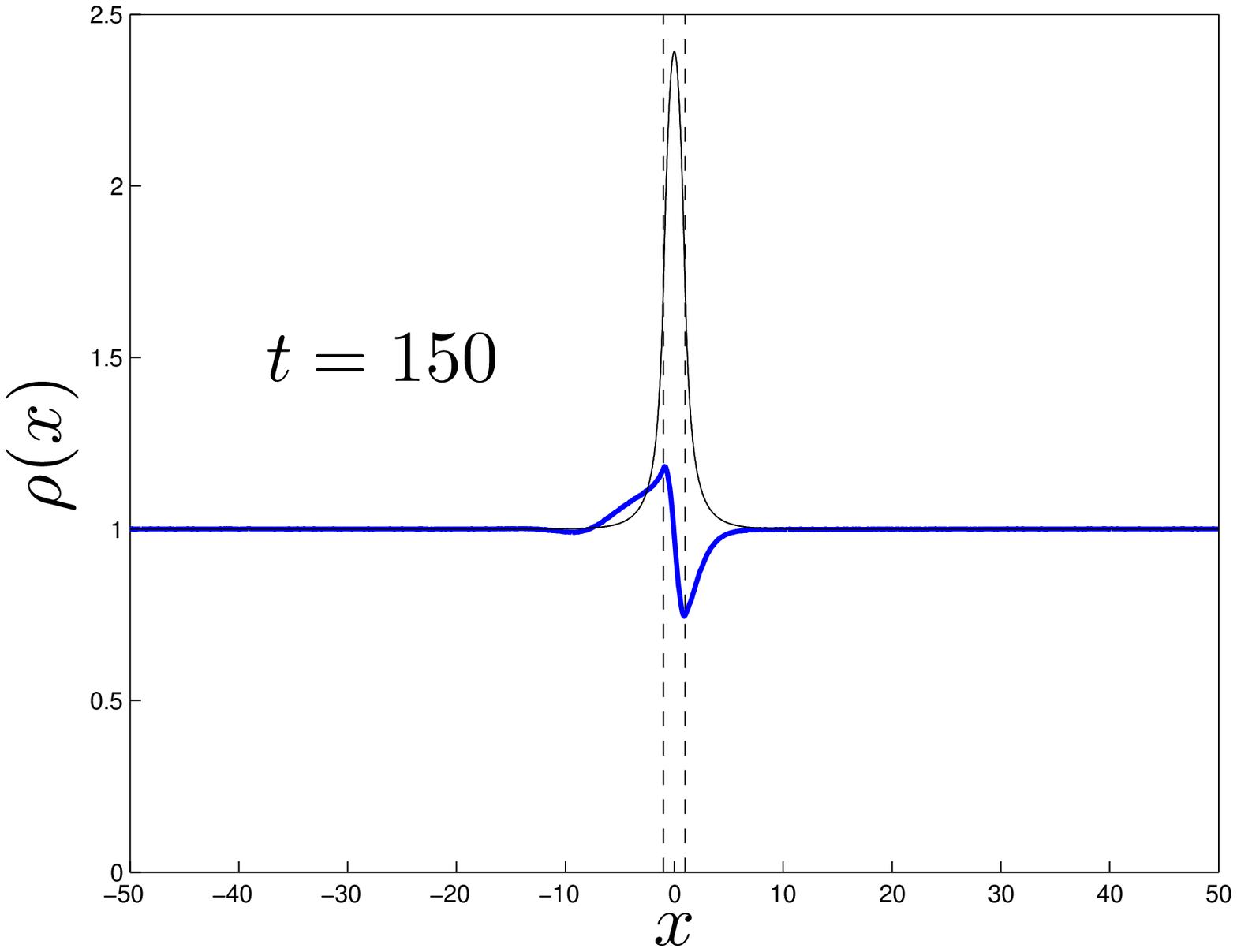} &
    \includegraphics[width=0.5\columnwidth]{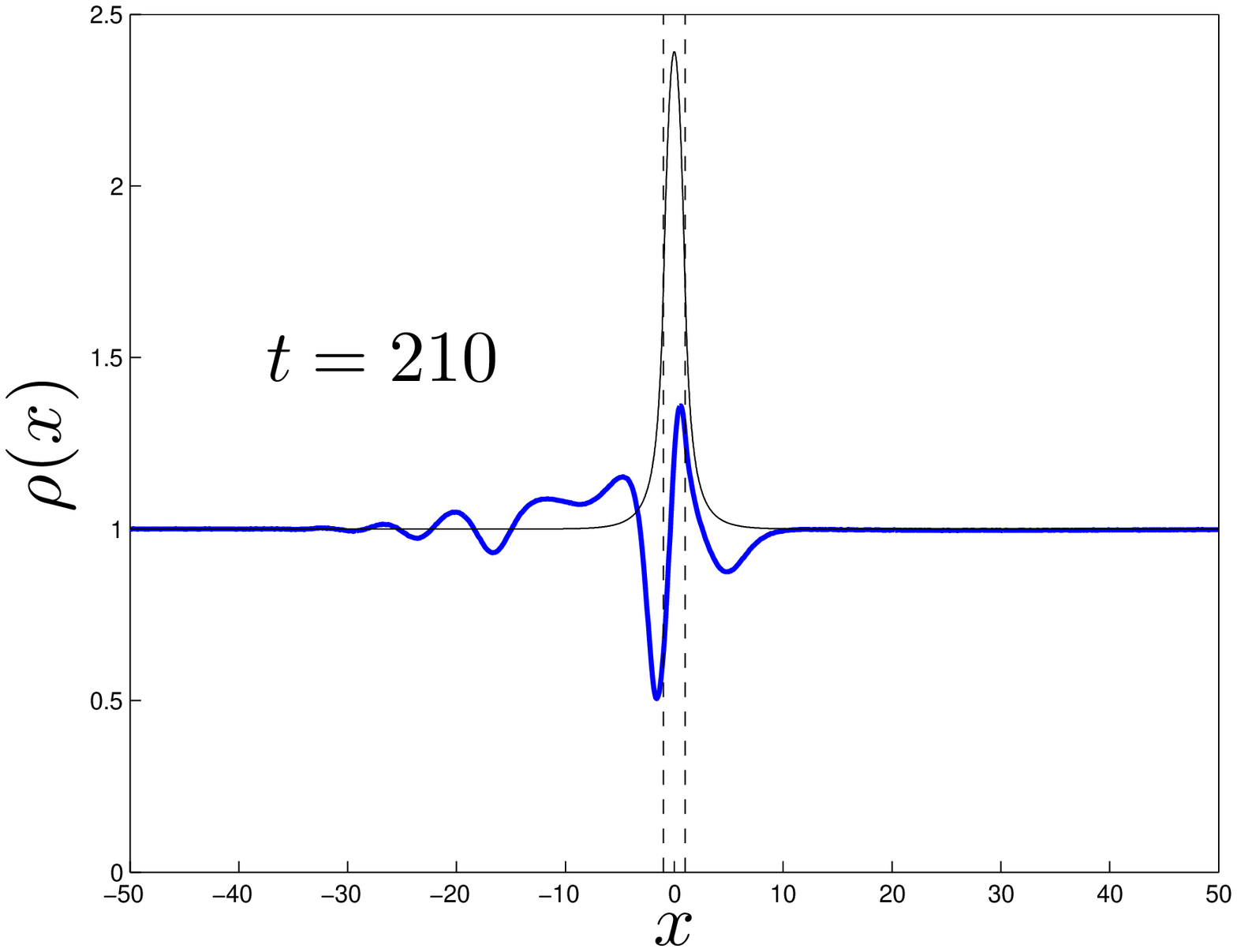} &
    \includegraphics[width=0.5\columnwidth]{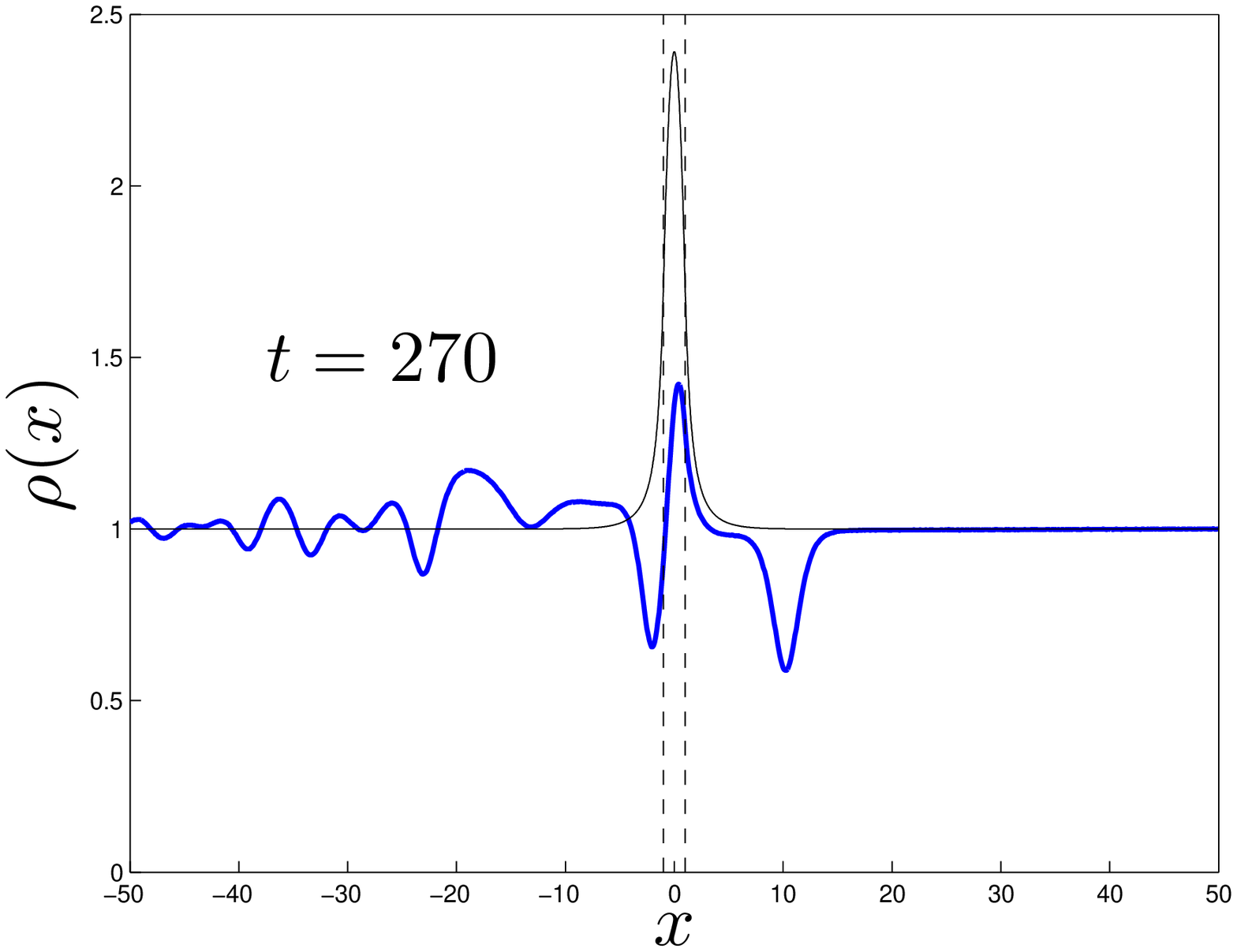} &
    \includegraphics[width=0.5\columnwidth]{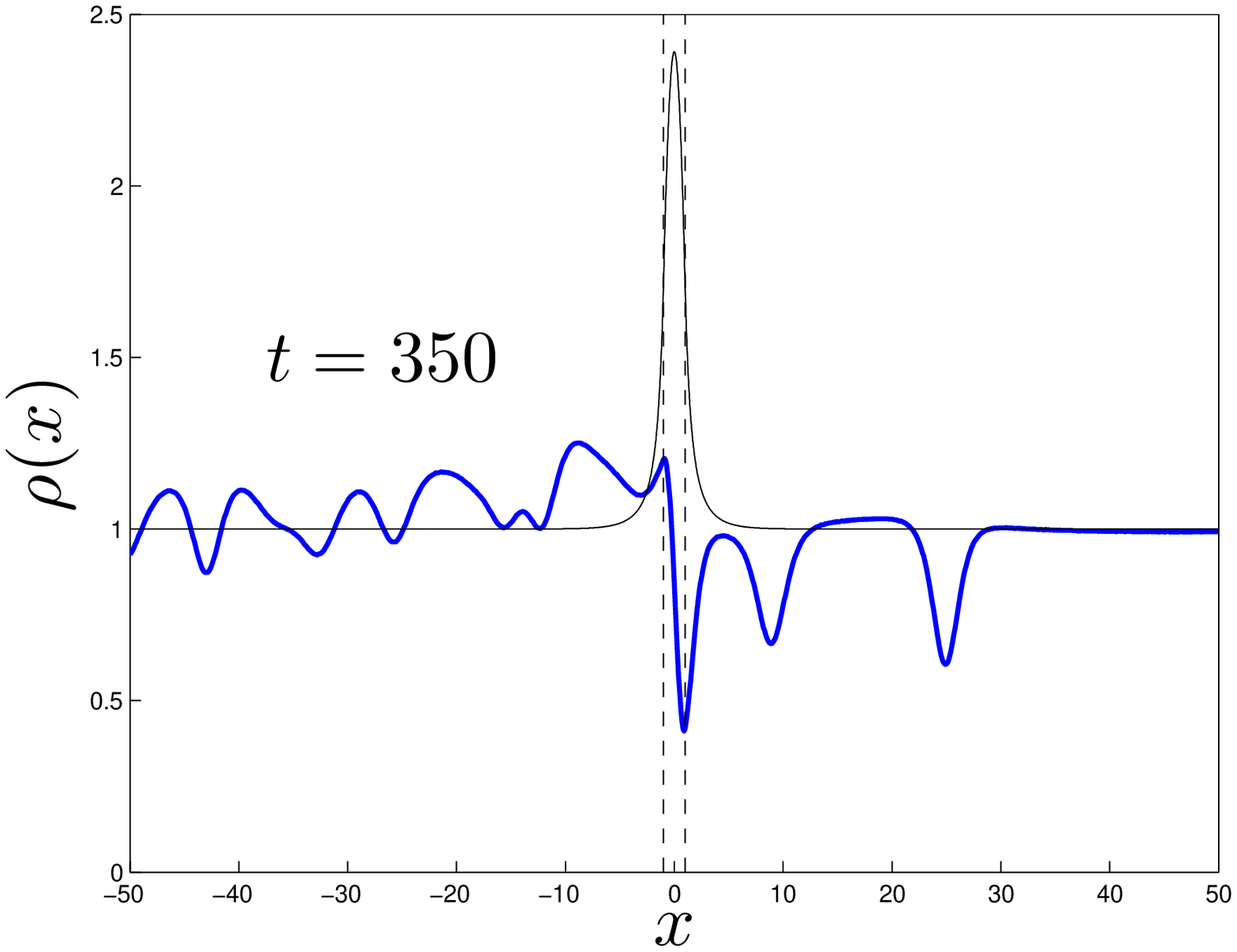} \\
    \includegraphics[width=0.5\columnwidth]{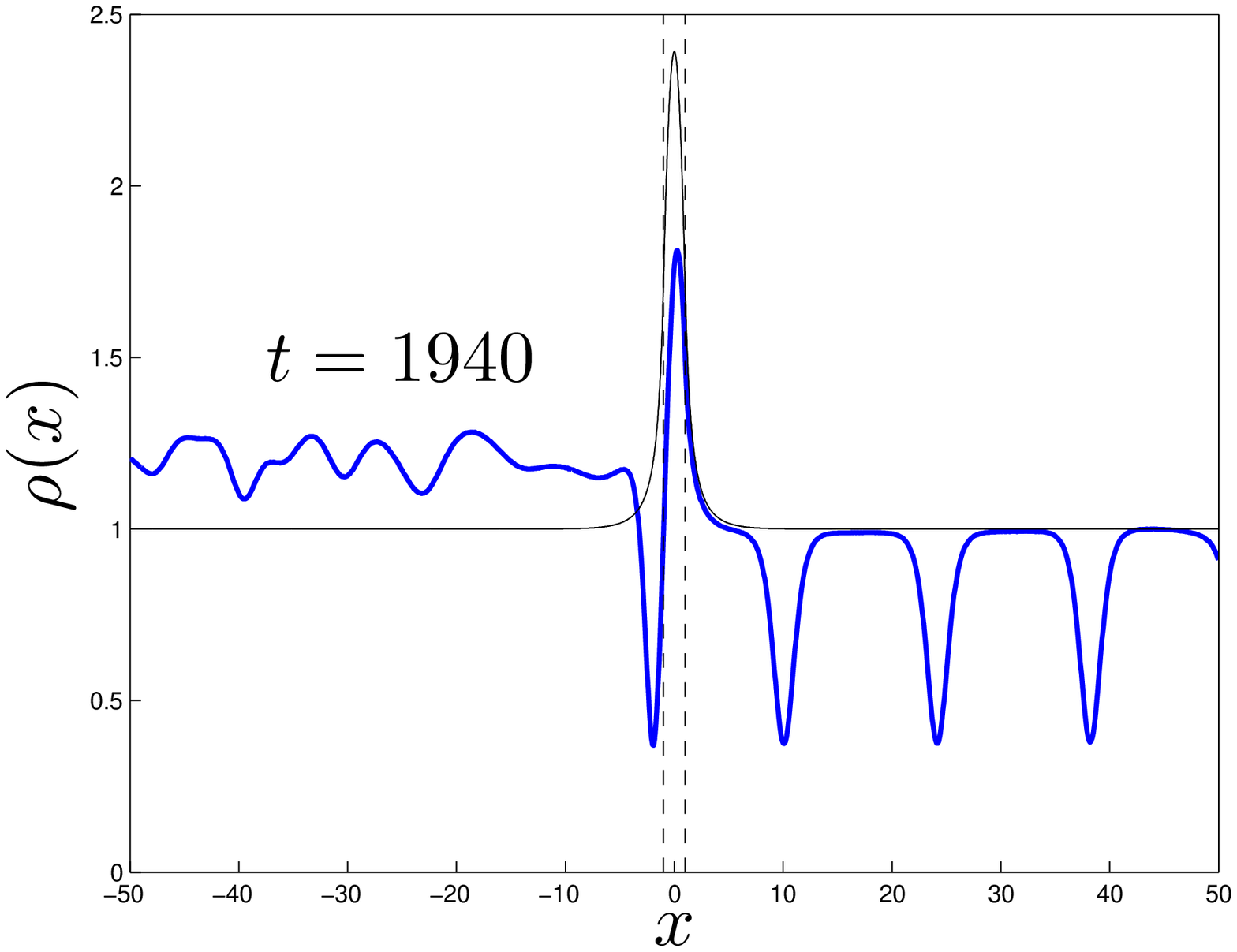} &
    \includegraphics[width=0.5\columnwidth]{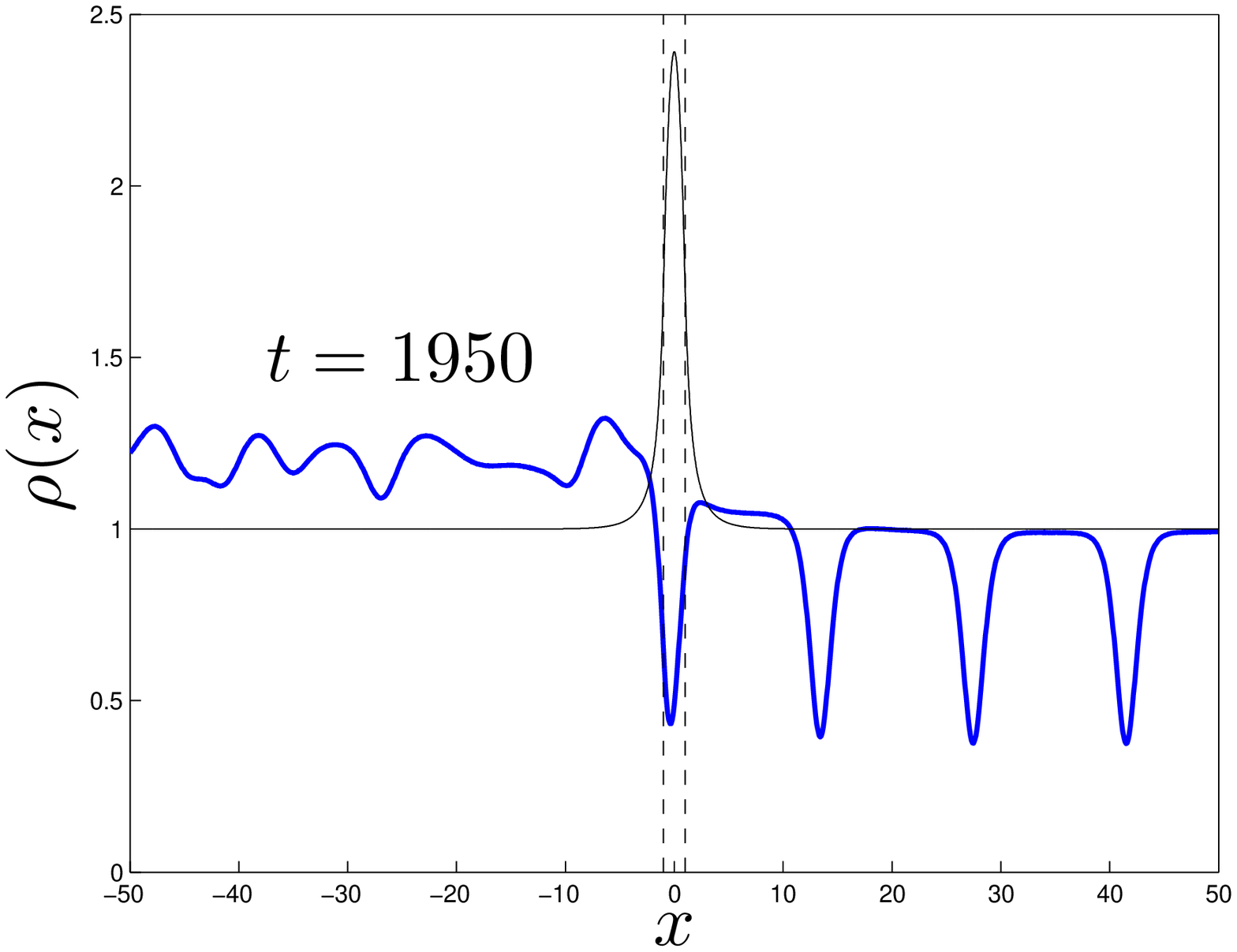} &
    \includegraphics[width=0.5\columnwidth]{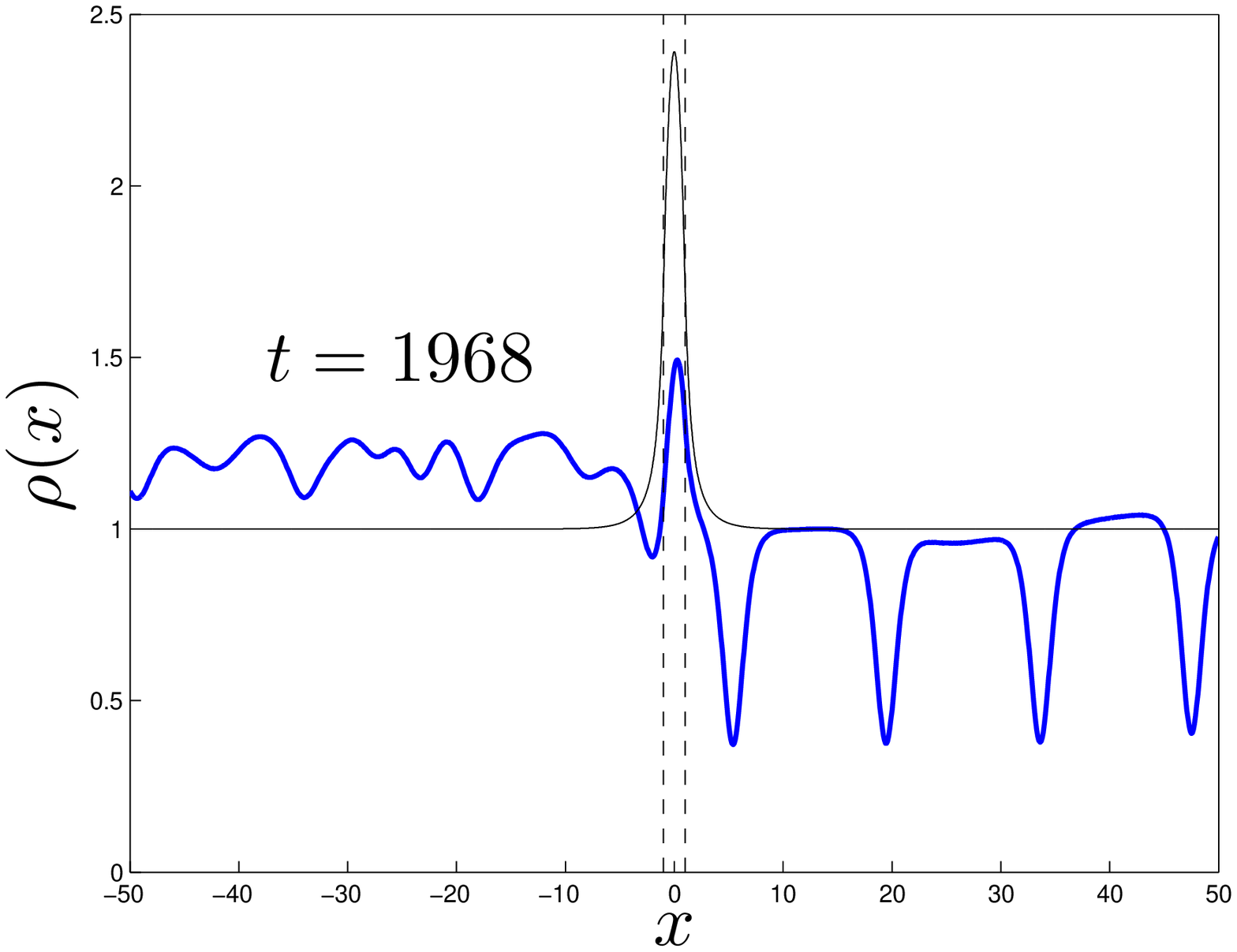} &
    \includegraphics[width=0.5\columnwidth]{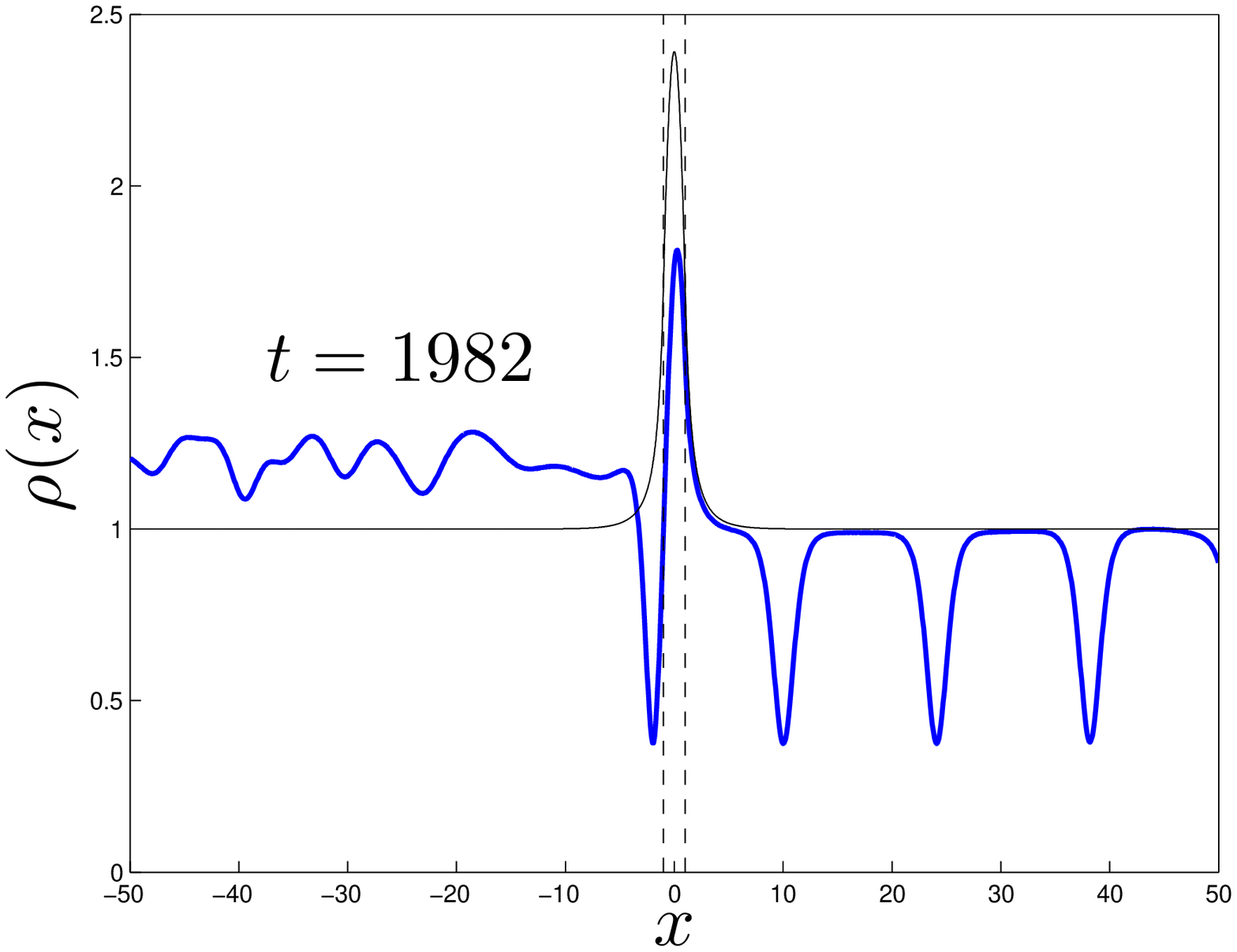}
\end{tabular}
\caption{Snapshots of the condensate density (solid blue line) at different evolution times (indicated in the panels) for a configuration with $v=0.95$, $c_2=0.4$ and $X=2$ that eventually reaches the CES regime. The horizontal black line represent the initial homogeneous condensate density. After some transient, the continuous emission of solitons begins. Upper row: initial evolution of the system when the initial linear instability sets in. The vertical black line marks the $x=0$ position. Lower row: periodic evolution at late times after the system has reached the CES regime. The vertical dashed lines marks the boundaries of the supersonic region. At $t=1940$, a soliton emerges near the black hole (subsonic-supersonic) interface at $x<0$. This soliton cannot travel upstream and then it bounces back, traveling to the downstream region and finally crossing the supersonic region, see plot at $t=1950$. After this soliton has left around $t=1968$, the density grows again and a new soliton starts growing near the interface. Finally, at $t=1982$, the system recovers the same state as at $t=1940$ and the process repeats. The complete time evolution is shown in \href{https://www.youtube.com/watch?v=NoOG_wJXKhYv}{Movie 7}.}
\label{fig:CESmechanism}
\end{figure*}

As done in the previous section, to understand the physics underlying the CES process we start by considering configurations with a single unstable mode where only the $n=0$ nonlinear stationary solution is present. Several snapshots of the corresponding time-evolution are shown in Fig. \ref{fig:CESmechanism} for a parameter choice that ends in the CES regime; the complete time evolution can be observed in \href{https://www.youtube.com/watch?v=NoOG_wJXKhYv}{Movie 7}. In the upper row of the figure, we represent the initial evolution of the system: the emission of solitons right after the onset of the initial instability. In the lower row, we analyze in detail the CES mechanism, which persists in a periodic way indefinitely for arbitrarily long times. The soliton that tries to be emitted in the upstream direction is dragged by the flowing condensate and bounces back. Eventually, it ends up in the downstream region and travels towards $x\to +\infty$. After the soliton has gone, the density modulation in the supersonic region begins to grow again until a new soliton is generated. Continuous periodic repetition of this process leads to the emission of a train of solitons into the downstream region.

\begin{figure}[!htb]
\includegraphics[width=\columnwidth]{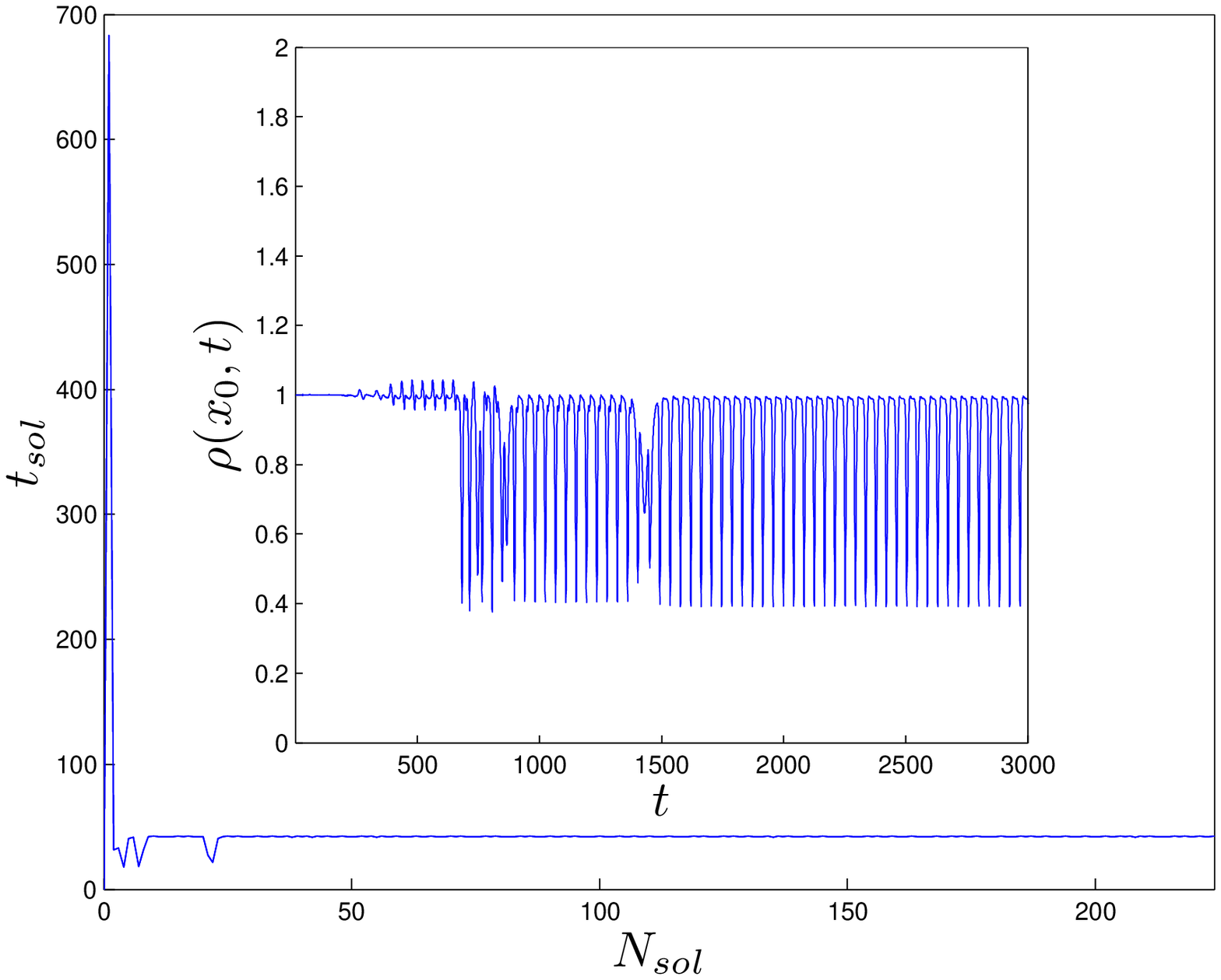}
\caption{Main panel: plot of $t_{sol}$, the time lapse between the arrival of consecutive solitons, as a function of the soliton number: after the emission of some solitons, the system reaches a regime of periodic soliton emission of period. Inset: time evolution of the density at a given point, $\rho(x_0,t)$ with $x_0=50$. The density minima correspond to the periodic passage of solitons. System parameters: $v=0.95$, $c_2=0.4$ and $X=2$.}
\label{fig:regularplot}
\end{figure}

In order to check quantitatively the periodicity of the soliton emission process, we introduce the quantity $t_{sol}(N)$, which is defined as the time lapse between the emission of the two consecutive $(N-1)$th soliton and $N$th soliton. If the emission of solitons is periodic, $t_{sol}$ should be constant and equal to the period of emission $T$, $t_{sol}=T$. For counting solitons, we monitor the density at a generic point $x_0\gg X/2$ in the downstream region as a function of time, as shown in the inset of Fig. \ref{fig:regularplot}. The minima of the density correspond to the passage of a soliton. In the main panel we represent the time-evolution of the time lapse $t_{sol}$ as a function of the number of emitted solitons, $N_{sol}$. After some irregular transient, we see that this quantity approaches a constant value $T$, meaning that the soliton emission process becomes an almost perfectly periodic one.

\begin{figure}[htb!]
\includegraphics[width=1\columnwidth]{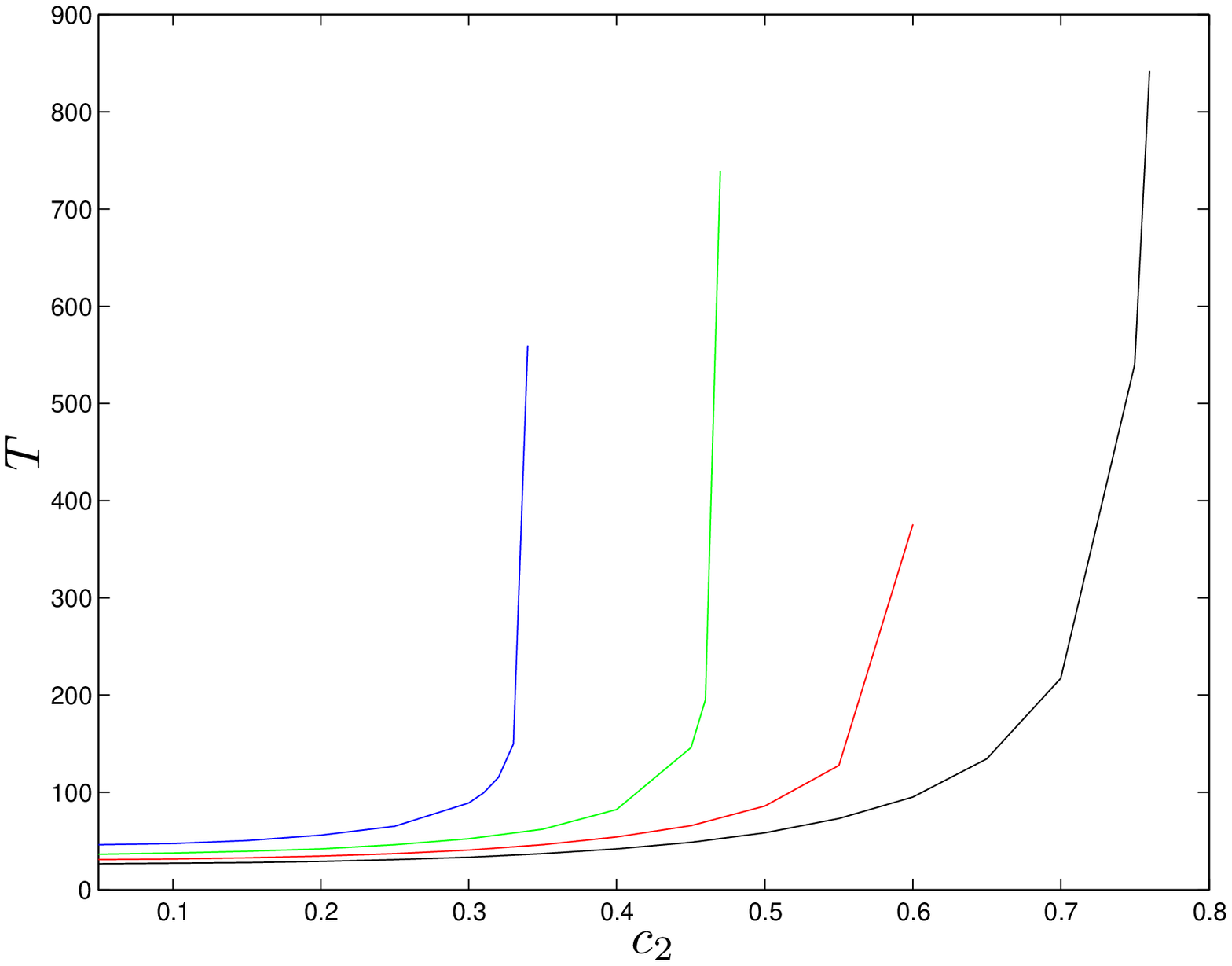}
\caption{Soliton emission period $T$ as a function of $c_2$ for (from top to bottom) $v=0.8,0.85,0.9,0.95$ and for fixed $X=2$.}
\label{fig:SolPeriod}
\end{figure}

Interesting light on the soliton emission process can be obtained by studying the dependence of the emission period $T$ on the system parameters. As an example, in Fig. \ref{fig:SolPeriod}, we plot the CES period as a function of $c_2$ for different values of $v$ for fixed $X=2$. Every curve stops at some critical value of $c_2$: after this critical value the system no longer performs a CES and a single soliton is in fact emitted to the upstream region while the system relaxes to the $n=0$ nonlinear solution. Below this critical value, the period increases with $c_2$ as expected since for larger values of $c_2$ the instability of the the supersonic region becomes weaker.  This physical interpretation also explains the decreasing value of $T$ for increasing $v$ seen in Fig. \ref{fig:SolPeriod}; this argument also agrees with the numerical observation of the decrease of $T$ for increasing $X$.

As a last ingredient before proposing a qualitative explanation of the CES mechanism, it is useful to remind a few crucial features of dark solitons in atomic condensates. As we have seen in the previous sections, dark solitons are often emitted into the upstream and/or downstream external regions to compensate for the increased density in the internal region. In the external region, the coupling constant is homogeneous $g(x)=1$ and the condensate flows at a constant speed $v$, so dark solitons are described by solutions of the GP equation of the form
\begin{equation}\label{eq:soliton}
\Psi(x)=e^{ivx}\left[ik+\sqrt{1-k^2}\tanh(\sqrt{1-k^2}\,x)\right],
\end{equation}
parametrized by a single real parameter $-1\leq k \leq 1$. The density profile shows a dip with a minimum density equal to $k^2$ and, apart from some irrelevant global time-dependent phase, it rigidly moves at a $k$-dependent velocity
\begin{equation}\label{eq:solitonvelocity}
v_s=v+k
\end{equation}
which results from the Galilean combination of an overall drag at the condensate speed $v$ and a relative motion at speed $k$ equal to the local value of the speed of sound at the density minimum: the deeper the soliton, the smaller its relative velocity with respect to the surrounding fluid.

A soliton is typically emitted into the upstream direction in order to conserve the total number of particles. We can expect that its minimum density $k^2$ is a roughly decreasing function of the amplitude of the non-linear stationary solution $n=0$. As the soliton moves in the upstream direction, the sign of $k$ is fixed to $k<0$. In particular, we have numerically observed that the $k$ parameter of the emitted soliton depends very weakly on the initial condition so we can consider it to be a function of the system parameters, $k=k(v,c_2,X)$. As the $k$ parameter fixes the relative velocity of the soliton with respect to the underlying fluid, a qualitative change of behavior is expected to occur around $v=|k(v,c_2,X)|$. For $v<|k(v,c_2,X)|$, the soliton can freely travel in the upstream direction. However, in the opposite case, $v>|k(v,c_2,X)|$, the flow speed $v$ is larger than the relative soliton velocity $|k|$, so the soliton cannot travel to the upstream region and is dragged by the condensate flow into the downstream direction, originating the CES mechanism previously described.

More insight on the transition to the CES regime can be obtained through Movies 8, 9 and 10. These movies report the result of numerical simulations for fixed $v=0.8,c_2=0.4$ and growing $X=2.2$ (\href{https://www.youtube.com/watch?v=HihIODRJ68k}{Movie 8}), $X=2.4$ (\href{https://www.youtube.com/watch?v=iGYMK9xftXc}{Movie 9}), and $X=2.5$ (\href{https://www.youtube.com/watch?v=pvlqOd1Vato}{Movie 10}). As $X$ is increased, the soliton must become deeper and deeper in order to allow the system to relax to the $n=0$ nonlinear stationary state, which implies a reduction of $k$. As a result, the soliton becomes slower and slower: while for $X=2.2$ and $X=2.4$ the dark soliton is able to form and escape to the upstream direction so to compensate for the higher density of the $n=0$ nonlinear stationary solutions, this can no longer happen for $X=2.5$, where we have in fact entered the $v>k$ regime and the soliton is forced to bounce back to the central region. As a result, it finally escapes in the downstream direction leaving the system again in an unstable configuration, so that the CES process keeps periodically repeating for indefinite times.

While this qualitative observation is in good agreement with the numerical results, it is far from offering a complete theoretical picture. In particular, finding a quantitative relation between the minimum amplitude of the soliton emitted upstream and the parameters of the system, i.e., the function $k(v,c_2,X)$, is still a theoretically unsettled problem. Obtaining and understanding a relation of this kind is an open question that should be addressed in future works.

The CES behavior in the  $X>X_1(v,c_2)$ regime with several unstable modes can be much more complex due to the presence of other $n>0$ nonlinear stationary solutions. Typically, the system emits several solitons before reaching the periodic CES regime described previously. We show this scenario in \href{https://www.youtube.com/watch?v=KnWXyr2Y9E4}{Movie 11}, which summarizes the result of a simulation for $v=0.65, c_2=0.3, X=8$. In other situations, the system first evolves towards a $n>0$ solution, remains in the vicinity of such solution for quite some time, then departs again from it and, after some transient, reaches the CES regime. We show an example of this latter case in \href{https://www.youtube.com/watch?v=Cj0v56KvReg}{Movie 12}: in this simulation, the system stays in the vicinity of the dynamically unstable $n=1$ solution for quite some time before entering in the CES regime. Finally, for a large number of unstable modes, the periodicity of the soliton emission may completely disappear, with the system strongly oscillating around different solutions with $n>0$. This chaotic scenario is illustrated in \href{https://www.youtube.com/watch?v=7NHbsb1JvVE}{Movie 13}, where the system shows a very complicated aperiodic emission of solitons.

\section{Discussion and comparison with optical lasers}\label{sec:LaserComparison}

On the basis of the results presented in the previous Sections, we are
now in a position to critically discuss the actual physical content of
the widely used expression {\em black-hole laser}.

Historically, when this expression was first introduced by Corley and
Jacobson~\cite{Corley1999}, the authors had in mind the linear dynamical instability of the
configuration with a pair of neighboring black- and white-hole
horizons. In terms of laser devices in quantum optics, this dynamical
instability corresponds to the linear instability of the electromagnetic vacuum state of the
cavity when the (unsaturated) gain exceeds losses. In this case, any weak signal due to,
e.g., spontaneous emission, is able to trigger the instability and quickly gets coherently amplified by stimulated
emission.

The actual operation of a continuous-wave laser device~\cite{QuantumNoise,QuantumOptics,Petruccione,MandelWolf} takes
place in a very different regime, where nonlinear gain saturation brings
the system to a nonlinear stationary state where (saturated) gain and losses
exactly compensate each other and the field keeps oscillating at a well-defined frequency with a well-defined amplitude.
This clean and coherent laser emission is attained at long times after switch-on once all (possibly complex) transient phenomena are gone: from the point of view of the field dynamics, the long-time limit is a limit cycle in the classical dynamics, and the only observable quantum effect are small fluctuations in the oscillation frequency, leading to a very long but finite coherence time of the emission: in the simplest cases, this slow decoherence of the laser emission goes under the name of Schawlow-Townes linewidth.

Even though from a quantum optics perspective the expression {\em black-hole laser} suggests a spontaneously oscillating behavior with no memory of the initial state and strongly determined by nonlinear effects, the complex nonlinear dynamics of analog models in the two-horizon configurations has began being investigated only very recently in~\cite{Michel2013,Michel2015}. Building on top of these works, we have seen in the previous sections that the long-time behavior of such configurations is much richer than the one of standard optical lasers.

In many cases (Sec.~\ref{subsec:longtimestat}), the system is in fact able to quickly find a new time-independent stationary state where the instability is quenched by a suitable redistribution of the atomic density which eliminates the supersonic character of the central region. During this rapid ``evaporation'' of the horizons, the extra energy and particle density are compensated by the emission of a short sequence of sound and soliton pulses in the condensate, whose properties strongly depend on the (classical or quantum) noise triggering the initial linear instability.

On the other hand, regimes (Sec.~\ref{subsec:CES}) where the density and the current keep oscillating in a periodic way in time can be considered as a hydrodynamic counterpart of a continuous-wave monochromatic optical laser oscillation, leading to a continuous and periodic emission of sound waves in the condensate in the form of a train of solitons. The excellent periodicity and insensitivity to initial conditions of this soliton train guarantees a very high degree of coherence of the emission, closely analogous to the one of the output beam of an optical laser device. On this basis, a rigorous reader may then argue that the {\em black-hole lasing} expression should only be used to refer to this latter case.

As a final remark, it is worth noting that in contrast to standard single-mode laser devices which emit light of a single frequency, the solitons emitted by the black-hole laser device are intrinsically nonlinear objects containing a number of frequency components: differently from pulsed lasers where the pulse-generating mechanism serves to modulate a much faster carrier frequency and creates tightly spaced sidebands, here the nonlinearity provides new frequencies at integer multiples of the fundamental one. In a qualitative picture, these components can be understood as resulting from a phase-locking of the different modes of the central, super-sonic region that are simultaneously oscillating. The energy for these oscillations is of course provided by the macroscopic flow of the condensate, and the mechanism responsible for the phase-locking of the different components originates from the hydrodynamic nonlinearities at large density modulations.

\section{Conclusions and outlook} \label{sec:conclu}

In this work we have reported the results of an extensive campaign of numerical simulations of the Gross-Pitaevskii equation describing the time-evolution of an atomic Bose-Einstein condensate. For the sake of conceptual simplicity, our numerical study was performed on a theoretically most convenient two-horizon configuration, which is theoretically convenient since it allows to suppress additional phenomena such as Bogoliubov-Cherenkov emission of sound waves. While this configuration may require a challenging experimental effort to simultaneously control in a spatially selective way both the atomic interaction constant and the external potential, we expect that the qualitative conclusions of our work extend to a wider range of black-hole laser configurations, e.g. the ones used for the experimental observation of the (linear) black hole instability in~\cite{Steinhauer2014}. Preliminary numerical work for this specific configuration~\cite{Tettamanti2016} suggests that, at least at short times, the black-hole lasing mechanism is taking place along very similar lines also in this case.

In agreement with existing theoretical and numerical results about linear instabilities and nonlinear stationary solutions ~\cite{Michel2013,Michel2015}, our simulations provide useful physical insight on the system behavior as a function of the initial system parameters. In some cases, the central unstable region is eventually evaporated away and the system at late times approaches a stationary stable configuration. In other cases, the system approaches a novel regime of continuous and periodic emission of solitons for indefinite times.
Interesting analogies and differences between the continuous soliton emission in black-hole lasers and the operation of an optical laser device can be drawn from our simulations. On this basis, a more restrictive and coherent use of the expression {\em black-hole laser} is proposed.

As a further remarkable result, our simulations highlight how the system typically forgets its initial state and, after possibly very different transients, the late time dynamics is only determined by the parameters of the configuration and not by the initial condition. This fact provides numerical evidence of the existence of a no-hair theorem also for analog black-hole lasers, in a similar fashion to the no-hair theorem for analog black holes \cite{Michel2015,Michel2016}.

While all our discussion has focussed on the classical black-hole laser dynamics, future work will address the effect of quantum fluctuations on these systems: taking again inspiration from quantum optics and laser theory, we may legitimately expect that quantum effects should reduce to weak fluctuations in the emission frequency and, therefore, to a slow decoherence of the soliton emission analogous to the long, but finite coherence time of optical laser devices~\cite{QuantumNoise,QuantumOptics,Petruccione,MandelWolf}.

From the experimental point of view, we can reasonably expect that an upgraded version of the ultracold atom experiment in~\cite{Steinhauer2014} with a longer condensate sample will soon be able to investigate the nonlinear dynamics of the black-hole laser at late times after the onset of the dynamical instability and hopefully characterize the rich phenomenology discussed in this work.

Even though all the discussion in this work has been carried out having an atomic implementation in mind, of course we expect that most conclusions can be directly transferred to analog models based on fluids of light in the so-called propagating geometry, for which theoretical works on black hole configurations have recently appeared~\cite{FleurovEPL2011,CarusottoPRSA2014} as well as first experimental evidences of superfluid behaviors~\cite{Vocke_Optica2015}.

Apart from analog gravitational purposes, the black-hole laser setup here described and, in particular, the characterized non-linear soliton laser regime, can provide an interesting scenario for the investigation of quantum transport, within the emergent field of atomtronics \cite{Seaman2007,Labouvie2015}.

\acknowledgments
We thank F. Michel, R. Parentani, J. Steinhauer, F. Sols and R. Balbinot for fruitful and stimulating discussions. We also thank F. Sols and I. Zapata for valuable comments on the manuscript. This work has been supported by MINECO (Spain) through grants FIS2010-21372 and FIS2013-41716, Comunidad de Madrid through grant MICROSERES-CM (S2009/TIC-1476). IC acknowledges financial support by the ERC through the QGBE grant, by the EU-FET Proactive grant AQuS, Project No. 640800, and by the Autonomous Province of Trento, partially through the project ``On silicon chip quantum optics for quantum computing and secure communications'' (``SiQuro'').

\appendix

\section{Numerical scheme}\label{app:NumericalScheme}

The GP equation, Eq. (\ref{eq:GPNumerical}), is integrated making use of a time-splitting finite-difference (TSFD) scheme \cite{Hua2012,Bao2013}. For this purpose, we write the GP equation as
\begin{eqnarray}\label{eq:GPsplitting}
\nonumber i\frac{\partial \Psi(x,t)}{\partial t}&=&H_{GP}(x,t)\Psi(x,t)\\
&=&\left[H_0(x,t)+H_C(x,t)\right]\Psi(x,t)
\end{eqnarray}
where $H_0$ is the kinetic term and $H_C=g(x)(|\Psi(x,t)|^2-1)$ is the term associated with the contact interaction. We divide the spatial interval $[-L_g/2,L_g/2]$ in steps of size $\Delta x$ and impose periodic boundary conditions. The time interval $[0,t]$ is divided in $n$ steps of size $\Delta t$ . Within this scheme, the time evolution operator for time $t_k=k\Delta t$ to $t_{k+1}=(k+1)\Delta t$ can be approximated as:
\begin{equation}\label{eq:TSFD}
U_k\simeq e^{-iH_0\frac{\Delta t}{2}}e^{-iH^k_C\Delta t}e^{-iH_0\frac{\Delta t}{2}}
\end{equation}
where $H^k_C=g(x)(|\tilde{\Psi}(x,t_k)|^2-1)$, $\tilde{\Psi}(x,t_k)=e^{-iH_0\frac{\Delta t}{2}}\Psi(x,t_k)$. The total evolution operator can be then written in the form:
\begin{widetext}
\begin{eqnarray}\label{eq:TimeOperator}
\Psi(x,t)&=&U(t)\Psi(x,0)\\
\nonumber U(t)&=&e^{-iH_0\frac{\Delta t}{2}}e^{-iH^{n-1}_C\Delta t}\left(\prod_{k=0}^{n-2}e^{-iH_0\Delta t}e^{-iH^k_C\Delta t}\right)e^{-iH_0\frac{\Delta t}{2}}
\end{eqnarray}
\end{widetext}
In the TSFD method, instead of using the usual Fourier transform to compute the kinetic evolution operator $e^{-iH_0\Delta t}$, we use the Crank-Nicolson method within a finite difference scheme \cite{Ames1977,Press2007,PhysRevA.76.063605,deNova2014a}. The main advantage of using the Crank-Nicolson method is that it is unconditionally unstable, which allows us to use higher values of $\Delta x, \Delta t$.

In addition, such a method keeps working even when the kinetic term involves derivatives with a spatially dependent coefficient. This feature will be crucial to implement the absorbing mechanism needed to suppress artifacts in long-time simulations: our strategy to this purpose is to add a diffusive term on the boundaries of the integration box so to absorb all emitted perturbations and avoid their return to the central region of interest. Specifically, we replace $H_{GP}$ by $H_T=H_{GP}+iH_A$ in Eq. (\ref{eq:GPsplitting}), where the absorbing term $H_A$ has the form:
\begin{equation}\label{eq:diffusion}
H_A(x,t)=G(x)\left(D_0\partial_{x}^{2}e^{-ivx}-F_0[|\Psi(x,t)|^2-1]\right)
\end{equation}
in terms of a function $G(x)$ localized on the boundary of the grid and sufficiently smooth in order to avoid reflections. In particular, this term vanishes when acting on the unperturbed homogeneous plane wave $e^{ivx}$. Within $H_A$, the first term between the brackets of the r.h.s is a diffusive term that absorbs the Fourier components outside the plane wave $e^{ivx}$ while the second term is a source term that serves to compensate the undesired loss of atoms introduced by the diffusive term.

The new diffusive GP equation can still be integrated using the mentioned TSFD method by just replacing $H_0$ and $H_C$ in Eqs. (\ref{eq:GPsplitting})-(\ref{eq:TimeOperator}) by $H_D$ and $H_X$, where $H_D$ is the term that contains the derivatives (the kinetic and diffusive terms) and $H_X$ is the term that takes into account the non-linearities (the interaction and the source term), i.e.,
\begin{eqnarray}\label{eq:DiffusiveOperators}
H_D&=&-\frac{1}{2}\frac{\partial^2}{\partial x^2}+iD_0G(x)\frac{\partial^2}{\partial x^2}e^{-ivx}\\
\nonumber H_X&=&[g(x)-iF_0G(x)][|\Psi(x,t)|^2-1]
\end{eqnarray}

As the source term does not conserves the norm, it is useful to refine the simple scheme discussed above by introducing a predictor-corrector method in order to provide a better estimation for the non-linear term in $H_X$, similar to that of Refs. \cite{PhysRevA.76.063605,deNova2014a}. At every step, we perform a first iteration to compute $\tilde{\Psi}'(x,t_k)=e^{-iH^k_X\Delta t}\tilde{\Psi}(x,t_k)$ where $\tilde{\Psi}(x,t_k)$ is now $\tilde{\Psi}(x,t_k)=e^{-iH_D\frac{\Delta t}{2}}\Psi(x,t_k)$ and $H^k_X=[g(x)-iF_0G(x)](|\tilde{\Psi}(x,t_k)|^2-1)$. After this first iteration, we replace $\tilde{\Psi}(x,t_k)$ by $\left[\tilde{\Psi}'(x,t_k)+\tilde{\Psi}(x,t_k)\right]/2$ in $H^k_X$ and we perform a second iteration to obtain the final value of $\tilde{\Psi}'(x,t_k)$.

Typically we take for $G(x)$ a Gaussian shape with a typical width on the order of $10^2$ and values in the range $1-10$ for $F_0$ and $D_0$. With parameter choices in such intervals, we have not seen any significant variation in the result of the simulations. As a further check, we have verified that the diffusive scheme does not introduce any spurious result by comparing with the result of integrating the plain GP equation without diffusion but on a much larger spatial grid.

It is worth noting that such a refined treatment of the boundaries is essential for the study of the CES regime of Sec. \ref{subsec:CES}. Indeed, the observation of the perfect periodicity of the soliton emission is in fact sensitive to small spurious reflections at the boundaries. As in all other cases, such artifacts are of course further suppressed with a sufficiently wide integration box.

\section{Movies}\label{app:Movies}

In this Appendix, we summarize the system parameters used for the Movies that are discussed in this work. In all the videos, the upper panel displays the time evolution of the density profile (solid blue line) while the lower profile shows that of the sound (solid blue) and flow (red dashed) velocities. The limits of the internal region are marked by two vertical dashed black lines.

\begin{itemize}
 \item \href{https://www.youtube.com/watch?v=gbt3tL9haD4}{Movie 1}: $v=0.75$, $c_2=0.3$ and $X=2$ satisfying $X_0(v,c_2)<X<X_{1/2}(v,c_2)$. The unstable mode initially grows in the same direction of the $n=0$ nonlinear stationary solution (solid black line in the upper panel).
 \item \href{https://www.youtube.com/watch?v=2wPY_D7Aou8}{Movie 2}: Same parameters as Movie 1 but now the unstable mode initially grows in the opposite direction.
 \item \href{https://www.youtube.com/watch?v=xI6LE5FXLnk}{Movie 3}: $v=0.75,c_2=0.5,X=5$, with $X_{1/2}(v,c_2)<X<X_1(v,c_2)$. We observe the oscillating behavior during the growth of the unstable mode, eventually reaching the $n=0$ ground state.
 \item \href{https://www.youtube.com/watch?v=kX0z8AZCzNk}{Movie 4}: $v=0.75,c_2=0.6,X=10$, with $X_{1}(v,c_2)<X<X_{3/2}(v,c_2)$. The unstable mode grows towards the $n=1$ stationary solution. Since that solution is dynamically unstable, the system stays in its vicinity for a large but finite time, then it departs from it to finally reach the $n=0$ ground state (solid black line in the upper panel).
 \item \href{https://www.youtube.com/watch?v=hqcxajEjDoE}{Movie 5}: Same parameters as Movie 4 but the initial unstable mode grows now in the opposite direction. The system emits some waves and solitons and then reaches the $n=0$ ground state in a shorter time.
 \item \href{https://www.youtube.com/watch?v=OfUsjMDDMJU}{Movie 6}: $v=0.9,c_2=0.75,X=20$, with $X_{5/2}(v,c_2)<X<X_3(v,c_2)$. After some transient, the system reaches a configuration near the $n=1$ stationary solution (black line in the upper panel), around which the system is still oscillating at the end of the simulation ($t=10000$).
 \item \href{https://www.youtube.com/watch?v=NoOG_wJXKhYv}{Movie 7}: $v=0.95,c_2=0.4,X=2$, with $X_{0}(v,c_2)<X<X_{1/2}(v,c_2)$. Instead of evolving to the ground state $n=0$ (solid black line in the upper panel), the system reaches a regime of continuous emission of solitons. A slow-motion effect has been introduced at large times in order to make easier the observation of the mechanism of periodic emission of solitons.
 \item \href{https://www.youtube.com/watch?v=HihIODRJ68k}{Movie 8}: $v=0.8,c_2=0.4,X=2.2$, with $X_{0}(v,c_2)<X<X_{1/2}(v,c_2)$. We observe that the system emits a soliton to the upstream region and reaches the ground state $n=0$ solution (solid black line in the upper panel).
 \item \href{https://www.youtube.com/watch?v=iGYMK9xftXc}{Movie 9}: Same parameters as Movie 8, but with $X=2.4$. The soliton is still able to travel in the upstream direction.
 \item \href{https://www.youtube.com/watch?v=pvlqOd1Vato}{Movie 10}:  Same parameters as Movie 8, but with $X=2.5$. In this case, the upstream traveling soliton bounces back and the system reaches the CES regime.
 \item \href{https://www.youtube.com/watch?v=KnWXyr2Y9E4}{Movie 11}: $v=0.65,c_2=0.3,X=8$, with $X_{1}(v,c_2)<X<X_{3/2}(v,c_2)$. Even when more than one unstable mode is present, after some transient, the system reaches the same CES regime as before.
 \item \href{https://www.youtube.com/watch?v=Cj0v56KvReg}{Movie 12}: $v=0.65,c_2=0.1,X=8$, with $X_{5}(v,c_2)<X<X_{11/2}(v,c_2)$. The system evolves towards the $n=1$ non-linear solution (black line in the upper panel). As this solution is dynamically unstable, after some time the system departs from it and eventually reaches the CES regime.
 \item \href{https://www.youtube.com/watch?v=7NHbsb1JvVE}{Movie 13}: $v=0.9,c_2=0.2,X=20$, with $X_{5}(v,c_2)<X<X_{11/2}(v,c_2)$. Due to the high number of unstable modes, the system is continuously emitting solitons but in a chaotic way, in contrast to the usual periodic CES regime. The non-linear $n=5$ stationary solution is depicted as a solid black line in the upper panel.
\end{itemize}

\bibliographystyle{apsrev}
\bibliography{CES}

\end{document}